\documentclass[12pt]{article}

\pdfoutput=1

\usepackage{graphicx}
\usepackage{xspace}
\usepackage{epsfig}
\usepackage{amssymb,amsmath}
\usepackage{epstopdf}
\def\be{\begin{equation}}
\def\ee{\end{equation}}
\def\bea{\begin{eqnarray}}
\def\eea{\end{eqnarray}}
\def\beal{\begin{align}}
\def\eeal{\end{align}}

\def\Tr{\mathop{\rm Tr}}

\def\hc{\ {\rm h.c}}
\def\nn{\nonumber}
\def\del{\partial}
\def\half{\frac{1}{2}}
\def\lam{\lambda}
\def\cb{c_{\beta}}
\def\sb{s_{\beta}}
\def\tb{t_{\beta}}
\def\ca{c_{\alpha}}
\def\sa{s_{\alpha}}

\def\cg{c_{\gamma}}
\def\sg{s_{\gamma}}

\newcommand{\Kahler}{\textrm{K\"{a}hler}~}

\newcommand{\CV}{\rm \mbox{\hspace{-1.2mm} \slash{\hspace{-2.1mm}c}}}
\newcommand{\CP}{\rm \mbox{c}}

\relax

\begin{document}
\topmargin-2.5cm
%
\begin{titlepage}
\begin{flushright}
\normalsize{  
FERMILAB-PUB-09-419-T\\
SLAC-PUB-13784  
  }  
\end{flushright}
\vskip 0.3in
\begin{center}{\Large\bf
Supersymmetric Higgs Bosons and Beyond}

\vskip .5in
{\bf Marcela Carena}$^{a,b}$,
{\bf Kyoungchul Kong}$^{a,c}$,
{\bf Eduardo Pont\'{o}n}$^{d}$,
{\bf Jos\'e Zurita}$^{a,e}$
\vskip.5in

$^{a}$ 
{\em Theoretical Physics Department, Fermilab, Batavia, IL 60510, USA}\\
$^{b}$ 
{\em Enrico Fermi Institute, Univ. of Chicago, 5640 Ellis Ave., Chicago, IL 60637, USA}\\
$^{c}$ 
{\em Theoretical Physics Department, SLAC, Menlo Park, CA 94025, USA} \\
$^{d}$ 
{\em Department of Physics, Columbia University,\\
538 W. 120th St, New York, NY 10027, USA}\\
$^{e}$ 
{\em Departamento de F\'isica, Universidad de Buenos Aires,\\
(1428) Pabell\'on 1 Ciudad Universitaria, Capital Federal, Argentina} \\
\end{center}
\vskip.5cm
\begin{center}
{\bf Abstract}
\end{center}
\begin{quote}
We consider supersymmetric models that include particles beyond the
Minimal Supersymmetric Standard Model (MSSM) with masses in the TeV
range, and that couple significantly to the MSSM Higgs sector.  We
perform a model-independent analysis of the spectrum and couplings of
the MSSM Higgs fields, based on an effective theory of the MSSM
degrees of freedom.  The \textit{tree-level} mass of the lightest
CP-even state can easily be above the LEP bound of $114~{\rm GeV}$,
thus allowing for a relatively light spectrum of superpartners,
restricted only by direct searches.  The Higgs spectrum and couplings
can be significantly modified compared to the MSSM ones, often
allowing for interesting new decay modes.  We also observe that the
gluon fusion production cross section of the SM-like Higgs can be
enhanced with respect to both the Standard Model and the MSSM.
\end{quote}
\vskip1.cm
\end{titlepage}
\setcounter{footnote}{0}
\setcounter{page}{1}
\newpage
%

\baselineskip18pt    

\noindent
\section{Introduction}
\label{sec:intro}

Supersymmetry (SUSY) offers an elegant solution to the hierarchy
problem, can explain electroweak symmetry breaking (EWSB) dynamically
through renormalization group running and, if established
experimentally, could open a window into the physics associated with
length scales much shorter than can be probed directly.

The most studied SUSY extension is the minimal supersymmetric standard
model (MSSM).  In the Higgs sector, the model incorporates two Higgs
doublets, $H_{u}$ and $H_{d}$.  After EWSB, 5 physical scalars remain
in the spectrum~\cite{Gunion:1984yn}.  Assuming no CP violation, these
can be classified as: two neutral CP-even scalars ($h^{0}$ the
lightest, $H^{0}$ the heaviest), one neutral CP-odd scalar ($A^{0}$),
and a charged Higgs pair ($H^{\pm}$).  The phenomenology in the Higgs
sector is largely determined by the masses of these particles, by a
mixing angle $\alpha$ that governs the relation between gauge and
mass eigenstates in the CP-even sector, and by the ratio of the Higgs
VEV's, $\tan\beta = v_{u}/v_{d}$.  In the MSSM only two of these
parameters are independent, and are conventionally chosen as $m_{A}$
and $\tan\beta$.  Dependence on other sectors of the theory enters
through radiative corrections, most notably in the mass of the
lightest CP-even scalar, $h^0$.  This scalar is found to be below
about $130~{\rm GeV}$~\cite{hmass}, and together with the direct
bounds imposed by LEP suggests the presence of some degree of
fine-tuning in the MSSM.

This has been one motivation to study extensions of the MSSM that can
either relax the upper bound on the lightest Higgs
mass~\cite{raising,SU2}, or alter its properties in a way that weakens
the LEP bounds~\cite{hiding}.  Often the extensions considered aim at
addressing theoretical issues such as the $\mu$-problem, i.e. how to
link the supersymmetric Higgs mass parameter to the scale of EWSB,
which in turn is set by the scale of SUSY breaking in the observable
sector.  Further theoretical constraints are also often imposed, such
as the requirement of perturbativity up to a very high scale, the
preservation of gauge coupling unification, or various simplifying
assumptions that allow to more easily constrain the low-energy
parameters.  However, when exploring the Higgs collider phenomenology,
it may be healthy to keep an open mind regarding such theoretical
assumptions.

If the constraints on the MSSM are to be relaxed, it is necessary to
introduce new degrees of freedom (that interact with the MSSM Higgs
sector) at or near the weak scale.  It has been observed that even if
the new particles in the Higgs sector are slightly heavier than the EW
scale, the lightest Higgs boson mass may receive important
contributions that can relax the LEP constraints~\cite{Dine:2007xi}
(see also \cite{Babu:2008ge} for examples with TeV scale vector-like
matter).  The new degrees of freedom may or may not be directly
accessible at the LHC, but in either case their presence could
potentially be inferred by studying the spectrum and couplings of the
lighter states.  In particular, the observation of a SM-like Higgs
with a mass significantly above $130~{\rm GeV}$, together with the
observation of other superpartners, would provide a clear hint that
the Higgs sector is more complicated than in the MSSM.

We concentrate on supersymmetric scenarios with particles beyond those
in the MSSM, under the assumption that they have order one couplings
to the MSSM Higgs sector, and that they are heavier than, but close
to, the weak scale.  This allows to perform a model-independent
analysis of the properties of the lighter states (i.e. those of the
MSSM) by encoding the effects of the heavy physics via
higher-dimension operators.  As pointed out in
\cite{Dine:2007xi},\footnote{The earlier Ref.~\cite{Brignole:2003cm}
also considered in detail the effects of higher-dimension operators on
the MSSM Higgs sector in the context of low-scale supersymmetry
breaking mediation.  They also considered effects similar to those we
include below, although they restrict to a study of the renormalizable
part of the scalar potential.} at leading order in $1/M$ (where $M$ is
the scale of the heavy physics) only two new parameters are introduced
(corresponding to two operators in the superpotential).  Therefore,
even if the MSSM extension turned out to include a large number of
degrees of freedom, their low-energy effects admit a rather simple
parametrization.  Furthermore, it was found that the $1/M$ effects can
give rather important contributions to the mass of the lightest Higgs
state.  That the effects of such $1/M$-suppressed operators can be as
important as those of the renormalizable terms can be understood by
considering the structure of the Higgs quartic couplings.  In terms of
the general two-Higgs doublet model (2HDM) parametrization of
\cite{Haber:1993an} (see Eq.~(\ref{eq:pot}) below), it is well known
that of the seven independent quartic couplings only $\lambda_{1}$,
$\lambda_{2}$, $\lambda_{3}$ and $\lambda_{4}$ receive contributions
in the MSSM, at tree-level.  The leading order higher-dimension
operators contribute to $\lambda_{5}$, $\lambda_{6}$ and
$\lambda_{7}$, so that they lead to qualitatively new effects in the
Higgs sector.  This is the same underlying reason that loop effects in
the MSSM, which turn on all possible quartic couplings, can give
sizable effects.  Thus, the presence of the heavy physics allows a
Higgs spectrum consistent with the LEP bounds, even if the SUSY
breaking terms (in the top-stop sector) are of order a couple hundred
GeV, thus alleviating the tensions found within the MSSM.

Working still at leading order in $1/M$, one finds not only
contributions to the quartic Higgs couplings, but also higher
dimension operators in the Higgs potential.  These are essential in
bounding the scalar potential from below.  It was observed in
\cite{Batra:2008rc}, that taking these operators into account leads to
the existence of new vacua that can be studied within the above
effective field theory (EFT), and that these vacua have distinct
properties.  For instance, electroweak symmetry breaking is not
necessarilly controlled by supersymmetry breaking but, unlike in the
MSSM, can occur already in the supersymmetric limit (this possibility
was considered in the early days of supersymmetry~\cite{oldsEWSBrefs},
though not in light of the EFT approach.  See also Appendix~B of
Ref.~\cite{Haber:1984rc}, and more recently~\cite{Harnik:2003rs}).
These were dubbed ``sEWSB vacua'' (or supersymmetric EWSB vacua)
in~\cite{Batra:2008rc}.  Typical features of the Higgs physics in the
sEWSB vacua are order one $\tan\beta$, a heavy CP-even Higgs ($H^0$)
with SM-like properties, and relatively light charginos and
neutralinos.

In this work, we further consider the next order in the $1/M$
expansion.  On the one hand these lead to the ``first order
corrections'' to the physics in sEWSB vacua~\cite{Batra:2008rc}.
Second, even in MSSM-like vacua their effects can be
phenomenologically relevant.  This can again be understood by
referring to the quartic couplings $\lambda_{1,2,3,4}$, whose size is
set at leading order by the electroweak (EW) gauge couplings squared,
hence are numerically small (the source of the lightness of the
SM-like Higgs within the MSSM).  The ${\cal O}(1/M)$ operators from
the superpotential do not contribute to these couplings, and therefore
the ${\cal O}(1/M^2)$ effects become the leading order contributions
from the heavy physics to the corresponding quartic operators in the
scalar potential.  Furthermore, up to numerical factors, the heavy
physics gives a contribution of order $v^2/M^2$ which is comparable to
the MSSM one for $M \sim v/g$, with $g$ an EW gauge coupling and $v$ a
Lagrangian parameter of order the EW scale.\footnote{Note that sizable
effects at order $1/M^2$ do not necessarily signal a breakdown of the
$1/M$ expansion.  In contrast, large effects at order $1/M^3$ in
general signal such a breakdown.  We take care to study parameter
points where the next order effects are expected to be small by a
simple criterion described in Subsection~\ref{EFTValidity}.} At second
order in the $1/M$ expansion several new operators appear.  Within a
given UV completion, the coefficients of these operators may or may
not be related to the coefficients of the leading order operators
(depending on the complexity of the MSSM extension).  Here we do not
impose any correlations, but vary the coefficients of the
higher-dimension operators (in the superpotential and \Kahler
potential) independently.  Our purpose is to survey the possible
signatures in the Higgs sector, which could suggest the presence of an
extended sector that might be more difficult to observe directly (for
instance if it consists of SM singlets).

Having developed the relevant formalism, we start in this paper a
study of the Higgs collider phenomenology.  We contrast the observed
features against both the SM and MSSM. For instance, we observe a
general enhancement of the gluon fusion production cross section of
the SM-like Higgs, which is interesting at hadron colliders.  Also
noteworthy is the fact that the Higgs spectrum can be altered
sufficiently to allow for new decay modes with rather significant
branching fractions.  Here we comment only on some of the possible
signals, and defer a more complete study of the Higgs collider
phenomenology to~\cite{CKPZ}.

Higher-dimension operators in the SUSY context were considered
in~\cite{Strumia:1999jm,Brignole:2003cm}, and a more complete
classification was presented in~\cite{Antoniadis:2007xc}, where field
redefinitions were used to reduce the number of independent operators.
The issue of the stability of the Higgs potential for MSSM-like minima
(as opposed to sEWSB minima~\cite{Batra:2008rc}) was considered
in~\cite{Blum:2009na}, while the implications for fine-tuning in such
scenarios was considered in~\cite{Casas:2003jx,Cassel:2009ps}.
Higher-dimension operators can also have interesting consequences for
the dark matter relic density~\cite{Cheung:2009qk,Berg:2009mq} as well
as for cosmology~\cite{Bernal:2009hd} and EW
baryogenesis~\cite{Grojean:2004xa,Bodeker:2004ws,Delaunay:2007wb,Noble:2007kk,Grinstein:2008qi}.

This paper is organized as follows.  In Section \ref{sec:extended}, we
define the effective theory to be studied and work out the general
expressions for the masses and couplings of the light Higgs degrees of
freedom to order $1/M^2$.  In Section~\ref{sec:analres} we give simple
analytic formulae in certain limits that allow to understand the
qualitative features of the effective theory.  In
Section~\ref{sec:definitions} we present the strategy to be used in
the numerical study, to be undertaken in Section~\ref{sec:results}.
There we comment on a selected number of features.  A more complete
study of the collider phenomenology will be presented in~\cite{CKPZ}.
We conclude in Section~\ref{sec:conclu}.  In
Appendix~\ref{app:UVcompletions}, we consider several possible UV
completions that illustrate how the higher-dimension operators
in the EFT can arise.  In Appendices~\ref{app:custodial}, \ref{sec:inos}
and \ref{app:couplings} we comment on custodial symmetry violation,
give the chargino/neutralino mass matrices, and collect several Higgs
trilinear couplings, respectively.

\section{Extended Supersymmetric Higgs Sectors}
\label{sec:extended}

We start by setting up the framework.  Our point of departure is a
generic supersymmetric theory with an extended Higgs sector, but where
the degrees of freedom beyond those in the MSSM have masses of order
$M$, assumed to be slightly larger than the EW scale.  In this case, a
model-independent effective field theory analysis is useful.  We
further assume that all supersymmetry breaking parameters are of order
a couple hundred GeV, so that the heavy spectrum is approximately
supersymmetric, with masses of order $M$.  In this case, it is useful
to write the effective theory in supersymmetric notation, keeping
track of supersymmetry breaking effects via a spurion superfield that
gets a VEV in its $F$ component (we do not consider $D$-term breaking
here).  In this section, we define the effective theory to be studied
--which describes explicitly only the MSSM Higgs degrees of freedom--
and work out the Higgs spectrum and their couplings.

\subsection{Generalized SUSY Two-Higgs Doublet Model}
\label{sec:SUSY2HDM}

The effects of heavy particles on the physical properties of the MSSM
Higgs fields can be described by a tower of higher-dimension operators
suppressed by powers of $M$.  Our ultimate goal is to study the
associated collider phenomenology, and we start by working out the
spectrum and couplings of the light states (light compared to $M$).
For the reasons spelled out in the introduction, we work up to
next-to-leading-order in the $1/M$ expansion.  Next-to-next-to-leading
order contributions are expected to be small, provided the $1/M$
expansion converges, a point we address in
Subsection~\ref{EFTValidity}.

At leading order in $1/M$, the superpotential reads
\be\label{genw}
W= \mu H_u H_d + \frac{\omega_{1}}{2M}  (H_u H_d)^2~,
\ee
where $H_u H_d = H_u^{0} H_d^{0} - H_u^+ H_d^-$ and $\omega_{1}$ is a
dimensionless parameter that we assume to be of order one.  Soft
supersymmetry breaking can be parametrized via a spurion superfield
$X=m_s \theta^2$, where $m_s$ sets the scale of SUSY breaking (note
that we choose $X$ to be dimensionless).  Each operator in the
superpotential (or \Kahler potential) leads to an associated SUSY
breaking operator through multiplication by the spurion, $X$.  We will
assume here that the coefficients of the SUSY breaking operators are
proportional to those of the corresponding supersymmetric terms.
Besides the $B\mu$ term, at leading order in $1/M$ one has
\be\label{WSpurion}
W_{\rm spurion}= \alpha_1  \frac{\omega_{1}}{2M} X (H_u H_d)^2
\ee
for a dimensionless parameter $\alpha_1$, also taken to be of order
one.  This term leads directly to a quartic interaction among the
Higgs scalar fields.

At order $1/M^2$ there are no operators in the superpotential, but
several operators enter through the \Kahler potential:
\bea
\label{genk}
K&=& H_d^{\dagger} \, e^{2V} H_d  + H_u^{\dagger} \, e^{2V} H_u + \Delta K^{\CV} + \Delta K^{\CP}~,
\eea
where
\bea
\Delta K^{\CV} &=& \frac{c_1}{2|M|^2} (H_d^{\dagger} e^{2V} H_d)^2   + \frac{c_2}{2|M|^2} (H_u^{\dagger} e^{2V} H_u)^2 + 
\frac{c_3}{|M|^2} (H_u^{\dagger} e^{2V} H_u) (H_d^{\dagger} e^{2V} H_d)~,
\label{DeltaKCV}
\\ [0.5em]
\Delta K^{\CP} &=& \frac{c_4}{|M|^2} |H_u H_d|^2 + \left[ \frac{c_6}{|M|^2} H_d^{\dagger} \, e^{2V} H_d   + 
\frac{c_7}{|M|^2} H_u^{\dagger} \, e^{2V} H_u \right] (H_u H_d)  + {\rm h.c.}
\label{DeltaKCP}
\eea
We separated the higher-dimension contributions in the \Kahler
potential into those that violate the custodial symmetry, $K^{\CV}$,
and those that respect it, $K^{\CP}$ (see Eqs.~(\ref{mw}) and
(\ref{alphaT}) below, as well as Appendix~\ref{app:custodial}).

The above dimension-6 operators lead to associated SUSY breaking
operators as follows:
\bea
K^{\CV}_{\rm spurion} &=& 
\frac{c_1}{|M|^2} \left[(\gamma_1 X + \gamma^*_1 X^{\dagger})  + \beta_1 X^{\dagger} X \right] \, (H^\dagger_d \, e^{2V} H_d)^{2}  
\nonumber \\ [0.4em]
&& \mbox{} + \frac{c_2}{|M|^2} \left[(\gamma_2 X + \gamma^*_2 X^{\dagger})  + \beta_2 X^{\dagger} X \right] \, (H^\dagger_u \, e^{2V} H_u)^{2}  
\label{KSpurionCV}
\\ [0.4em]
& &
\mbox{} + \frac{c_3}{|M|^2} \left[(\gamma_3 X + \gamma^*_3 X^{\dagger})  + \beta_3 X^{\dagger} X \right] \, (H^\dagger_u \, e^{2V} H_u) (H^\dagger_d \, e^{2V} H_d)~,
\nonumber \\[0.5em]
K^{\CP}_{\rm spurion} &=& 
\frac{c_4}{|M|^2} \left[(\gamma_4 X + \gamma^*_4 X^{\dagger})  + \beta_4 X^{\dagger} X \right] \, |H_u H_d|^{2}  
\nonumber \\ [0.4em]
&& \mbox{} + \left[ \frac{c_6}{|M|^2} ( \gamma_6 X + \delta_6 X^{\dagger} + \beta_6 X^{\dagger} X) H_d^{\dagger} \, e^{2V} H_d  \right.
\label{KSpurionCP}
\\ [0.4em]
& & \hspace{7mm} \left.
\mbox{} + \frac{c_7}{|M|^2} ( \gamma_7 X + \delta_7 X^{\dagger} + \beta_7 X^{\dagger} X) H_u^{\dagger} \, e^{2V} H_u \right] (H_u H_d)  + {\rm h.c.} 
\nonumber
\eea
Here we used the fact that operators like $X H_{d}^{\dagger} \, e^{2V}
H_{d}$ or $X H_{u}^{\dagger} \, e^{2V} H_{u}$ can be set to zero by a
superfield redefinition~\cite{Antoniadis:2007xc}.  We did not write
explicitly the usual soft breaking masses $m^{2}_{H_{u}}$ and
$m^{2}_{H_{d}}$, which correspond to operators of the form
$X^{\dagger}X H_{u}^{\dagger} \, e^{2V} H_{u}$ and $X^{\dagger}X
H_{d}^{\dagger} \, e^{2V} H_{d}$, although such terms are understood.
We assumed that each type of soft breaking operator vanishes in the
absence of the corresponding SUSY preserving one, i.e. we define the
parameters $\beta_i, \gamma_i, \delta_{i}$ by factoring out the
associated $c_{i}$.  This will be the case if the corresponding
statement holds in the UV theory that induces the higher-dimension
operators we are considering, and is a property of several realistic
SUSY breaking mediation mechanisms (in simple extensions one obtains a
strict proportionality, as illustrated in the examples described in
Appendix~\ref{app:UVcompletions}).  Notice also that operators like
$X^\dagger (H_{u}H_{d})^2$ or $X^\dagger X (H_{u}H_{d})^2$ in the
\Kahler potential are equivalent to the superpotential operators of
Eqs.~(\ref{genw}) and (\ref{WSpurion}), and since we are taking
$\omega_{1}$ and $\alpha_1$ as free parameters, there is no loss of
generality in omitting them from Eq.~(\ref{KSpurionCP}).

It is straightforward to work out the scalar potential that follows
from Eqs.~(\ref{genw}), (\ref{WSpurion}), (\ref{genk}),
(\ref{KSpurionCV}) and (\ref{KSpurionCP}) which, at the renormalizable
level, takes the form
\bea
\label{eq:pot}
V_{\rm ren.} &=& m_{u}^2 H_u^{\dagger} H_u + m_{d}^2 H_d^{\dagger} H_d -
\left[ b H_u H_d + \hc \right]  
\nonumber \\  [0.4em]
&& \mbox{} +  \frac{1}{2} \lambda_1 ( H_d^{\dagger} H_d )^2 + \frac{1}{2} \lambda_2 ( H_u^{\dagger} H_u )^2 +   \lambda_3
( H_u^{\dagger} H_u )( H_d^{\dagger} H_d ) +
\lambda_4 ( H_u H_d )( H_u^{\dagger} H_d^{\dagger} )
\nonumber \\ [0.4em]
&& \mbox{} +  \left\{ \frac{1}{2} \lambda_5 ( H_u H_d)^2 +
  \left[ \lambda_6 (H_d^{\dagger} H_d ) + \lambda_7 (
      H_u^{\dagger} H_u )\right] (H_u  H_d) +
  \hc. \right\}~,
\eea
where we added the soft masses $m_{H_{u}}^2$, $m_{H_{d}}^2$ and $b$,
and $m^2_{u,d} \equiv m_{H_{u,d}}^2 + |\mu|^2$ also include the
supersymmetric mass term, $|\mu|^2$.  The quartic operators are
defined in such a way that the $\lambda_i, i=1,\ldots,7$, map exactly
to those used in \cite{Haber:1993an} with their alternate convention
for the hypercharges of the two Higgs-doublet fields.  In the MSSM
limit, the quartic couplings are
\be
\lambda_1^{(0)} =\lambda_2^{(0)} = \frac{1}{4} \left( g^2 + g'^2 \right)~, \hspace{5mm} 
\lambda_3^{(0)} = \frac{1}{4} \left( g^2 - g'^2 \right)~,  
\hspace{5mm} 
\lambda_4^{(0)} = - \frac{g^2}{2}~, 
\label{MSSMLambdas}
\ee
with the rest vanishing.  At order $1/M$ (henceforth referred to as
dimension-5), only $\lambda_{5}$, $\lambda_{6}$ and $\lambda_{7}$
receive contributions:
\be
\label{dim5}
\Delta \lam^{(5)}_5 \,=\,   - \alpha_1 \, \omega_{1}  \frac{m_s}{M}~,
\hspace{1cm}
\Delta \lam^{(5)}_6 \,=\, \Delta \lam^{(5)}_7 \,=\,  \omega_{1} \frac{\mu}{M}~.
\ee
At order $1/M^{2}$ all the quartic couplings receive contributions:
\be\label{dim6}
\begin{split}
& \Delta \lam^{(6)}_1 \,  =  - 2 (c_{3} + c_4) \frac{\mu^2}{M^2} + 4 c_6 \delta_6 \frac{m_s \mu}{M^2} - 2 c_{1} \beta_1 \frac{m_{s}^2}{M^2}~, \\
& \Delta \lam^{(6)}_2 \,  =  - 2 (c_{3} + c_4) \frac{\mu^2}{M^2} + 4 c_7 \delta_7 \frac{m_s \mu}{M^2}  - 2 c_{2} \beta_2 \frac{m_{s}^2}{M^2}~, \\
& \Delta \lam^{(6)}_3 \,  =  - (c_{1} + c_{2} + 2c_4) \frac{\mu^2}{M^2} + 2 ( c_6 \delta_6 + c_7 \delta_7) \frac{m_s \mu}{M^2} - c_{3} \beta_3 \frac{m_{s}^2}{M^2}~, \\
& \Delta \lam^{(6)}_4 \,  =  - (c_{1} + c_{2} + 2c_3) \frac{\mu^2}{M^2} + 2 (c_6 \delta_6 + c_7 \delta_7) \frac{m_s \mu}{M^2} - c_4 \beta_4 \frac{m_s^2}{M^2}~,  \\
& \Delta \lam^{(6)}_5 \,  =  2 (c_6 \gamma_6 + c_7 \gamma_7)  \frac{m_s \mu}{M^2}~,  \\
& \Delta \lam^{(6)}_6 \,  = - (c_6 + 2c_7) \frac{\mu^2}{M^2} + (2c_1 \gamma_1 + c_{3} \gamma_3 + c_4 \gamma_4) \frac{m_s \mu}{M^2} - c_6 \beta_6 \frac{m_s^2}{M^2}~,  \\
& \Delta \lam^{(6)}_7 \,  = - (c_7 + 2c_6) \frac{\mu^2}{M^2} + (2c_2 \gamma_2 + c_{3} \gamma_3 + c_4 \gamma_4) \frac{m_s \mu}{M^2} - c_7 \beta_7 \frac{m_s^2}{M^2}~,
\end{split}
\ee
where we assumed, as we will do for simplicity in the rest of the
paper, that all parameters are real.  At this order we should consider
also dimension-6 operators in the scalar potential that give
parametrically comparable effects.  The dimension-6 operators
generated by integrating out the $F$- and $D$-terms take the form
\bea
\label{eq:honest}
V_{\rm non-ren.} 
&=& \frac{1}{M^{2}} \left\{ \lvert H_u H_d \rvert ^2 [(\lam_8 H_d^{\dagger} H_d + \lam'_8 H_u^{\dagger} H_u))] 
+ \left( \lam_9 \lvert H_u H_d \rvert^2 + \lam_{10} (H^{\dagger}_d H_d)^2
\right. \right.
\nonumber \\ [0.4em]
&& \hspace{0mm} \left. \mbox{}  + \lam_{11} (H^{\dagger}_u H_u)^2 + \lam_{12} (H^{\dagger}_d H_d) (H^{\dagger}_u H_u) \right) [H_u H_d + H^{\dagger}_u H^{\dagger}_d ] \\ [0.4em]
&& \hspace{0mm} \left. \mbox{}  + \lam_{13} (H^{\dagger}_d H_d)^3 + \lam_{14} (H^{\dagger}_d H_d)^2 (H^{\dagger}_u H_u)  + \lam_{15} (H^{\dagger}_d H_d) (H^{\dagger}_u H_u)^2 + \lam_{16} (H^{\dagger}_u H_u)^3 \right\}~,
\nonumber
\eea
where
\be\label{lambda816}
\begin{split}
& \lam_8 \,  =  \omega^2_{1} - (c_1 + c_{3}) \frac{g^2}{2}~,  \qquad \qquad \hspace{1mm}  \lam'_8 \,  =  \omega^2_{1} - (c_2 + c_{3}) \frac{g^2}{2}~,\\
& \lam_9 \,  = - (c_6+c_7)\frac{g^2}{2}~, \qquad \qquad \qquad \hspace{-1mm} \lam_{10} \,  = c_6 \frac{(g^2 + g'^2)}{4}~, \\
& \lam_{11} \,  =  c_7 \frac{(g^2 + g'^2)}{4}~, \qquad \qquad \qquad \lam_{12} \,  =  (c_6+c_7) \frac{(g^2-g'^2)}{4}~, \\
& \lam_{13} \,  =  c_1 \frac{(g^2 + g'^2)}{4}~, \qquad \qquad \qquad \lam_{14} \,  = c_{3} \frac{g^2}{2} - c_1 \frac{(g^2-g'^2)}{4}~, \\
& \lam_{15} \,  = c_{3} \frac{g^2}{2} - c_2 \frac{(g^2-g'^2)}{4}~, \qquad \hspace{3mm} \lam_{16} \,  =  c_2 \frac{(g^2 + g'^2)}{4}~. \\
\end{split}
\ee
The operators with coefficients $\lambda_{8}$ and $\lambda'_{8}$ are
essential in stabilizing the sEWSB vacua discussed in
\cite{Batra:2008rc}.  At order $1/M^{2}$, higher-dimension operators
involving two derivatives need also be included, since after EWSB they
lead to contributions to the Higgs kinetic terms.  After canonical
normalization, these give additional contributions to the masses and
couplings of the Higgs fields.  These operators are found to be
\bea
\label{eq:kinmix}
{\cal L} &\supset& - \frac{1}{M^2} 
\left\{ c_1  \left( \left[(D^2 H_d)^{\dagger} H_d \right] (H^\dagger_{d} H_{d})
+ [(D_{\mu} H_{d})^\dagger H_{d}] [(D^{\mu} H_{d})^\dagger H_{d}] \right) \right.
\nonumber
\\[0.4em]
&&  \mbox{} + c_2 \left( \left[(D^2 H_u)^{\dagger} H_u \right] (H^\dagger_{u} H_{u}) 
+ [(D_{\mu} H_{u})^\dagger H_{u}] [(D^{\mu} H_{u})^\dagger H_{u}] \right)
\nonumber \\[0.4em]
&& \mbox{} + c_3 \left(\left[(D^2 H_u)^{\dagger} H_u \right] (H^\dagger_{d} H_{d})
 + \left[(D^2 H_d)^{\dagger} H_d \right] (H^\dagger_{u} H_{u}) + 
2 [(D_{\mu} H_{u})^\dagger H_{u}] [(D^{\mu} H_{d})^\dagger H_{d}] \right)
\nonumber
\\[0.4em]
&& \left. \mbox{} - c_4 \, \partial_{\mu} (H_u^{\dagger} H_d^{\dagger}) \partial ^{\mu} (H_u H_d) + 
\left( \left[ c_6 \, (D^2 H_d)^{\dagger} H_d
 + c_7 \, (D^2 H_u)^{\dagger} H_u \right] (H_u H_d) + {\rm h.c.} \right) \right\}~,
\eea
where $D^{2}=D_{\mu}D^{\mu}$ and $D_{\mu} = \partial_{\mu} + i g
W_{\mu} + i g^\prime Y B_{\mu}$ is the gauge covariant derivative
($Y=+1/2$ for $H_{u}$ and $Y=-1/2$ for $H_{d}$).

Finally, although there can be ${\cal O}(1/M^2)$ contributions to the
gauge kinetic terms, we do not consider them here (but
see~\cite{Brignole:2003cm}).

\subsection{Higgs Spectrum}
\label{sec:spectrum}

We obtain the spectrum and couplings of the Higgs fields in two steps:
first we neglect the non-canonical kinetic terms that follow from
Eq.~(\ref{eq:kinmix}), minimize the potential (which does not require
canonical normalization), and obtain general expressions for the (at
this point unphysical) masses and couplings in terms of
\be
\lam_i=\lam^{(0)}_i+\Delta \lam^{(5)}_i+\Delta \lam^{(6)}_i+\Delta \lam^{\rm 1-loop}_i~,
\label{lam}
\ee
where the last term represents the 1-loop corrections (to be included
in the numerical analysis of Section~\ref{sec:results}), which will
be discussed in Subsection~\ref{loops}.  In this section we
concentrate on the tree-level contributions [i.e. the first three
terms in Eq~(\ref{lam})].  Second, we make a field redefinition to
arrive at canonical normalization in the presence of the operators of
Eq.~(\ref{eq:kinmix}), and rediagonalize to obtain the physical
spectrum.  The first step defines CP eigenstates $h^0, H^0, A^0,
H^{\pm}$, with mass parameters $m^2_{h^0}, m^2_{H^0},m^2_{A^0},
m^2_{H^{\pm}}$ as follows,
\bea
\left( \begin{array}{c}  H_u^{0} \\  H_d^{0} \end{array}  \right) = 
\frac{1}{\sqrt{2}} 
\left( \begin{array}{c}  v_u \\  v_d \end{array}  \right) + 
\frac{1}{\sqrt{2}}   
\left( \begin{array}{cc} c_{\alpha} & s_{\alpha} \\ -s_{\alpha} & c_{\alpha} \end{array} \right)  \left( \begin{array}{c}  h^{0} \\  H^{0} \end{array} \right) +  
\frac{i}{\sqrt{2}}  
\left( \begin{array}{cc} \sb  & \cb  \\ -\cb  & \sb  \end{array} \right)  \left( \begin{array}{c}  G^0 \\  A^0 \end{array} \right)~,
\label{NeutralFluctuations}
\eea
\bea
\left( \begin{array}{c}  H_u^{+} \\  H_d^{-*} \end{array}  \right) =   
\left( \begin{array}{cc} \sb  & \cb  \\ -\cb  & \sb  \end{array} \right)     \left( \begin{array}{c}  G^+ \\  H^+ \end{array} \right)~,
\eea
where $\sb = \sin\beta$ and $\cb = \cos\beta$ with $\tb = v_u/v_d$ and
$v^2= v_u^2+v_d^2 \approx (246\ {\rm GeV})^2$.  For small
fluctuations, $G^{0,\pm}$ are the eaten Goldstone bosons that will be
set to zero in the following.  Minimization of the potential requires
\bea
m^2_{u} &=&  b \, t_{\beta} ^{-1} - \frac{v^2}{2} \sb^2 \, [\lam_2 + 3 \lam_7 t_{\beta} ^{-1} + \tilde{\lam}_3 t_{\beta} ^{-2} + \lam_6 t_{\beta} ^{-3} ] 
\nonumber \\ 
& & \mbox{} - \frac{v^4}{4M^2} \sb^4 \, [3 \lam_{16} + 5 \lam_{11} \tb^{-1} + 2\tilde{\lam}'_8 t_{\beta}^{-2} + 3 \tilde{\lam}_9 \tb^{-3} + \tilde{\lam}_8 t_{\beta}^{-4} + \lam_{10} \tb^{-5}]~,   
\label{m2hu}
\\ [0.4em]
m^2_{d} &=&  b \, t_{\beta}  - \frac{v^2}{2}  \tb \sb^2 \, [\lam_7 + \tilde{\lam}_3 \tb^{-1} + 3 \lam_6 \tb^{-2} + \lam_1 \tb^{-3}] 
\nonumber\\ 
& & \mbox{} - \frac{v^4}{4M^2} \tb \sb^4 \, [\lam_{11} + \tilde{\lam}'_8 \tb^{-1} + 3 \tilde{\lam}_9 \tb^{-2} + 2\tilde{\lam}_8 t_{\beta}^{-3} + 5 \lam_{10} \tb^{-4} + 3 \lam_{13} \tb^{-5}]~,
\label{m2hd}
\eea
where $\widehat{\lambda_3} = \lambda_3 + \lambda_4$,
$\tilde{\lambda}_3 = \widehat{\lambda_3} + \lam_5$,
$\tilde{\lambda}_{8} = \lambda_{8} + \lambda_{14}$,
$\tilde{\lambda}'_{8} = \lambda'_{8} + \lambda_{15}$ and
$\tilde{\lam}_9=\lam_9 + \lam_{12}$.  The masses of the CP-odd and
charged Higgses, ${A}^0$ and ${H}^\pm$, can then be written as
\bea
{m}^2_{{A}^0} &=& \frac{b \, \tb}{\sb^{2} }-\frac{v^2}{2} \tb \left(\lambda_7 + 2 \lambda_5 \tb^{-1}  + \lambda_6 t_{\beta}^{-2} \right) - \frac{v^4}{4M^2} \,  \tb \sb^2 \, [\lambda_{11} + \tilde{\lam}_9 \tb^{-2} + \lambda_{10} \tb^{-4} ]~,  
\label{mA} \\ [0.4em]
{m}^2_{{H}^\pm} &=& m^2_{{A}^0}+ \frac{v^2}{2} \left( \lambda_5 - \lambda_4 \right) - \frac{v^4}{4M^2} \, \sb^2 \, \left(\lam'_8 + 2\lam_9 \tb^{-1} + \lambda_{8} \tb^{-2} \right)~.
\label{mHCharged}
\eea
The mass matrix for the CP-even states in the $(h_{d},h_{u}) =
(c_{\alpha} H^{0} - s_{\alpha} h^{0}, c_{\alpha} h^{0} + s_{\alpha}
H^{0})$ basis is
\bea
{\cal M}^2&=&m^2_{A^0} \sb^2 
\left( \! 
\begin{array}{cc} 
1  & - \tb^{-1} \\ [0.4em]
- \tb^{-1}  & \tb^{-2}  
\end{array} \! \right)  
+  v^2 \sb^2 
\left( \! \begin{array}{cc}
\lambda_5 +  2 \lambda_6 \tb^{-1} + \lambda_1 \tb^{-2}  
&  \lambda_7  + \widehat{\lambda_3} \tb^{-1}  +\lambda_6 \tb^{-2}   \\ [0.4em]
\lambda_7 + \widehat{\lambda_3} \tb^{-1} +\lambda_6 \tb^{-2}   
&   \lambda_2 + 2 \lambda_7 \tb^{-1} + \lambda_5 \tb^{-2} 
\end{array} \! \right)  
\label{CPevenMassMatrix}
 \\ [0.5em]
&+&  \frac{v^4 \sb^4}{M^2}  
\left( \! \begin{array}{cc}
\tilde{\lam}_9 \tb^{-1} + \tilde{\lam}_8 \tb^{-2} + 4  \lam_{10} \tb^{-3} + 3 \lam_{13} \tb^{-4} 
& \lam_{11} + \tilde{\lam}'_8 \tb^{-1} + 2 \tilde{\lam}_9 \tb^{-2} + \tilde{\lam}_8 \tb^{-3} + \lam_{10} \tb^{-4} \\  [0.4em]
\lam_{11} + \tilde{\lam}'_8 \tb^{-1} + 2 \tilde{\lam}_9 \tb^{-2} + \tilde{\lam}_8 \tb^{-3} + \lam_{10} \tb^{-4} 
& 3 \lambda_{16} + 4  \lam_{11} \tb^{-1} + \tilde{\lam}'_8 \tb^{-2} + \tilde{\lam}_9 \tb^{-3}  \end{array} \! \right)
\nonumber
\eea
which leads to mass eigenvalues and a mixing angle $\alpha$ given by
\bea
{m}^2_{{H}^0, \hat{h}^0} &=& \half \left[ \Tr {\cal M}^2 \pm \sqrt{\left( \Tr {\cal M}^2\right)^2 - 4 \det {\cal M}^2 }\right]~,  
\label{m2hH}
\\
{s}_{2 \alpha} &=& \frac{2 {\cal M}^2_{12}}{\sqrt{\left( \Tr {\cal M}^2\right)^2 - 4 \det {\cal M}^2 }}~,  
\label{eq:alpangle1} \\
{c}_{2 \alpha} &=& \frac{{\cal M}^2_{11}-{\cal M}^2_{22} }{\sqrt{\left( \Tr {\cal M}^2\right)^2 - 4 \det {\cal M}^2 }}~.
\label{eq:alpangle2}
\eea
In order to find the physical masses at order $1/M^{2}$ we need to
include the operators in Eq.~(\ref{eq:kinmix}), that contribute to the
kinetic terms of $h^0$, $H^0$, $A^0$ and $H^{\pm}$.  Canonical
normalization is achieved by the field redefinitions
\be\label{eq:fieldredef}
\begin{split}
h^0 &\rightarrow \left( 1- \half A_1  \right) h^0 - \half B_1  H^0~, \qquad   A^0 \rightarrow \left( 1- \half  E_1  \right) A^0~, \\
H^0 &\rightarrow \left( 1 -  \half D_1   \right) H^0 - \half   B_1    h^0~,  \qquad
H^{\pm} \rightarrow \left( 1 - \half F_1  \right) H^{\pm}~,
\end{split}  
\ee
where 
\bea
A_1&=& \frac{v^2}{2M^2} \sb^2 \left\{ 
c_{2} (1+c_{2\alpha}) + c_{1}  \tb^{-2} (1-c_{2\alpha}) 
+ \frac{1}{2} c_{3} \, \left[ \sb^{-2} - \left(1 - \tb^{-2} \right) c_{2\alpha} - 2 \tb^{-1} s_{2\alpha} \right] \right.
\nonumber \\ [0.3em]
& & \hspace{1.4cm} \left. \mbox{} + c_4 \sb^{-2} c_{\alpha + \beta}^2 
+ c_7 \left[-s_{2\alpha} + 2 \tb^{-1} (1 + c_{2\alpha}) \right]
+ c_6 \, \tb^{-1} \left[2 (1 - c_{2\alpha}) - \tb^{-1} s_{2\alpha} \right]  \right\}~,
\nonumber\\ [0.5em]
B_1&=& \frac{v^2}{2M^2} \sb^{2} \left\{
\left( c_{2} - c_{1} \tb^{-2} \right) s_{2\alpha} 
- \frac{1}{2} c_{3} \, \left[ \left(1 - \tb^{-2} \right) s_{2\alpha} - 2 \tb^{-1} c_{2\alpha} \right] \right.
\nonumber \\ [0.3em]
& & \hspace{1.4cm} \left. \mbox{} + c_4  \sb^{-2} c_{\alpha + \beta} s_{\alpha + \beta} 
+ c_7 \left( c_{2\alpha} + 2 \tb^{-1} s_{2 \alpha} \right)
+ c_6 \, \tb^{-1} \left( - 2 s_{2 \alpha} + \tb^{-1} c_{2\alpha} \right) \right\}~,
\nonumber\\ [0.5em]
D_1&=& \frac{v^2}{2M^2} \sb^2 \left\{ 
c_{2} (1-c_{2\alpha}) + c_{1}  \tb^{-2} (1+c_{2\alpha}) 
+ \frac{1}{2} c_{3} \, \left[  \sb^{-2} + \left(1 - \tb^{-2} \right) c_{2\alpha} + 2 \tb^{-1} s_{2\alpha} \right] \right.
\label{ABDEF}
 \\ [0.3em]
& & \hspace{1.4cm} \left. \mbox{} + c_4 \sb^{-2} s_{\alpha + \beta}^2 
+ c_7 \left[s_{2\alpha} + 2 \tb^{-1} (1 - c_{2\alpha}) \right] 
+ c_6 \, \tb^{-1} \left[2 (1 + c_{2\alpha}) + \tb^{-1} s_{2\alpha} \right]
\right\}~,
\nonumber\\ [0.5em]
E_1&=& \frac{v^2}{2M^2} \left\{
\frac{1}{8} (c_{1} + c_{2}) \left[ 1 - \sb^4 \left(1 - 14 \tb^{-2} + \tb^{-4} \right) \right] 
+ \frac{1}{4} c_{3} \left[ 3 + \sb^4 \left( 1 + 2 \tb^{-2} + \tb^{-4} \right) \right]
\right.
\nonumber \\ [0.3em]
& & \hspace{1.1cm} \left. \mbox{} + c_4 + 2\tb^{-1} \sb^{2} \left[c_6 \, (1+ \sb^2) + c_7 \, (1+ \tb^{-2} \sb^2) \right] \rule{0mm}{5.2mm} \right\}~,
\nonumber\\ [0.5em]
F_1&=& \frac{v^2}{M^2} \sb^{4} \left[ \frac{1}{8} c_{3} \left( 3\sb^{-4} + 1- 6 \tb^{-2} + \tb^{-4} \right) +  \tb^{-1} (c_6 + c_7 \, \tb^{-2}) + \frac{1}{2} (c_{1} + c_{2}) \tb^{-2} \right]~.
\nonumber
\eea
At this point, one needs to perform a further rotation (by an angle
$\gamma$) in the CP even sector, that rediagonalizes the corresponding
mass matrix:
\be
\begin{split}
h^0 &\rightarrow c_{\gamma } h^0 + s_{\gamma} H^0~,  \\
H^0 &\rightarrow c_{\gamma} H^0 -s_{\gamma} h^0~,
\end{split}
\ee
where, to order $1/M^2$,
\bea
t_{2\gamma} &=& -\frac{m_{H^0}^2 + m_{h^0}^2 }{m_{H^0}^2(1-D_{1})-m_{h^0}^2(1-A_{1})} \, B_1~. 
\label{gamma}
\eea
We do not expand the denominator to cover cases where $m_{H^0}^2
\approx m_{h^0}^2$.  The physical CP-even squared masses are finally
given by
\bea
m^2_{h} &=& m^2_{h^0} \left( 1- A_1  \right) \cg^2 + m^2_{H^0} \left( 1 -   D_1   \right) \sg^2 + B_{1} (m^2_{H^0} + m^2_{h^0}) \cg \sg~, 
\nonumber \\ [0.4em]
\hspace{7mm}
m^2_{H} &=& m^2_{H^0} \left( 1 -   D_1   \right) \cg^2 + m^2_{h^0} \left( 1- A_1  \right) \sg^2 - B_{1} (m^2_{H^0} + m^2_{h^0}) \cg \sg~, 
\label{FinalMasses}
\eea
while the physical CP-odd and charged Higgs masses are given by
$m^2_{A^0} ( 1- E_1)$ and $m^2_{H^{\pm}} ( 1 - F_1)$, respectively.
Whenever $\gamma \ll 1$ we have $m^2_{h} \approx m^2_{h^0} \left( 1-
A_1 \right)$ and $m^2_{H} \approx m^2_{H^0} \left( 1- D_1 \right)$.
However, there are regions where the denominator in Eq.~(\ref{gamma})
is small and all orders in $\gamma$ should be kept, in spite of it
formally being of order $1/M^2$.

There is one additional source of corrections at order $1/M^{2}$ that
affect the spectrum, as well as the Higgs couplings to be discussed in
the next subsection.  The two-derivative operators in
Eq.~(\ref{eq:kinmix}) give contributions to the gauge boson masses as
follows:
\bea
m^2_{Z} &=& \frac{1}{4} g^{2}_{Z} v^{2} \left[ 1 + \frac{v^{2}}{M^{2}} s^4_{\beta} \left( c_{2} + \tb^{-1} c_{7} + \tb^{-3} c_{6} + \tb^{-4} c_{1} \right) \right]~,
\label{mz} \\ [0.4em]
m^2_{W} &=& m^2_{Z}c^{2}_{W} \left( 1 + \alpha \tilde{T} \right)~,
\label{mw}
\eea
where
\bea
\alpha \tilde{T} &=& - \frac{v^{2}}{2M^{2}} \sb^4 \left[ c_{2} - 2 \tb^{-2} c_{3} + \tb^{-4} c_{1}  \right]
\label{alphaT}
\eea
is the contribution from the higher-dimension operators to the
Peskin-Takeuchi $T$-parameter.  Note that, as mentioned in
Section~\ref{sec:SUSY2HDM}, only the operators in Eq.~(\ref{DeltaKCV})
contribute to $T$.  At loop level, there are other contributions to
the $T$-parameter from the distorted Higgs spectrum (heavier SM-like
Higgs in addition to mass splittings among the non-standard Higgses),
as well as from custodially violating mass splittings in the
superpartner spectrum~\cite{Medina:2009ey}.  For the time being, we
note that keeping $m_{Z}$ fixed at the observed value, Eq.~(\ref{mz})
implies a shift in $v$ at order $1/M^2$:
\bea
\frac{\Delta v}{v} &\approx& -\frac{v^{2}}{2M^{2}} \sb^4 \left( c_{2} + \tb^{-1} c_{7} + \tb^{-3} c_{6} + \tb^{-4} c_{1} \right)~,
\label{Deltav}
\eea
so that the VEV $v$ in this model differs from the SM value of
$246~{\rm GeV}$ by terms of order $1/M^{2}$.  At this order, only the
terms in Eq.~(\ref{mHCharged}) and (\ref{CPevenMassMatrix}) that are
proportional to $\lambda_{1}$, $\lambda_{2}$, $\lambda_{3}$ and
$\lambda_{4}$ --which are non-vanishing in the MSSM limit-- are
affected by the shift in the VEV (in particular, $m^{2}_{A}$ is not
affected).  In practice, at this order in the $1/M$ expansion it is
sufficient to use $v = (1 + \Delta v/v) \times 246~{\rm GeV}$ in
Eqs.~(\ref{NeutralFluctuations})--(\ref{FinalMasses}), with $\Delta
v/v$ given in Eq.~(\ref{Deltav}).

\subsection{Higgs Couplings to Fermions and Gauge Bosons}
\label{sec:couplings}

Here we focus on the couplings of the Higgs scalars to gauge bosons
and fermions.  Recall that in the MSSM the couplings of the neutral
Higgs bosons to fermion pairs, relative to the SM value, $g
m_{f}/2m_{W}$, read (for up-type and down-type quarks)
\bea
\begin{array}{lclcl}
h\bar{t}t : \ca / \sb  & \hspace{5mm} & H\bar{t}t : \sa / \sb & \hspace{5mm} & A\bar{t}t : 1/ \tb
\\ [0.4em]
h\bar{b}b : - \sa / \cb & & H\bar{b}b : \ca / \cb  & &  A\bar{b}b : \tb ~,
\end{array}
\label{MSSMhff}
\eea
while the couplings to a pair of gauge bosons relative to the SM value
($g m_{W}$ for $V=W$ and $g m_{Z}/c_{W}$ for $V=Z$) are
\be
hVV : s_{\beta - \alpha} \qquad HVV : c_{\beta - \alpha}~.
\label{MSSMhVV}
\ee
These couplings are changed by the higher-dimension operators in two
ways: indirectly through the mixing angle $\alpha$ [see
Eqs.~(\ref{CPevenMassMatrix}), (\ref{eq:alpangle1}) and
(\ref{eq:alpangle2})], and directly due to the effects associated with
the operators of Eq.~(\ref{eq:kinmix}), which include the field
redefinitions of Eqs.~(\ref{eq:fieldredef}).  The latter involve a
rescaling and therefore cannot be parametrized as a rotation by an
effective angle.  These effects correspond to the mixing of the MSSM
Higgs fields with heavy degrees of freedom that are not included
explicitly in the effective theory, and appear first at order
$1/M^{2}$.

Since it is convenient to give the Higgs couplings relative to the SM
values, we need to take into account the shift in the Higgs VEV,
Eq.~(\ref{Deltav}), which implies a shift in the Yukawa couplings
relative to the SM values: $\Delta y_{f}/y_{f} = -\Delta v/v$.
Together with the Higgs field redefinition of
Eqs.~(\ref{eq:fieldredef}) these induce a shift in the
Higgs-$\bar{f}f$ couplings at order $1/M^{2}$ compared to
Eqs.~(\ref{MSSMhff}) (on top of the shifts implicit through the mixing
angle $\alpha$).

Similarly, the normalization of the Higgs-gauge boson couplings to the
SM value needs to take into account the shifts in the gauge boson
masses given in Eqs.~(\ref{mz}) and (\ref{mw}).  In addition, the
two-derivative operators in Eq.~(\ref{eq:kinmix}) also give direct
corrections to these vertices from the terms quadratic in the gauge
fields and linear in $h^{0}$ or $H^{0}$:
\bea
&& \frac{g^{2} v}{4 c^{2}_{W}} \left( \delta g_{hZZ} h^0 + \delta g_{HZZ} H^0 \right) Z_{\mu}Z^{\mu} + \frac{g^{2} v}{2} \left( \delta g_{hWW} h^0 + \delta g_{HWW} H^0 \right)  W^{+}_{\mu} W^{-\mu}~,
\nonumber
\eea
where
\bea
\delta g_{hZZ} &=& \frac{v^2}{2M^2} \left\{ 4 \sb^3 \left( c_{2} c_{\alpha} - c_{1} \tb^{-3} s_{\alpha} \right) + \sb^2 \left[ c_{7} \left( 2c_{\alpha+\beta} + c_{\alpha-\beta} \right) + c_{6} \tb^{-2} \left( 2c_{\alpha+\beta} - c_{\alpha-\beta} \right) \right] \right\}~,
\nonumber \\[0.4em]
\delta g_{HZZ} &=& \frac{v^2}{2M^2} \left\{ 4 \sb^3 \left( c_{2} s_{\alpha} + c_{1} \tb^{-3} c_{\alpha} \right) + \sb^2 \left[ c_{7} \left( 2s_{\alpha+\beta} + s_{\alpha-\beta} \right) + c_{6} \tb^{-2} \left( 2s_{\alpha+\beta} - s_{\alpha-\beta} \right) \right] \right\}~,
\nonumber \\[0.4em]
\delta g_{hWW} &=& \frac{v^2}{2M^2} \left\{ 2 \sb^3 \left( c_{2} c_{\alpha} - c_{1} \tb^{-3} s_{\alpha} \right) + 2 c_{3} \tb^{-1} \sb^2 c_{\alpha+\beta} \right.
\nonumber \\[0.3em]
&& \hspace{1cm} \left. \mbox{} + \sb^2 \left[ c_{7} \left( 2c_{\alpha+\beta} + c_{\alpha-\beta} \right) + c_{6} \tb^{-2} \left( 2c_{\alpha+\beta} - c_{\alpha-\beta} \right) \right] \right\}~,
\label{deltagVV} \\[0.4em]
\delta g_{HWW} &=& \frac{v^2}{2M^2} \left\{ 2 \sb^3 \left( c_{2} s_{\alpha} + c_{1} \tb^{-3} c_{\alpha} \right) + 2 c_{3} \tb^{-1} \sb^2 s_{\alpha+\beta} \right.
\nonumber \\[0.3em]
&& \hspace{1cm} \left. \mbox{} + \sb^2 \left[ c_{7} \left( 2s_{\alpha+\beta} + s_{\alpha-\beta} \right) + c_{6} \tb^{-2} \left( 2s_{\alpha+\beta} - s_{\alpha-\beta} \right) \right] \right\}~.
\nonumber \eea
These should be combined with the field redefinitions
Eqs.~(\ref{eq:fieldredef}).  Note that the corrections to the $Z$ and
$W$ couplings are different only due to the custodially violating
coefficients $c_{1}$, $c_{2}$ and $c_{3}$.

Putting all these ingredients together, the couplings of
Eqs.~(\ref{MSSMhff}) and (\ref{MSSMhVV}) are then generalized as
follows.  For the light CP-even Higgs, $h^{0}$, we get
\bea
h\bar{t}t &:& \frac{1}{\sb} \left[c_{\alpha+\gamma} \left( 1 - \frac{\Delta v}{v} - \frac{1}{2} A_1 \right)-  \frac{1}{2} B_1 s_{\alpha-\gamma} + \frac{1}{2} (D_1-A_1) \sa \sg \right]~,
\nonumber \\ [0.4em]
h\bar{b}b &:& - \frac{1}{\cb} \left[s_{\alpha+\gamma} \left(1 - \frac{\Delta v}{v} - \frac{1}{2} A_1 \right) + \frac{1}{2} B_1 c_{\alpha-\gamma} - \frac{1}{2} (D_1-A_1) \ca \sg \right]~,
\nonumber\\ [0.4em]
hZZ &:& s_{\beta-\alpha-\gamma} \left( 1 + \frac{\Delta v}{v} - \frac{1}{2} A_1 \right) - \frac{1}{2} B_1 c_{\beta-\alpha + \gamma} + \frac{1}{2} (D_1-A_1) c_{\beta - \alpha} \sg 
\label{hcouplings} \\ [0.3em]
&& \mbox{} + \cg \delta g_{hZZ} - \sg \delta g_{HZZ}~,
\nonumber \\ [0.4em]
hWW &:& s_{\beta-\alpha-\gamma} \left( 1 + \frac{\Delta v}{v} - \frac{1}{2} \alpha \tilde{T} - \frac{1}{2} A_1 \right) - \frac{1}{2} B_1 c_{\beta-\alpha + \gamma} + \frac{1}{2} (D_1-A_1) c_{\beta - \alpha} \sg 
\nonumber \\ [0.3em]
&& \mbox{} + \cg \delta g_{hWW} - \sg \delta g_{HWW}~,
\nonumber 
\eea
while for the heavy CP-even Higgs, $H^{0}$, we get
\bea
H\bar{t}t &:& \frac{1}{\sb} \left[s_{\alpha+\gamma} \left(1 - \frac{\Delta v}{v} - \frac{1}{2} A_1 \right)-  \frac{1}{2} B_1 c_{\alpha-\gamma} - \frac{1}{2} (D_1-A_1) \sa \cg \right]~,
\nonumber \\ [0.4em]
H\bar{b}b &:& \frac{1}{\cb} \left[ c_{\alpha+\gamma} \left(1 - \frac{\Delta v}{v} - \frac{1}{2} A_1 \right)+  \frac{1}{2} B_1 s_{\alpha-\gamma} - \frac{1}{2} (D_1-A_1) \ca \cg \right]~,
\nonumber \\ [0.4em]
HZZ &:& c_{\beta-\alpha- \gamma} \left(1 + \frac{\Delta v}{v} - \frac{1}{2} A_1 \right) - \frac{1}{2} B_1 s_{\beta-\alpha + \gamma} - \frac{1}{2} (D_1-A_1) c_{\beta - \alpha} \cg
\label{Hcouplings} \\ [0.3em]
&& \mbox{} + \cg \delta g_{HZZ} + \sg \delta g_{hZZ}~,
\nonumber \\ [0.4em]
HWW &:& c_{\beta-\alpha- \gamma} \left(1 + \frac{\Delta v}{v} - \frac{1}{2} \alpha \tilde{T} - \frac{1}{2} A_1 \right) - \frac{1}{2} B_1 s_{\beta-\alpha + \gamma} - \frac{1}{2} (D_1-A_1) c_{\beta - \alpha} \cg
\nonumber \\ [0.3em]
&& \mbox{} + \cg \delta g_{HWW} + \sg \delta g_{hWW}~,
\nonumber 
\eea
where $\alpha$ is determined by Eqs.~(\ref{eq:alpangle1}) and
(\ref{eq:alpangle2}).  The couplings to fermion pairs of the CP-odd
and charged Higgses, $A^{0}$ and $H^{\pm}$, which are independent of
$\alpha$, are obtained from those in the MSSM by multiplication by $(1
- \Delta v/v - \frac{1}{2} E_{1})$ and $(1 - \Delta v/v - \frac{1}{2}
F_{1})$, respectively.  Here $\gamma$ is defined by Eq.~(\ref{gamma}),
$A_{1}$, $B_{1}$, $D_{1}$, $F_{1}$ and $E_{1}$ are as defined in
Eq.~(\ref{ABDEF}), $\Delta v/v$ is given in Eq.~(\ref{Deltav}),
$\alpha \tilde{T}$ is given in
Eq.~(\ref{alphaT}),\footnote{Eqs.~(\ref{hcouplings}) and
(\ref{Hcouplings}) are the tree-level expressions.  There are loop
level contributions to $\alpha T$, discussed in Section~\ref{EWPT},
that have to be added.  Since experimentally $\alpha T$ is bound to be
small, one can neglect it in $hWW$ and $HWW$ for the
phenomenologically allowed points.} and $\delta g_{hVV}$ and $\delta
g_{HVV}$ are given in Eqs.~(\ref{deltagVV}).  The $hVV$ and $HVV$
couplings in Eqs.~(\ref{hcouplings}) and (\ref{Hcouplings}) are given
relative to $g m_{Z}/c_{W}$ and $g m_{W}$ respectively, where $m_{Z}$
and $m_{W}$ include the order $1/M^{2}$ corrections as in
Eqs.~(\ref{mz}) and (\ref{mw}).  The Higgs couplings to fermion pairs
are given relative to $g m_{f}/2m_{W}$, which is the SM Yukawa
coupling.


Trilinear interactions involving one gauge boson and two Higgses are
also of great phenomenological relevance, and receive corrections from
the higher-dimension operators of Eq.~(\ref{genk}):
\bea
{\cal L}_3 &\supset& 
- { g \over 2 c_W} Z^\mu \left\{ (\eta_{ZHA} H^0 \del_\mu A^0 - \eta_{ZAH} A^0 \del_\mu{H^0}) - (\eta_{ZhA} h^0 \del_\mu A^0 - \eta_{ZAh} A^0 \del_\mu {h^0})\right\}
\nonumber \\[0.5em]
&& \mbox{} -
{ i g c_{2W} \over 2c_{W}} \, \eta_{ZH^+H^-} Z^\mu \left( H^+  \del_\mu {H^-} - H^-  \del_\mu {H^+} \right)
\nonumber \\[0.5em]
&& \mbox{} -
{i g  \over 2} W^{+\mu} \left\{ \left(\eta_{W^{\pm}h H^\mp} h^0\del_\mu{H^-} - \eta_{W^{\pm}H^{\mp}h} H^-\del_\mu{h^0} \right) \right.
\nonumber \\[0.5em]
&& \hspace{2cm} \left. \mbox{} -
\left(\eta_{W^{\pm} H H^{\mp}} H^0\del_\mu{H^-}  - \eta_{W^{\pm} H^{\mp}H} H^-\del_\mu{H^0}\right)  \right\} + {\rm h.c.} 
\nn \\[0.5em]
&& \mbox{} -
{g  \over 2} W^{+\mu} \left\{ \eta_{W^{\pm} H^{\mp}A} H^- \del_\mu A^0 - \eta_{W^{\pm} A H^{\mp}} A^0\del_\mu{H^-} \right\} + {\rm h.c.},   
\label{eq:trilinear} 
\eea
We give the detailed form of the coefficients $\eta_{ZhA}$,
$\eta_{ZAh}$, $\eta_{ZHA}$, $\eta_{ZAH}$ and $\eta_{ZH^+H^-}$, as well
as those for the interactions with a single $W$, in
Appendix~\ref{app:couplings}.

\section{Generic Features}
\label{sec:analres}

In the previous section we presented the effective theory in the Higgs
sector up to ${\cal O}(1/M^{2})$, where $M$ is the scale of the heavy
physics.  It is useful to obtain simple analytical expressions that
hold at order $1/M$, since these will determine the qualitative
features induced by the heavy physics.  In this section we perform
such an analysis, and consider several limiting cases that clarify the
generic features of the numerical study to be undertaken in the
next section.

We start by considering the masses of the CP-even Higgs states,
$m^{2}_{h^{0}}$ and $m^{2}_{H^{0}}$, as given in Eq.~(\ref{m2hH}) with
Eqs.~(\ref{MSSMLambdas}) and (\ref{dim5}).  Formally expanding to
leading order in $1/M$ one has
\bea m^2_{h^{0},H^{0}} &=&
(m^2_{h^{0},H^{0}})^{\rm MSSM} + (\Delta
m^2_{h^{0},H^{0}})^{\textrm{Dim-5}} + \cdots~,
\label{mhExpansion}
\eea
where the well-known MSSM contributions, $(m^2_{h^{0}})^{\rm MSSM}$
and $(m^2_{H^{0}})^{\rm MSSM}$, and the leading order corrections due
to the heavy physics, $(\Delta m^2_{h^{0}})^{\textrm{Dim-5}}$ and
$(\Delta m^2_{H^{0}})^{\textrm{Dim-5}}$, are given by
\bea
(m^2_{h^{0},H^{0}})^{\rm MSSM} &=& \frac{1}{2} (m_A^2 + m_Z^2 \mp D  )~,
\label{mhHDim5} \\ [0.4em]
(\Delta m^2_{h^{0},H^{0}})^{\textrm{Dim-5}} &=& \frac{1}{2} \omega_{1} v^2 \left[ 2 s_{2 \beta} \left( \frac{\mu}{M} \right) \left(1 \pm \frac{m_A^2+m_Z^2}{D } \right)  - \left( \frac{\alpha_1 m_s}{M} \right) \left(1 \mp \frac{c^2_{2 \beta} (m_A^2-m_Z^2)}{D } \right)  \right]~,
\nonumber
\eea
provided one has the inequality
\bea
D^2 &>& \omega_{1} v^{2} \left[ 2 s_{2\beta} \left(\frac{\mu}{M} \right) \left( m^{2}_{A} + m^{2}_{Z} \right) + c^{2}_{2\beta} \left( \frac{\alpha_{1} m_{s}}{M} \right) \left( m^{2}_{A} - m^{2}_{Z} \right) \right]~,
\label{inequality} \\ [0.4em]
D &\equiv& \sqrt{(m_A^2+m_Z^2)^2-4 \, c^2_{2 \beta} m_A^2 m_Z^2}~.
\nonumber
\eea
For the opposite inequality, the expressions for $m^{2}_{h^{0}}$ and
$m^{2}_{H^{0}}$ in Eq.~(\ref{mhHDim5}) should be interchanged.  For
typical values of the parameters we are interested in, the inequality
Eq.~(\ref{inequality}) is generally obtained for large $m_{A}$ while
the opposite inequality results for small $m_{A}$.  Near the point
where the two sides of (\ref{inequality}) coincide, the formal
expansion to order $1/M$ can be rather inaccurate due to the near
degeneracy of the two states (the mixing angle $\alpha$ can receive
large corrections).  Nevertheless the explicit expressions in
Eq.~(\ref{mhHDim5}) are useful, for instance showing how the sign of
the correction to the MSSM result depends on the signs of $\omega_{1}
\mu$ and $\omega_{1} \alpha_1$.

It is interesting to consider the $\tb$ dependence of the correction
to $m^{2}_{h^{0}}$ and $m^{2}_{H^{0}}$.  In the following we assume
that the inequality Eq.~(\ref{inequality}) holds.  As discussed above,
the complementary region can be obtained by simply interchanging
$h^{0}$ and $H^{0}$ in the relevant statements below.  For $\tb$ of
order one ($s_{2 \beta} \approx 1, \, c_{2 \beta} \approx 0$) one has
\bea
\label{eq:c1lowtb}
(\Delta m^2_{h^{0}})^{\textrm{Dim-5}} &\approx& \frac{1}{2} \omega_{1} v^2 \left(\frac{4 \mu - \alpha_1 m_s}{M} \right)~,
\eea
while for large $\tb$ (i.e. $s_{2 \beta} \approx 0, \, |c_{2 \beta}|
\approx 1$) one has 
\bea
\label{eq:c1largetb}
(\Delta m^2_{h^{0}})^{\textrm{Dim-5}} &\approx& \left\{
\begin{array}{ccc}
0  & & m_{A} \geq m_{Z}~, 
\\ [0.4em]
-\omega_{1}  v^2 \left(\frac{\alpha_1 m_s}{M} \right)  & & m_{A} < m_{Z}~.
\end{array}
\right.
\eea
Thus, the corrections are most important in the small $\tb$ regime (if
$m_{A} \geq m_{Z}$), and this region will allow to more easily evade
the LEP bound on $m_{h^{0}}$.  This is a well-appreciated feature in
models extended by singlet fields, such as the NMSSM and its
relatives.  We can easily estimate how large $m_{h^{0}}$ can be for
$\tb \approx 1$, where the MSSM contribution vanishes.  Setting
$\alpha_1 \sim -1$ and $\mu \sim m_{s}$, we have
\bea
m_{h^{0}} &\sim& \sqrt{\omega_{1}} \frac{v}{\sqrt{2}} \times {\cal O}(5\mu/M)^{1/2}~.
\label{upperboundmh}
\eea
For $\omega_{1} \sim 1$ this can easily be above the LEP bound.  For
the case of large $\tb$ the dimension-6 corrections can play a crucial
role in lifting the lightest Higgs mass from the MSSM limit, as will
be illustrated in the numerical analysis of the next section (notice
that dominant dimension-6 effects do not signal a breakdown of the EFT
analysis if the dimension-5 contributions are small due to a $\tb$
suppression or an accidentally small coupling $\omega_{1}$).

It is also interesting to consider the decoupling limit.  Expanding in
powers of $m_Z^2 / m_A^2 $, one gets
\bea
\label{eq:c1decoupl}
(\Delta m^2_{h^{0}})^{\textrm{Dim-5}} &\approx& \frac{1}{2} \omega_{1} v^2 s_{2 \beta} \left( \frac{4 \mu - s_{2 \beta} \alpha_1 m_s}{M} \right)~,
\eea
which shows that in a large region of parameter space the
leading-order correction to $m_{h^{0}}$ is only suppressed by about
$s_{2\beta}$ compared to Eq.~(\ref{upperboundmh}) [in general, this
correction has to be added in quadrature to the MSSM result, according
to Eq.~(\ref{mhExpansion})].

\begin{figure}[t]
\begin{center}
\includegraphics[width=8cm]{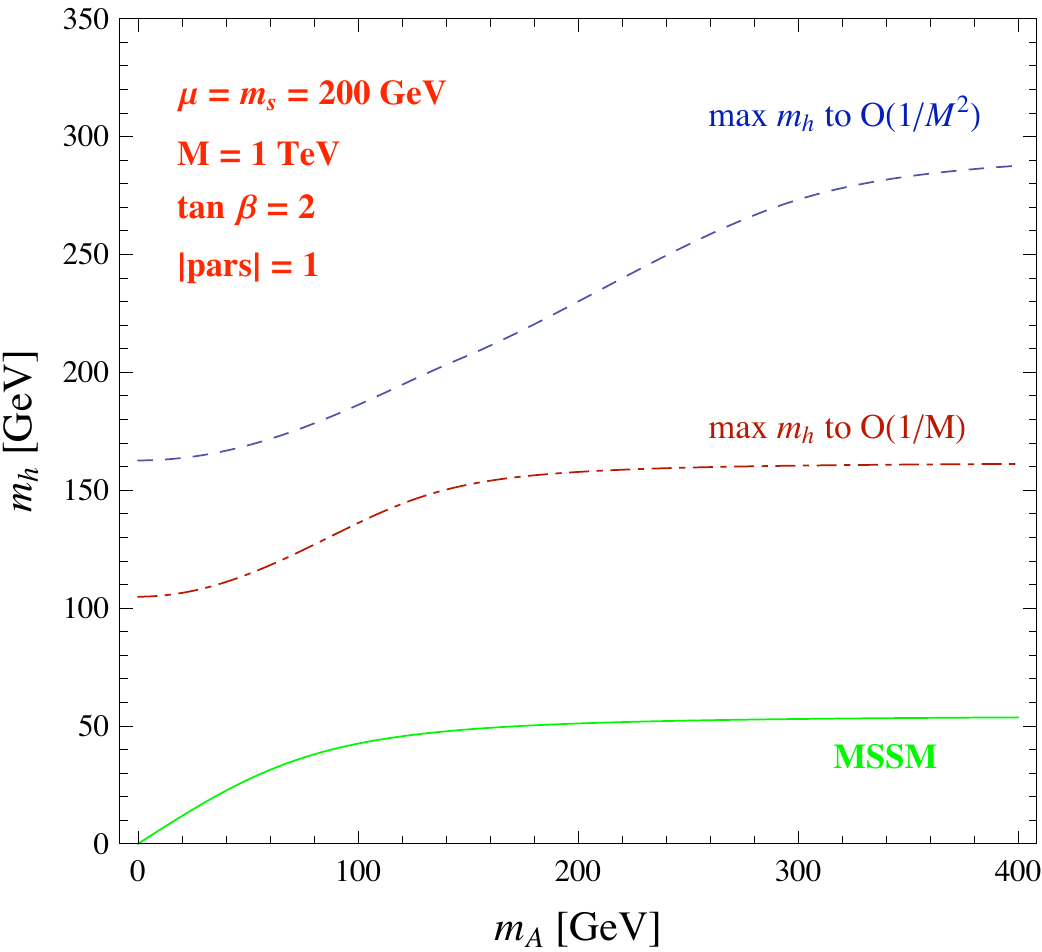}
\end{center}
\caption{\label{fig:maxmh}{\em Lightest CP-even Higgs tree-level mass,
$m_h$, as a function of $m_A$, for $\tan\beta = 2$.  The dashed blue
line corresponds to the maximum value of $m_{h}$ to ${\cal O}(1/M^2)$
when the dimensionless coefficients of the higher-dimension operators
are allowed to be as large as 1 (in absolute magnitude).  The
dashed-dotted red line corresponds to the maximum value of $m_{h}$ to
${\cal O}(1/M)$ under the same assumption.  The solid green line
corresponds to the tree-level MSSM result.}}
\end{figure}
In Fig.~\ref{fig:maxmh} we show the maximum tree-level value of the
lightest Higgs mass $m_{h}$ as a function of $m_{A}$, for small
$\tan\beta$, fixed representative values of the dimensionful
parameters, and assuming that all dimensionless parameters take values
at most equal to one in absolute value.  We show the tree-level value
of $m_{h}$ up to order $1/M$, which is obtained for $\omega_{1} =
-\alpha_{1}$ (dashed-dotted curve).  We also show the maximal values
of $m_{h}$ up to $1/M^2$ effects (dashed curve), and the MSSM (solid)
curve for comparison.  We see that the effects of the higher-dimension
operators can be quite substantial.  We emphasize that the effects of
order $1/M^3$ or higher are expected to be much smaller, as mentioned
in the introduction and discussed in Subsection~\ref{EFTValidity}.

For the heavy CP-even state we find that $(\Delta
m^2_{H^{0}})^{\textrm{Dim-5}} \approx - \frac{1}{2} \omega_{1} v^2
\alpha_1 m_{s}/M$ for $\tan\beta \approx 1$.  For $\tan\beta \gg 1$,
$(\Delta m^2_{H^{0}})^{\textrm{Dim-5}} \approx - \omega_{1} v^2
\alpha_1 m_{s}/M$ if $m_{A} \geq m_{Z}$, and $(\Delta
m^2_{H^{0}})^{\textrm{Dim-5}} \approx0$ if $m_{A} < m_{Z}$.  In the
decoupling limit $m^2_{A} \gg m^2_{Z}$, we get
\bea
(\Delta m^2_{H^{0}})^{\textrm{Dim-5}} &=& - \frac{1}{4} \omega_{1} v^2 [3 + \cos(4 \beta)] \left( \frac{\alpha_1 m_{s}}{M} \right)~.
\eea
Hence, the heavy CP-even state receives corrections at the leading
order in the $1/M$ expansion only due to the SUSY breaking operator,
Eq.~(\ref{WSpurion}).

Similarly, the charged Higgs mass takes the form
\bea
m^2_{H^{\pm}} &=& m^2_{W} + m^2_{A} - \frac{1}{2} \omega_{1} v^2 \left( \frac{\alpha_1 m_{s}}{M} \right)~,
\eea
and, at leading order in $1/M$, gets corrections only from the SUSY
breaking operator, Eq.~(\ref{WSpurion}).

Lastly, we consider the corrections to the Higgs couplings at leading
order in $1/M$.  We start from the mixing angle $\alpha$ which, at
this order is given by
\bea
t_{2 \alpha} &=& t_{2 \beta} \left(\frac{m_A^2+ m_Z^2}{m_A^2-m_Z^2}\right) + \frac{\omega_{1} v^2  \bigl(  - 2 \mu (m_A^2 - m_Z^2) + s_{2 \beta}  (m_A^2+m_Z^2) \alpha_1 m_s  \bigr)  }{M c_{2 \beta} (m_A^2-m_Z^2)^2}~.
\eea
In the decoupling limit this simplifies to
\bea
t_{2 \alpha} &=&  t_{2 \beta} + 2  t_{2 \beta} x + \frac{\omega_{1} v^2}{m_A^2 c_{2 \beta}} \frac{1}{M} \Bigl[ -2 \mu + s_{2 \beta} \alpha_1 m_s \Bigr]~,
\eea
where $x=m_Z^2 / m_A^2$.  Taking into account that $\alpha$ is in the
fourth quadrant, while $\beta$ belongs to the first one, one gets
\bea
\alpha &=& \beta-\frac{\pi}{2} + s_{2 \beta} c_{2 \beta} x + \frac{\omega_{1} v^2}{2m_A^2} \frac{c_{2 \beta}}{M} \Bigl[ -2 \mu + s_{2 \beta} \alpha_1 m_s \Bigr]
\nonumber \\
&\equiv& \beta-\frac{\pi}{2} + A^{(1)} x + A^{(2)}/M~.
\eea
This implies that the couplings of the CP-even Higgs fields to two
gauge bosons [see Eqs.~(\ref{MSSMhVV})] are
\bea
\begin{array}{ccccccc}
hVV &:& 1+ {\cal O}(x^{2},v^2/M^{2})~,  & \hspace{5mm} & 
HVV &:& A^{(1)} x + A^{(2)}/M~.
\label{hHVVLargemA}
\end{array}
\eea
Note that the couplings of the light state to gauge bosons do not
receive corrections at order $1/M$ and are not expected to deviate
very much from the MSSM ones, while those of the heavy state are
expected to get larger corrections.  The couplings to fermion pairs
relative to the SM are [see Eqs.~(\ref{MSSMhff})]
\bea
\begin{array}{ccccccc}
h\bar{t}t &:& 1 + t^{-1}_{\beta} (A^{(1)} x + A^{(2)}/M)~,
& \hspace{5mm} & 
H\bar{t}t &:& - \tb^{-1} \left[1 - \tb (A^{(1)} x + A^{(2)}/M) \right]~,
\\ [0.4em]
h\bar{b}b &:& 1 - \tb (A^{(1)} x + A^{(2)}/M)~,
& \hspace{5mm} & 
H\bar{b}b &:& - \tb \left[ 1 + \tb^{-1} (A^{(1)} x + A^{(2)}/M) \right]~.
\end{array}
\eea
Thus, in the decoupling limit there could be important variations of
the light state couplings to the up and down sectors with respect to
the SM predictions.  However, the $h\bar{t}t$ coupling remains SM-like
in the large $\tan\beta$ regime.  Similarly important variations occur
in the couplings of the heavy Higgs to the up-type and down-type
fermions, except for the $H\bar{b}b$ coupling in the large $\tan\beta$
regime, where the variations are small.

\section{Numerical analysis: Parameters and constraints}
\label{sec:definitions}

In the previous sections we worked out the spectrum and couplings of
the Higgs sector in a softly broken supersymmetric theory, under the
assumption that there is a set of particles that couple to the MSSM
Higgs fields, but that have masses parametrically larger than the weak
scale.  We also assumed that the SUSY breaking mass splittings in the
heavy supermultiplets are small, so that their masses have a nearly
supersymmetric origin.  In this section we define the regions or
parameter space and discuss several constraints that will be used in
the numerical exploration of the effects of the heavy physics on the
MSSM Higgs sector, performed in Section~\ref{sec:results}, and further
expanded in \cite{CKPZ}.

\subsection{Parameter space}
\label{parameters}

We start by defining our region of parameter space.  We assume that
all SUSY breaking mass parameters (as well as $\mu$) are of order the
EW scale (a couple hundred GeV).  For definiteness, we take $\mu =
m_{s} = 200~{\rm GeV}$, where $m_{s}$ is the F-component of the
spurion superfield.  We also set $M = 1~{\rm TeV}$.  In addition, we
scan over the dimensionless parameters defined in
Eqs.~(\ref{genw})--(\ref{KSpurionCP}) as follows:
\begin{itemize}
\item $|\omega_{1}|$, $|c_{1}|$, $|c_{2}|$, $|c_{3}|$, $|c_{4}|$, $|c_{6}|$, $|c_{7}|$ $\in [0,1]$.
\item $|\alpha_{1}|$, $|\beta_{i}|$, $|\gamma_{i}|$, $|\delta_{i}|$  $\in [1/3,1]$ for $i = 4,6,7$.
\end{itemize}
Recall that the coefficients $\omega_{1}$, $c_{1}$, $c_{2}$, $c_{3}$,
$c_{4}$, $c_{6}$ and $c_{7}$ set the size of the $1/M$ and $1/M^2$
suppressed operators (both SUSY-preserving and SUSY-breaking).  To be
definite we assume that their values are at most one in absolute
magnitude, but it should be clear that if they were larger the
physical effects would be correspondingly larger.  Our choice for the
ranges of $|\alpha_{2}|$, $|\beta_{i}|$, $|\gamma_{i}|$,
$|\delta_{i}|$ reflects our assumption that the SUSY-breaking
operators are proportional to the corresponding SUSY preserving ones,
so that these are parameters of order one.

We consider two representative values of $\tan\beta$: $\tan\beta = 2$
and $\tan\beta = 20$.  We also vary the CP-odd mass up to $400~{\rm
GeV}$, which is still below $M$, ensuring a proper separation between
the light and heavy scales, as required by the EFT analysis.

Note that scaling $\mu$, $m_{s}$ and $M$ by a common factor leaves the
corrections due to the higher-dimension operators, Eqs.~(\ref{dim5})
and (\ref{dim6}), unchanged [though not those of
Eqs.~(\ref{eq:honest}) and (\ref{eq:kinmix})].  In particular, the
leading order (dim-5) operators depend on $\mu$, $m_{s}$ and $M$ only
through the ratios $\mu/M$ and $m_{s}/M$.  Thus, these effects could
be relevant even if the scale of new physics is much higher than we
envision here, if SUSY breaking and/or $\mu$ are correspondingly
larger.  Even though we do not scan over the values of $\mu$, $m_{s}$
and $M$ (but rather fix them as specified above), our results should
be qualitatively applicable when all these scales are higher, keeping
the ratios fixed [the difference arises at ${\cal O}(1/M^2)$ through
the operators of Eqs.~(\ref{eq:honest}) and (\ref{eq:kinmix})].

\subsection{Uncertainty from higher orders and the EFT expansion}
\label{EFTValidity}

As was mentioned in the introduction (see also Fig.~\ref{fig:maxmh}),
the contributions of order $1/M$ and $1/M^2$ can be phenomenologically
sizable: the $1/M$ effects turn on Higgs quartic couplings not present
in the MSSM at tree-level, while the $1/M^2$ effects can easily be
comparable to the MSSM contribution which is set by the weak gauge
couplings.  However, one should make sure that the next order can be
reasonably expected to give a small contribution, since otherwise it
could signal the need to resum the $1/M$ effects to all orders (in
which case the details of the UV completion are essential and the EFT
approach ceases to be useful).  It can also happen (and does happen in
our random numerical scans) that there are accidental cancellations
between the MSSM contributions proportional to $g^2$ and those at
order $1/M^2$ [see Eq.~(\ref{dim6})], or also, if $\omega_{1}$ is
somewhat suppressed, between the order $1/M$ and $1/M^2$ effects.  In
such cases, the next order corrections can have a larger impact than
naively expected, and our $1/M^2$ analysis would fail to capture the
quantitative properties of such points in parameter space.

We perform a simple test to assess the robustness of a given point in
parameter space against higher-order corrections that have not been
computed (and that would depend on new coefficients that are arbitrary
from the EFT point of view).  Since the most important effects are
expected to enter through the Higgs quartic couplings, we model the
next order corrections as follows:\footnote{Notice that at order
$1/M^3$ there are only two new operators: the superpotential term
$(H_{u} H_{d})^3$ and the associated SUSY breaking operator.  There
are no new \Kahler operators at this order.  However, there are
$1/M^3$ effects associated with the lowest order coefficients, e.g.
proportional to $\omega_{1} c_{i}$.} keeping all the Lagrangian
parameters fixed ($m^2_{u}$, $m^2_{d}$, $b$, $\omega_{1}$,
$c_{i}$, $\alpha_{1}$, $\gamma_{i}$, $\beta_{i}$, $\delta_{6}$,
$\delta_{7}$, for $i=1,2,3,4,6,7$), we modify the quartic couplings by
hand as:~
\bea
\lambda_{i} \rightarrow \lambda_{i} \pm 2 \, {\rm Max}\left\{ 
|\omega_{1}|, |c_{1}|, |c_{2}|, |c_{3}|, |c_{4}|, |c_{6}|, |c_{7}| \right\} 
\left( \frac{\mu}{M} \right)^3~,
\hspace{1cm} i = 1, \ldots 7~,
\label{newlambdas}
\eea
i.e. we allow for $1/M^3$-suppressed operators with dimensionless
coefficients as large as those of the leading two orders in the $1/M$
expansion [the factor of $2$ models numerical factors that may appear,
as we have seen in the order $1/M^2$ expressions, Eqs.~(\ref{dim6})].
We then solve the minimization equations, (\ref{m2hu}) and
(\ref{m2hd}), with these new $\lambda_{i}$'s, which leads to values of
$v$ and $\tan\beta$ that are slightly different from their values in
the absence of the modification.  The amount by which these two
observables change should give a reasonable measure of the sensitivity
of the given parameter point (truncated at order $1/M^2$) to effects
from the next order.\footnote{Notice that the way to interpret a VEV
different from $246~{\rm GeV}$ in the presence of the modification
Eq.~(\ref{newlambdas}) is as follows: one should rescale \textit{all}
mass parameters by an appropriate factor so that one recovers the
``observed'' value for $v$.  This corresponds to normalizing to the
measured $Z$ mass, and does not change the physics since all mass
ratios are kept fixed.  Since all mass scales, in particular the
physical Higgs masses, are rescaled by the same factor, we see that
the change in the VEV due to the modification Eq.~(\ref{newlambdas})
corresponds directly to a change in the spectrum.  There can be
additional contributions to the spectrum from the higher-dimension
operators, not associated with the shift in the VEV, but we expect
that these are of the same order as the effect from the shift in the
VEV we have described.  We therefore consider the above a reasonable
estimate of the uncertainties associated with the higher order
operators.} We restrict to parameter points for which the above
procedure leads to a change of no more than $10\%$ in $v$.  This
should be taken as no more than an order of magnitude estimate of the
uncertainties associated with the truncation of the tower of
higher-dimension operators at order $1/M^2$.  We impose a looser
constraint on $\tan\beta$: for points with $\tan\beta = 2$, the point
is allowed if $1.5 < \tan\beta < 2.5$ after the modification
Eq.~(\ref{newlambdas}); for points with $\tan\beta = 20$, the point is
allowed if $15 < \tan\beta < 25$ after the modification of the
$\lambda$'s.  The rationale for a looser constraint in the shift in
$\tan\beta$ is that these shifted values would still be representative
small and large $\tan\beta$ cases, respectively, and no dramatic
change in the Higgs properties is expected within the above ranges.

\subsection{Global versus local minima}
\label{global}

The potential given by Eqs.~(\ref{eq:pot}) and (\ref{eq:honest}), $V =
V_{\rm ren.} + V_{\rm non-ren.}$, has in general several minima, which
may also include CP-violating or charge breaking VEV's.  Without loss
of generality, we can parameterize all minima by $\langle H_{u}
\rangle = (0,v_{u})$ and $\langle H_{d} \rangle = (v_{d} + i
v_{\delta},v_{CB})$, where $v_{u}$, $v_{d}$, $v_{\delta}$ and $v_{CB}$ are
real.  We can choose this form for $\langle H_{u} \rangle$ by
performing an appropriate $SU(2)_{L}$ rotation.  It is also clear from
the fact that the scalar potential depends only on $H^\dagger_{u}
H_{u}$, $H^\dagger_{d} H_{d}$, and $H_{u} H_{d}$ that, having set
$H^{+}_{u} = 0$, it depends only on $|H^{-}_{d}| \equiv v_{CB}$, so we
can assume that this latter VEV is real and non-negative.  We look
numerically for all solutions of the minimizations conditions
$\partial_{v_{u}} V = \partial_{v_{d}} V = \partial_{v_{\delta}} V =
\partial_{v_{CB}} V = 0$, which are polynomial in the variables, in
order to check that the minimum being considered is the global one,
and in particular preserves charge and CP. In this work we do not
consider the possibility of explicit nor spontaneous CP violation.
Also, we do not consider cases where the minimum is metastable but
long-lived, although such a case can in principle occur and could be
phenomenologically viable.

We also recall here the possibility of having different types of
(global) minima, as emphasized in~\cite{Batra:2008rc}.  The point is
that at the level of quartic couplings, the $\lambda_{6}$ and
$\lambda_{7}$ operators in Eq.~(\ref{eq:pot}) lead to runaway
directions.  These are stabilized by dimension-6 operators in the
scalar potential, Eq.~(\ref{eq:honest}).  As a result there can appear
minima that arise from balancing renormalizable versus
non-renormalizable terms. A simple way to
characterize them is to test the scaling $v \propto \sqrt{M}$ in the
large $M$ limit, keeping all other Lagrangian parameters fixed~(see
\cite{Batra:2008rc} for more details).  In contrast, for MSSM-like
minima the VEV reaches a finite constant in this limit.  We will use
this criterion to characterize the type of vacua.

\subsection{Electroweak constraints}
\label{EWPT}

Lastly, to assess the viability of a given parameter point, we also
estimate the contributions to the oblique parameters in order to
select points that are in reasonable agreement with the EW precision
constraints.  As seen in Eq.~(\ref{alphaT}), the heavy physics can
induce tree-level contributions to the Peskin-Takeuchi $T$
parameter~\cite{Peskin:1990zt} (see also
Appendices~\ref{app:UVcompletions} and \ref{app:custodial}).  If one
requires $|T| \sim 0.2$ and constrains $c_{1}$, $c_{2}$ and $c_{3}$
one at a time, one sees that for $\tan\beta > 1$ and $M = 1~{\rm
TeV}$, the strongest constraint is $c_{2} < 0.1$.  Due to the
$\tb^{-4}$ and $\tb^{-2}$ suppression in Eq.~(\ref{alphaT}), $c_{1}$
and $c_{3}$ are less constrained for $\tan\beta > 1$.  For instance,
if $\tan\beta = 2$ one has $c_{1} < 1.3$ and $c_{3} < 0.15$ (and are
virtually unconstrained for $\tan\beta = 20$).  Notice also that there
can be partial cancellations in Eq.~(\ref{alphaT}).~\footnote{As shown
in Appendix~\ref{app:UVcompletions}, Higgs triplets give $c_{1},
c_{2}, c_{3} > 0$.  This corresponds to a positive tree-level
contribution to $T$ from triplets with zero hypercharge and negative
from triplets with $Y = \pm 1$.  Ref.~\cite{DiChiara:2008rg} points
out that for lighter triplets it is also possible to have a
cancellation in $T$ arising from the interplay between the triplet
mass and the $\mu$-term.  } Thus, even if $c_{1}$, $c_{2}$ and $c_{3}$
are of order one, provided $M \sim 1~{\rm TeV}$, these contributions
to $T$ are not necessarily much larger than the experimental limit for
a reference Higgs mass of $m_{H_{\rm ref}} = 117~{\rm GeV}$.

There are other contributions to $T$ that can be comparable to the
tree-level one --which is suppressed by $v^2/M^2$-- from loops of the
MSSM Higgs fields (SM-like Higgs heavier than $m_{H_{\rm ref}}$, and
mass splittings among the non-standard Higgses) as well as from
custodially violating mass splittings in the (light) superpartner
sector.  The former give the following contributions to $T$ and
$S$~\cite{Inami:1992rb,Choudhury:2002qb}:
\bea
\alpha T^{\rm Higgs} &\approx& \frac{\alpha}{16\pi s^2_{W} m^2_{W}} \left\{ \rule{0mm}{5.5mm}
(\eta^{\rm eff}_{W^\pm H^\mp A})^2 f(m_{H^\pm}, m_{A}) 
+ (\eta^{\rm eff}_{W^\pm H^\mp h})^2 f(m_{H^\pm}, m_{h})  
\right.
\nonumber \\[0.4 em]
&& \left. \mbox{}
+ (\eta^{\rm eff}_{W^\pm H^\mp H})^2 f(m_{H^\pm}, m_{H}) 
- (\eta^{\rm eff}_{Z h A})^2 f(m_{h}, m_{A}) 
- (\eta^{\rm eff}_{Z H A})^2 f(m_{H}, m_{A}) 
\rule{0mm}{5.5mm} \right\}
\nonumber \\[0.4em]
&& \mbox{} +
\Delta \rho_{\rm SM}(m_{h}) + \Delta \rho_{\rm SM}(m_{H}) - \Delta \rho_{\rm SM}(m_{H_{\rm ref}})~,
\label{TMSSMHiggses}
\\[0.5em]
S^{\rm Higgs} & \approx& \frac{1}{12\pi} \left\{ \rule{0mm}{5.5mm}
(\eta^{\rm eff}_{ZhA})^2 F(m_{h},m_{A}) + (\eta^{\rm eff}_{ZHA})^2 F(m_{H},m_{A})
- (\eta^{\rm eff}_{ZH^\pm H^\mp})^2 \ln m^2_{H^\pm} 
\right.
\nonumber \\[0.4em]
&& \left. \mbox{}
+ hZZ^2 \ln m^2_{h} + HZZ^2 \ln m^2_{H} - \frac{5}{6} - \ln m^2_{H_{\rm ref}}
 \right\}~,
\label{SMSSMHiggses}
\eea
where $\alpha$ is the fine structure constant, $H_{\rm ref}$ is the
reference SM Higgs mass used in the fit to the EW precision
measurements, $hVV/HVV$ with $V=Z,W$ are defined in
Eqs.~(\ref{hcouplings}), $\eta^{\rm eff}_{W^\pm H^\mp A} \equiv
\frac{1}{2} (\eta_{W^\pm H^\mp A} + \eta_{W^\pm A H^\mp})$, with
$\eta_{W^\pm H^\mp A}$, $\eta_{W^\pm A H^\mp}$ as defined in
Eqs.~(\ref{eq:trilinear}) [and similar definitions for the other
$\eta^{\rm eff}$'s], and
\bea
f(x,y) &=& \frac{1}{2} \left( x^2 + y^2 \right) - \frac{x^2 y^2}{x^2 - y^2} \, \ln \left( \frac{x^2}{y^2} \right)~,
\nonumber \\[0.5em]
\Delta \rho_{\rm SM}(H_{i}) &=& \frac{3\alpha}{16\pi s^2_{W} m^2_{W}} \left[ H_{i}ZZ^2 f(m_{H_{i}}, m_{Z}) - H_{i}WW^2 f(m_{H_{i}}, M_{W}) \right]
- \frac{\alpha}{8\pi c^2_{W} m^2_{W}}~,
\nonumber\\[0.5em]
F(m_{1}, m_{2}) &=& \ln (m_{1} m_{2}) + \frac{2m^2_{1} m^2_{2}}{(m^2_{1} - m^2_{2})^2} + \frac{(m^2_{1} + m^2_{2})(m^4_{1} + m^4_{2} - 4 m^2_{1} m^2_{2})}{(m^2_{1} - m^2_{2})^3} \, \ln\left( \frac{m_{1}}{m_{2}} \right)~,
\nonumber
\eea
where $H_{i} = h, H$.  It should be noticed that the expressions
(\ref{TMSSMHiggses}) and (\ref{SMSSMHiggses}) are only approximate
since there can be logarithmically divergent terms in the loop
computation of the $T$ and $S$ parameters, in the effective theory
with higher-dimension operators, that scale like $v^2/M^2$ (the
quadratically divergent contributions to the gauge boson self-energies
vanish by gauge invariance).  However, due to the loop suppression,
these contributions are expected to be much smaller than the
tree-level one, given by Eq.~({\ref{alphaT}}), and are therefore
negligible for points where the latter is within the experimental
limits (the logarithmic enhancement is small for the scale of new
physics we consider).

As remarked in~\cite{Medina:2009ey} when the SUSY particles are light
there can be additional relevant contributions to the $T$ parameter.
These depend on parameters that do not affect directly the Higgs
sector.  As a result, we do not perform here a detailed fit to the EW
data.  However, we use Eqs.~(\ref{TMSSMHiggses}) and
(\ref{SMSSMHiggses}) plus the tree-level contribution,
Eq.~(\ref{alphaT}), to estimate whether a given point can be
reasonably expected to agree with the precision constraints if
appropriate SUSY contributions were added.  In
Ref.~\cite{Medina:2009ey} it was found that such SUSY contributions
are positive and easily as large as $ \Delta T^{SUSY} \sim 0.2$.
Taking $-0.2 < T^{\rm tot} < 0.3$ ($95\%$ C.L.),\footnote{We use the
code of Ref.~\cite{Han:2004az} with $m_{t} = 173.1 \pm 1.3~{\rm
GeV}/c^2$~\cite{:2009ec}, $M_{W} = 80.432 \pm 0.039~{\rm
GeV}/c^2$~\cite{:2008ut}, and $m_{H_{\rm ref.}} = 117~{\rm GeV}$.} we
therefore allow only points with $-0.4 < \tilde{T} + T^{\rm Higgs} <
0.3$.  After this cut we find that all points in the scan satisfy
$-0.05 < S^{\rm Higgs} < 0.08$, so that we do not impose any further
cuts on $S$.

\subsection{Loop effects}
\label{loops}

We have implemented our tree-level expressions for the spectrum and
Higgs couplings in HDECAY v3.4~\cite{Djouadi:1997yw}.  This allows us
to also take into account the QCD radiative corrections, that are
known to be sizable.  In addition, we include the radiative
corrections derived from the 1-loop RG improved effective potential
due to the supersymmetric particles, as computed
in~\cite{Carena:1995bx}.  Loop contributions from the heavy physics
that has been integrated out are suppressed by both a loop factor and
by powers of $M$, hence they are expected to be negligible.

As we will argue in the next section (see also Fig.~\ref{fig:maxmh}),
already at tree-level the Higgs spectrum satisfies the bounds from LEP
in large regions of parameter space.  Hence, the motivation for taking
rather large stop masses is absent in the extensions that we study.
Keeping with the philosophy that the SUSY breaking scale is small
compared to $M$, we consider a relatively light superpartner spectrum.
For concreteness we take the superpartner soft parameters to have a
common value $M_{SUSY} = 300~{\rm GeV}$ and $A_{t} = A_{b} =
0$.~\footnote{We denote the soft breaking masses by $M_{SUSY}$.  We
evaluate the scale inside the logarithms associated with SUSY loops at
$\sqrt{M^2_{SUSY} + m^2_{t}} \approx 347~{\rm GeV}$.} Thus, the SUSY
loop contributions to the Higgs masses are modest, but the loop
contributions to the Higgs couplings are more important and sensitive
to the details of the SUSY
spectrum~\cite{Carena:1994bv,Dawson:1996xz}.  The point to remember is
that the relevant loop-level effects can be fully computed given the
MSSM superpartner spectrum, and are only mildly dependent on the
details of the UV theory that gives rise to the EFT that we study.

\section{Numerical analysis: Results}
\label{sec:results}

In this section we present the results for the Higgs spectrum and
couplings that arise from a scan over the parameter region defined in
Subsection~\ref{parameters}.  We make sure that we concentrate on
parameter points that are expected to be relatively insensitive to
higher orders in the $1/M$ expansion, that they correspond to global
minima, and that they can be in agreement with the EW precision
constraints, given the uncertainties from the SUSY spectrum, as
discussed in Subsections~\ref{EFTValidity}, ~\ref{global}
and~\ref{EWPT}.  Since our choice of points depends on the
``robustness'' criterion described in Section~\ref{EFTValidity}, we
start in Subsection~\ref{robusteness} by commenting on the
consequences of this prescription.  We present these results in the
\textit{tree-level} approximation, so as to emphasize the effects of
the higher-dimension operators.

Next we present the results for some selected observables, such as the
gluon fusion production cross section relevant at hadron colliders,
and comment on certain ``exotic'' branching fractions.  For these we
include the supersymmetric radiative effects to the Higgs masses and
couplings, as described in Subsection~\ref{loops}.  We will present a
more complete study of the Higgs phenomenology in~\cite{CKPZ}, where
we will also include the detailed bounds from LEP and the Tevatron on
the Higgs spectrum.

\subsection{Sensitivity against higher-order corrections}
\label{robusteness}

%
\begin{figure}[t]
\begin{center}
\includegraphics[width=7.9cm]{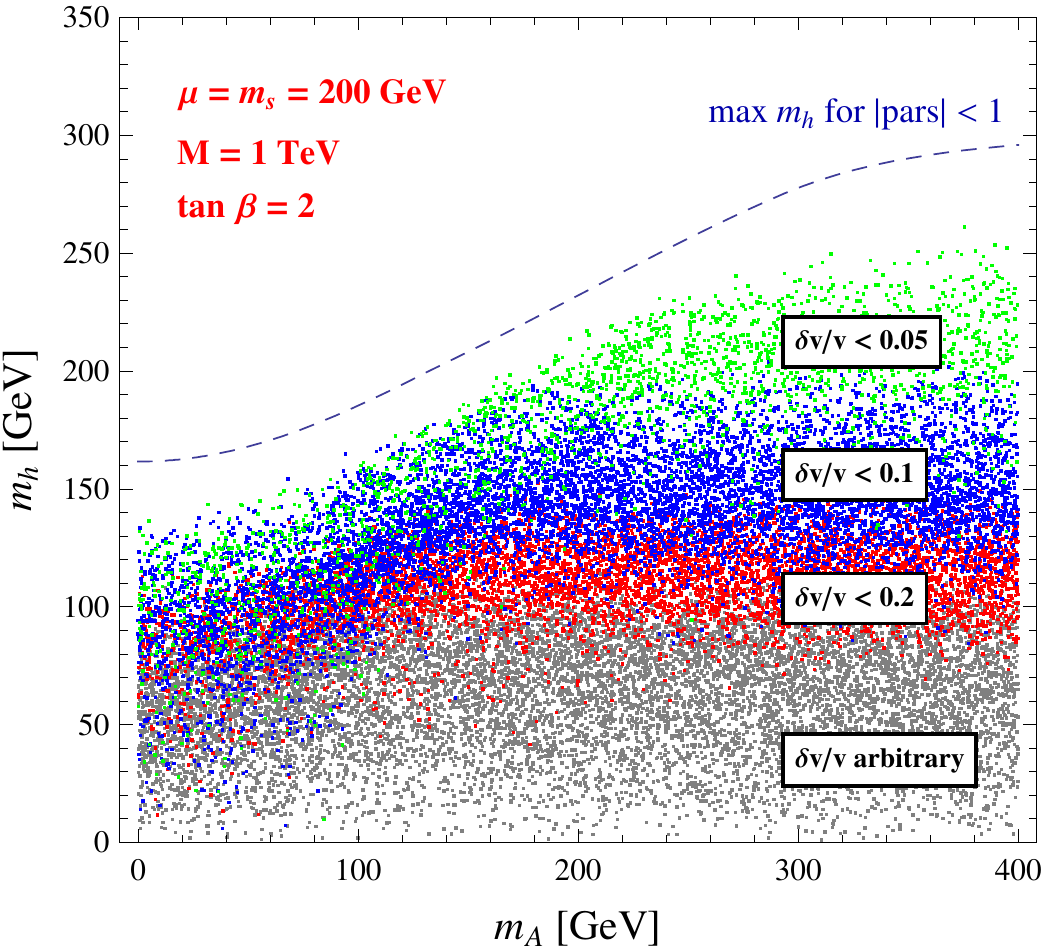}
\hspace{3mm}
\includegraphics[width =7.9cm]{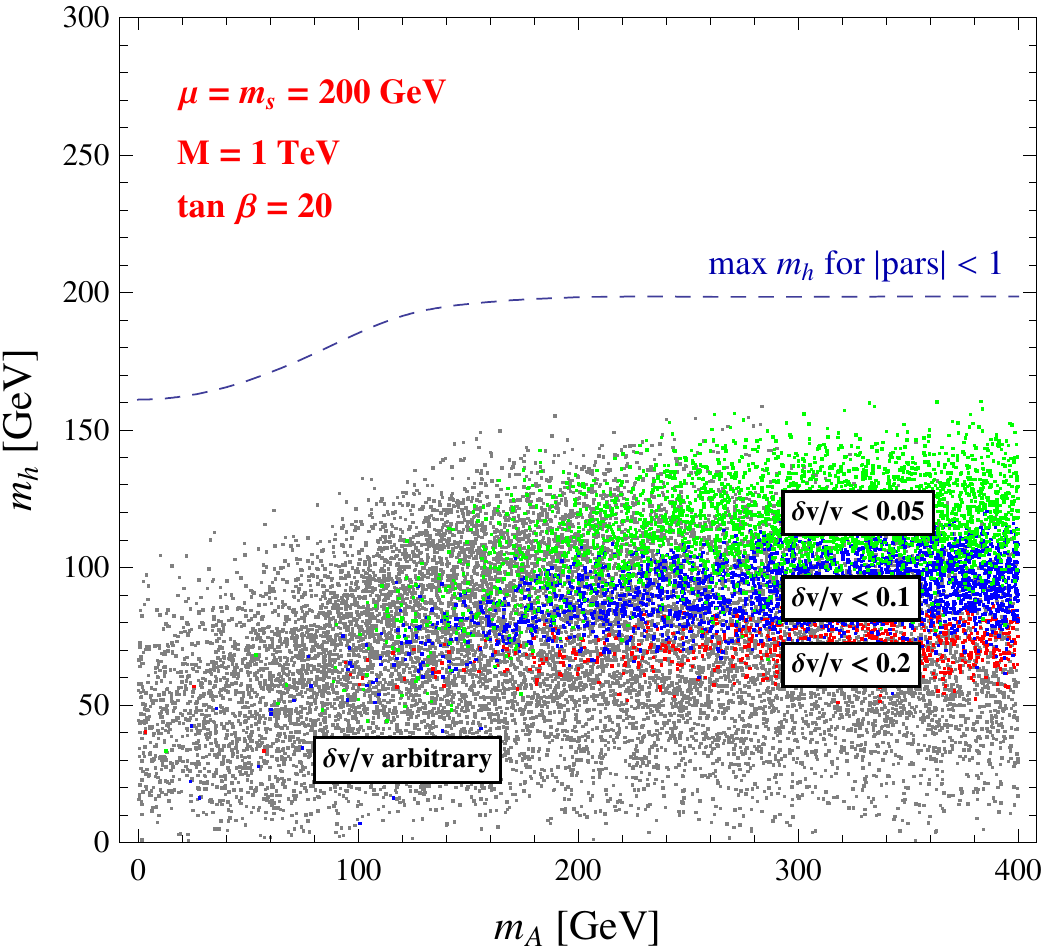}
\end{center}
\caption{\label{fig:mhmASensitivity}{\em Illustration of the
sensitivity of $m_h$ against higher order effects in the $1/M$
expansion, as a function of $m_A$, for $\tan\beta = 2$ (left panel)
and for $\tan\beta = 20$ (right panel).  The regions in (green, green
$+$ blue, green $+$ blue $+$ red) correspond to the requirement
($\delta v/v < 0.05$, $\delta v/v < 0.1$, $\delta v/v < 0.2$),
according to the prescription described in
Subsection~\ref{EFTValidity}.  The gray points are the additional
points in the scan that do not obey any of these three requirements.
The dashed blue line corresponds to the maximum tree-level value of
$m_{h}$ when the dimensionless coefficients of the higher-dimension
operators are allowed to be as large as 1 (in absolute magnitude).
The region of parameter space in the scan is described in the main
text.  }}
\end{figure}

We described in Subsection~\ref{EFTValidity} a simple prescription to
estimate the sensitivity of a given point in parameter space against
${\cal O}(1/M^{3})$ effects.  In this subsection we illustrate the
dependence on the allowed variation in $v$ by requiring that the VEV
change by less than $5\%$, $10\%$ and $20\%$ if the next order
corrections take a ``typical'' size, as estimated in
Subsection~\ref{EFTValidity}.  As an example, we show in
Fig.~\ref{fig:mhmASensitivity} $m_{h}$ as a function of $m_{A}$ for
$\tan\beta = 2$ (left panel) and $\tan\beta = 20$ (right panel).  We
have scanned over a total of $10^{5}$ points, but show only those
points that correspond to a global minimum and that can be in
reasonable agreement with the EW precision constraints, as explained
in Subsection~\ref{EWPT}.  For $\tan\beta = 2$ ($\tan\beta = 20$), we
find that about $70\%$ ($80\%$) of the points in the scan correspond
to global minima of the potential $V = V_{\rm ren.} + V_{\rm
non-ren.}$, defined by Eqs.~(\ref{eq:pot}) and (\ref{eq:honest}),
while the rest are only local minima that we discard.  The EW
precision constraints further reduce the number of potentially viable
points in the scan by $70\%$ ($80\%$).  In
Fig.~\ref{fig:mhmASensitivity} we exhibit how the number of points is
further reduced by requiring $\delta v/v < 0.2$ (green $+$ blue $+$
red regions), $\delta v/v < 0.1$ (green $+$ blue) and $\delta v/v <
0.05$ (green region), following the prescription of
Subsection~\ref{EFTValidity} (in the figures we apply the requirements
on $\delta \!  \tan\beta$ described in that subsection).  Recall that
this gives a measure of the sensitivity of a given point against
higher orders in the $1/M$ expansion.  The points shown in gray are
the additional points that would change by $\delta v/v > 0.2$ under
the modification of Eq.~(\ref{newlambdas}), and without any
restriction on $\delta \!  \tan\beta$ (i.e. in
Fig.~\ref{fig:mhmASensitivity} we show \textit{all} the points in the
scan that are global minima and obey the EW precision constraints).

We observe that the points with smaller $m_{h}$ are more easily
affected by order $1/M^3$ corrections (for instance, they can lead to
no EWSB after such a perturbation).  This is not to say that there can
be no models with small $m_{h}$, but only that points were $m_{h}$ is
heavier are relatively insensitive to those higher-order corrections.
The reason is that larger $m_{h}$ indicates that both the $1/M$ and
$1/M^2$ effects are contributing fully, without major accidental
cancellations (see Subsection~\ref{EFTValidity}).  The next order is
then suppressed by order $v/M$, as expected (again, note that all our
dimensionless couplings are at most one).  Note also that for
$\tan\beta = 20$, the points with small $m_{A}$ are found to be rather
sensitive to the higher-order corrections, and tend to be discarded
using our prescription.  Again, this is not to say that viable models
with large $\tan\beta$ and small $m_{A}$ do not exist, but only that
their properties may not be correctly captured at the order we are
working, so we choose not to concentrate on such cases.  In the
following, we restrict to points that satisfy $\delta v/v < 0.1$,
which should be interpreted as points for which the higher-order
corrections introduce an uncertainty of at most order $10\%$.
However, for many points the expected uncertainty should be smaller.

In the figure we also show the maximal value of $m_{h}$ for the
parameter region defined in Subsection~\ref{parameters}.  This
envelope was obtained by optimizing the values of the dimensionless
model parameters so as to maximize $m_{h}$.  The reason that the
points in the scan itself do not reach such large values of $m_{h}$ is
that there is a low probability  that \textit{all} the model parameters
simultaneously attain the optimal values that maximize $m_{h}$.
 
\subsection{Higgs Masses: Comparison to the MSSM}
\label{masses}

%
\begin{figure}[t]
\begin{center}
\includegraphics[width=7.9cm]{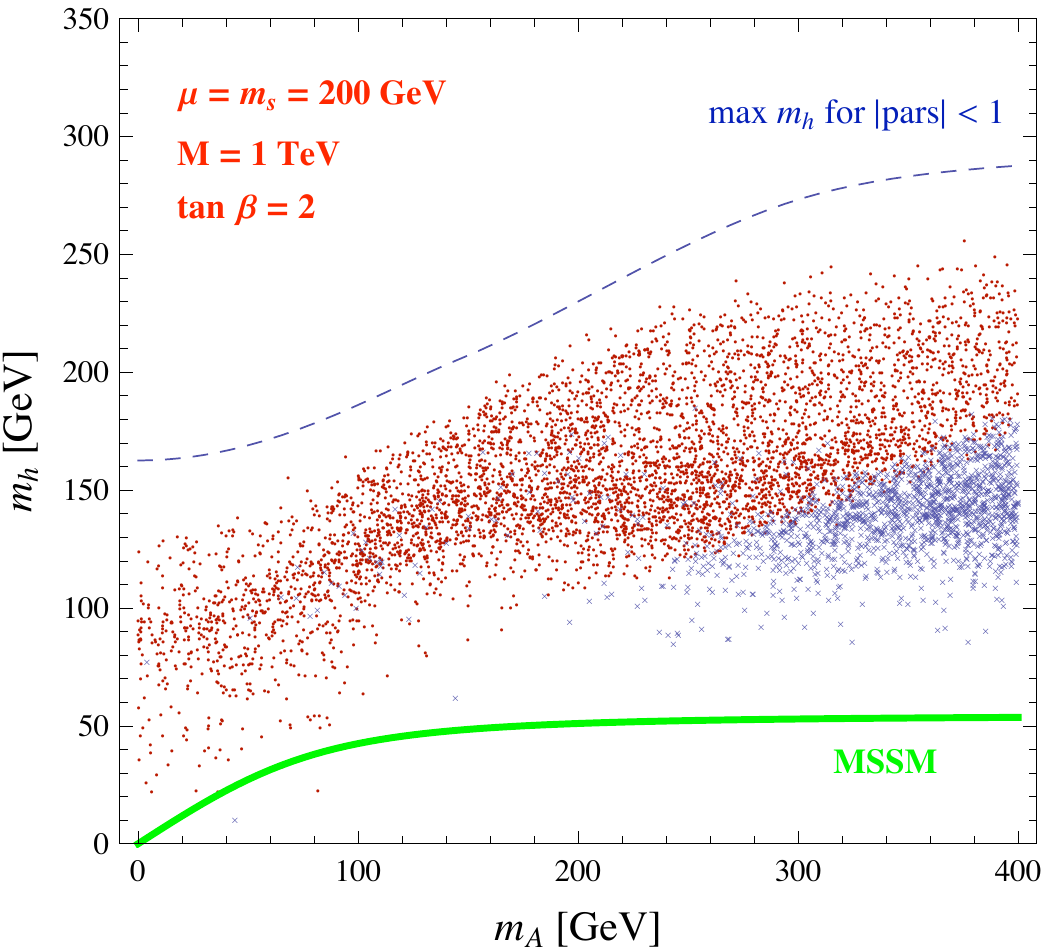}
\hspace{3mm}
\includegraphics[width =7.9cm]{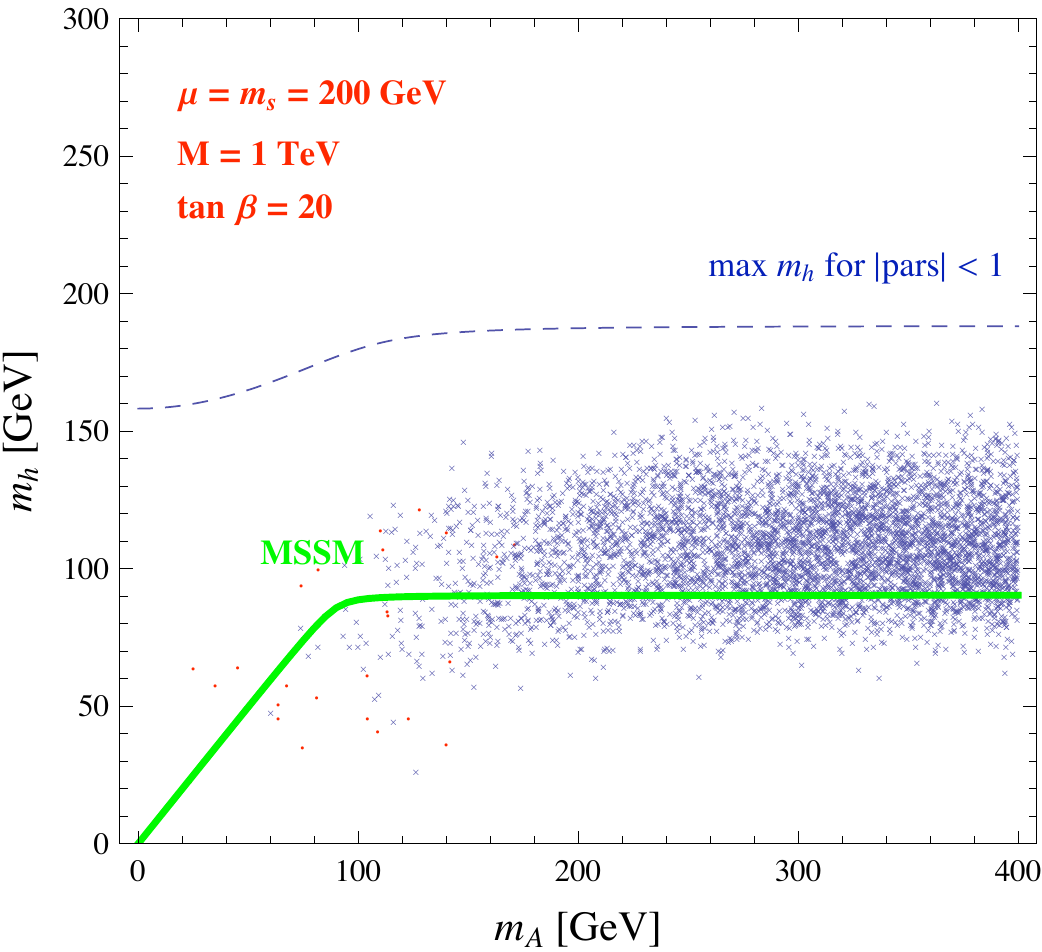}
\end{center}
\caption{\label{fig:mhmA}{\em Lightest CP-even Higgs tree-level mass,
$m_h$, as a function of $m_A$, for $\tan\beta = 2$ (left panel) and
for $\tan\beta = 20$ (right panel).  The dashed blue line corresponds
to the maximum value of $m_{h}$ when the dimensionless coefficients of
the higher-dimension operators are allowed to be as large as 1 (in
absolute magnitude).  The region of parameter space in the scan is
described in the main text.  The solid green line corresponds to the
tree-level MSSM result.  Red points correspond to sEWSB vacua, while
blue crosses correspond to MSSM-like vacua.}}
\end{figure}
\begin{figure}[!htp]
\begin{center}
\includegraphics[width=7.9cm]{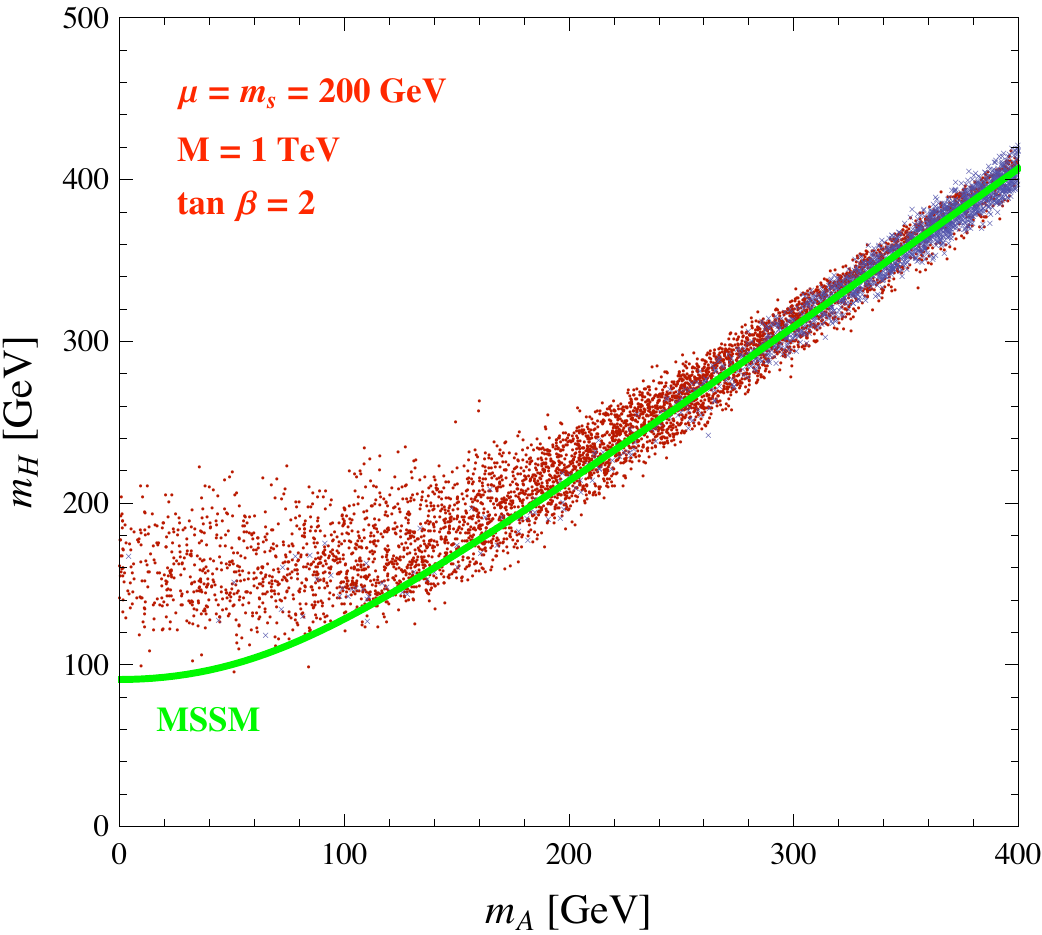}
\hspace{3mm}
\includegraphics[width=7.9cm]{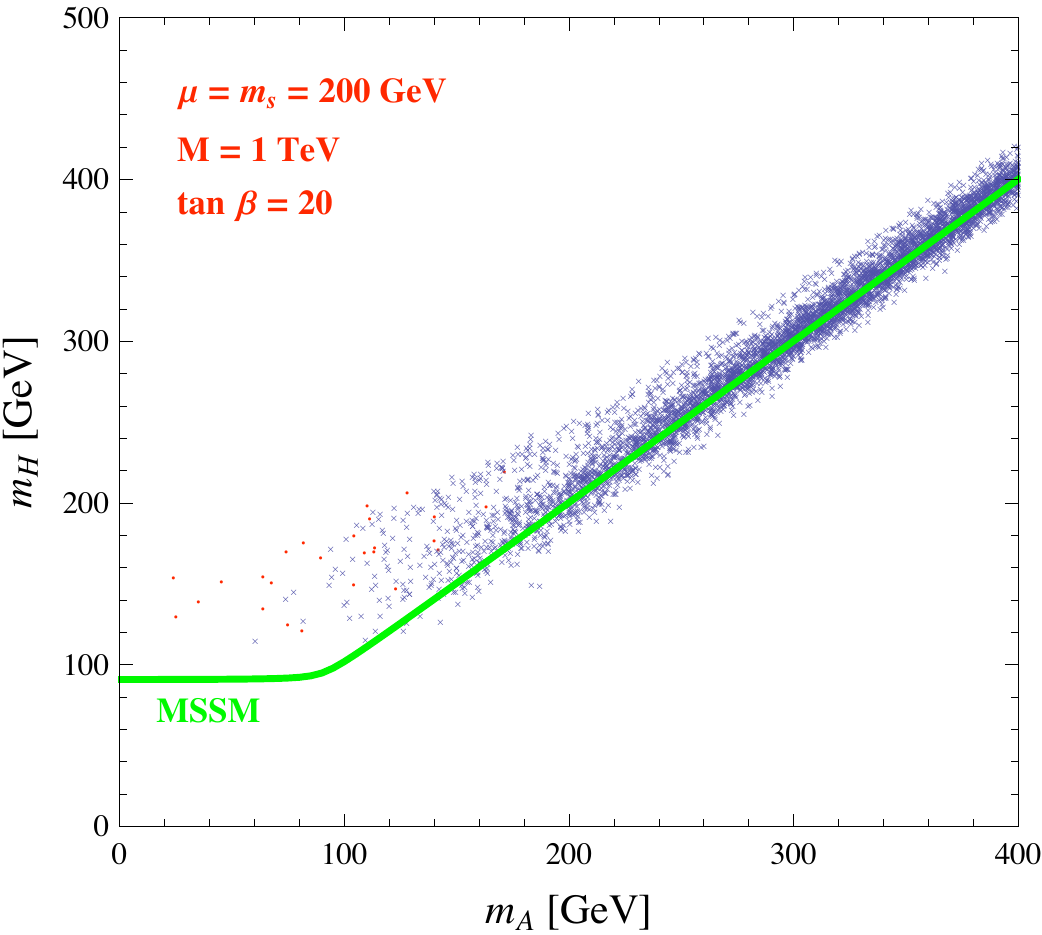}
\end{center}
\caption{\label{fig:mHmA}{\em Heavy CP-even Higgs tree-level mass,
$m_H$, as a function of $m_A$, for $\tan\beta = 2$ (left panel) and
for $\tan\beta = 20$ (right panel).  The solid green line corresponds
to the tree-level MSSM result.  The region of parameter space in the
scan is described in the main text.  Red points correspond to sEWSB
vacua, while blue crosses correspond to MSSM-like vacua.}}
\end{figure}
\begin{figure}[!htp]
\begin{center}
\includegraphics[width=7.9cm]{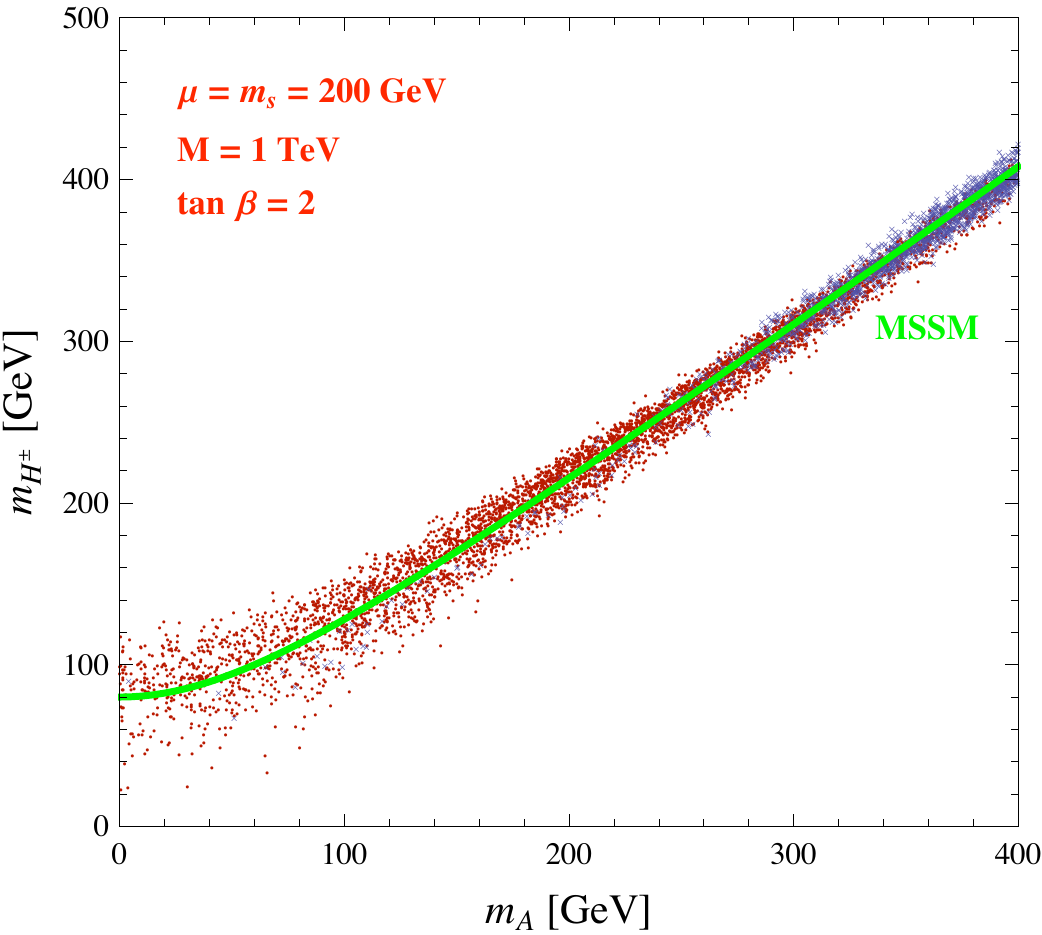}
\hspace{3mm}
\includegraphics[width=7.9cm]{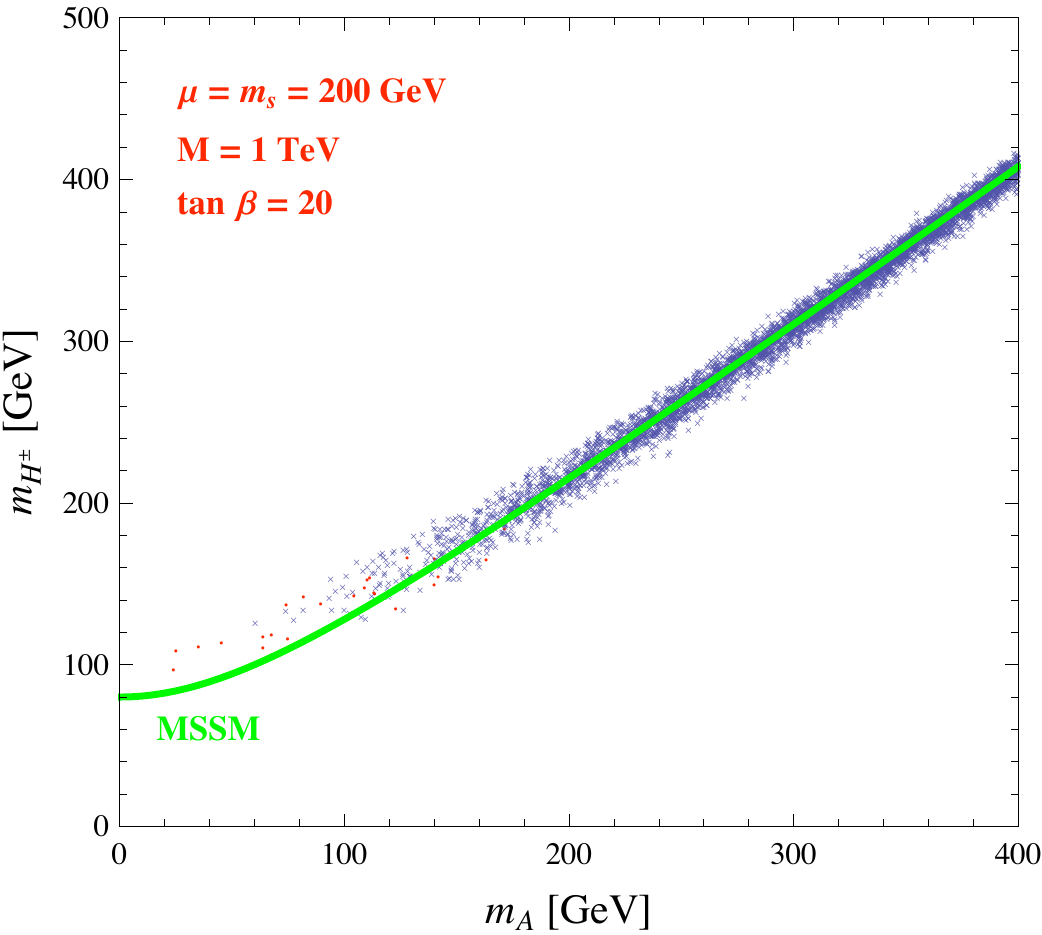}
\end{center}
\caption{\label{fig:mChargedmA}{\em Charged Higgs tree-level mass,
$m_{H^\pm}$, as a function of $m_A$, for $\tan\beta = 2$ (left panel)
and for $\tan\beta = 20$ (right panel).  The solid green line
corresponds to the tree-level MSSM result.  The region of parameter
space in the scan is described in the main text.  Red points
correspond to sEWSB vacua, while blue crosses correspond to MSSM-like
vacua.}}
\end{figure}
We start by presenting our results for the Higgs spectrum, and
comparing it to the MSSM one.  We do this at tree-level only.  Recall
that the radiative corrections are ``common'' in the MSSM and in the
effective theory under study, arising mainly from QCD and the MSSM
superpartner sector.  The observed differences can therefore be
interpreted as arising directly from the heavy physics through the
higher-dimension operators.  We present in
Figs.~\ref{fig:mhmA}-\ref{fig:mChargedmA} the results of the scan for
$m_{h}$, $m_{H}$ and $m_{H^\pm}$ as a function of $m_{A}$, for
$\tan\beta = 2$ (left panels) and $\tan\beta = 20$ (right panels).
The tree-level MSSM curve is shown for the corresponding $m_{A}$ and
$\tan\beta$.  We also indicate which points correspond to sEWSB vacua
(red points) and MSSM-like vacua (blue crosses), as described at the
end of Subsection~\ref{loops}.  We use the criterion described
in~\cite{Batra:2008rc}, which is based on the fact that for large $M$
with all other parameters fixed, the VEV in sEWSB vacua scales like $v
\propto \sqrt{M}$.  We see that for small $\tan\beta$ a large number
of points are of the sEWSB type, while for large $\tan\beta$ most
points correspond to MSSM-like vacua.

As expected from our discussion of Section~\ref{sec:analres}, the
corrections to $m_{h}$ are more important at low $\tan\beta$.
However, the scan shows that they can also be relevant at large
$\tan\beta$.  As already remarked in Subsection~\ref{robusteness}
there are significantly fewer points with small $m_{h}$, which is a
consequence of the ``robustness against higher-order corrections''
criterion described in Section~\ref{EFTValidity}.

It is also interesting that most points in Fig.~\ref{fig:mhmA} present
significant deviations from the corresponding MSSM values.  Our
parameter region includes the case where all higher-dimension
operators vanish, and therefore includes the MSSM limit.  However, in
the scan it is unlikely that all of them become small simultaneously,
which explains why there tends to be more points that exhibit
important deviations in $m_{h}$ with respect to the MSSM. In the large
$\tan\beta$ case, the overlap with the MSSM for sufficiently large
$m_h$ is possible mainly because many of the operators are 1/$\tan
\beta$ suppressed, and hence, at large $\tan\beta$ the number of
relevant coefficients that contribute to the departure of the Higgs
spectrum from the MSSM one is smaller than at low $\tan\beta$.  This
implies that at large $\tan\beta$ there is a higher probability to
effectively turn off all the effects from the higher-dimension
operators, and reproduce the MSSM values for $m_h$.  Our study simply
encapsulates the picture of heavy physics of mass $M$ ($=1~{\rm TeV}$)
with couplings of order one to the MSSM Higgs sector.  For instance,
the light CP-even Higgs can become sufficiently heavy for the
$W^{+}W^{-}$ and $ZZ$ channels to be kinematically accessible, thus
potentially allowing for a SM search of the light supersymmetric Higgs
in the dilepton plus missing energy as well as the four-lepton
channels.

Figs.~\ref{fig:mHmA} and \ref{fig:mChargedmA} show the $H$ and $H^\pm$
spectra as a function of the CP-odd Higgs mass for two values of $\tan
\beta$, respectively.  The deviations in $m_{H}$ and $m_{H^\pm}$ from
the MSSM values are less dramatic in the large $m_{A}$ region, but
they can be phenomenologically significant for lower values of
$m_{A}$.  Notice for instance that for $m_{A} \sim 100~{\rm GeV}$,
$m_{H}$ can be sufficiently large for the decay $H\rightarrow AA$ to
be kinematically open.  Similarly, the decay $H^{\pm} \rightarrow
W^{\pm} A$ can be open.  We will further comment on these decay
channels in the next section.

\subsection{Higgs couplings to gauge bosons and fermions}
\label{couplings}

%
\begin{figure}[!htp]
\begin{center}
\includegraphics[height=5.2cm]{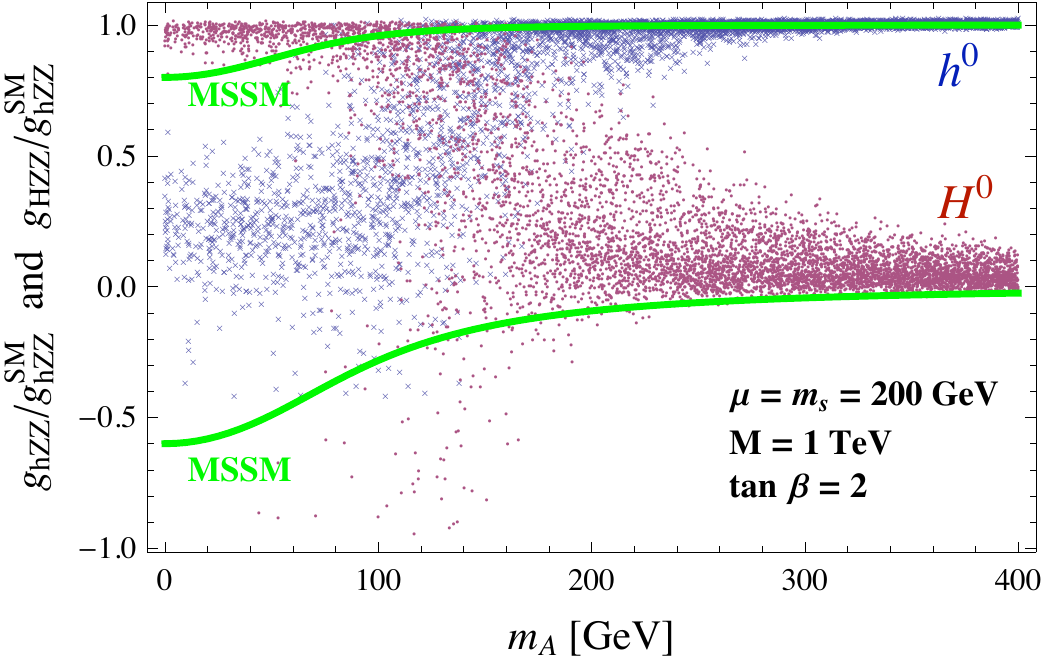}
\vspace*{8mm}
\includegraphics[height=5.2cm]{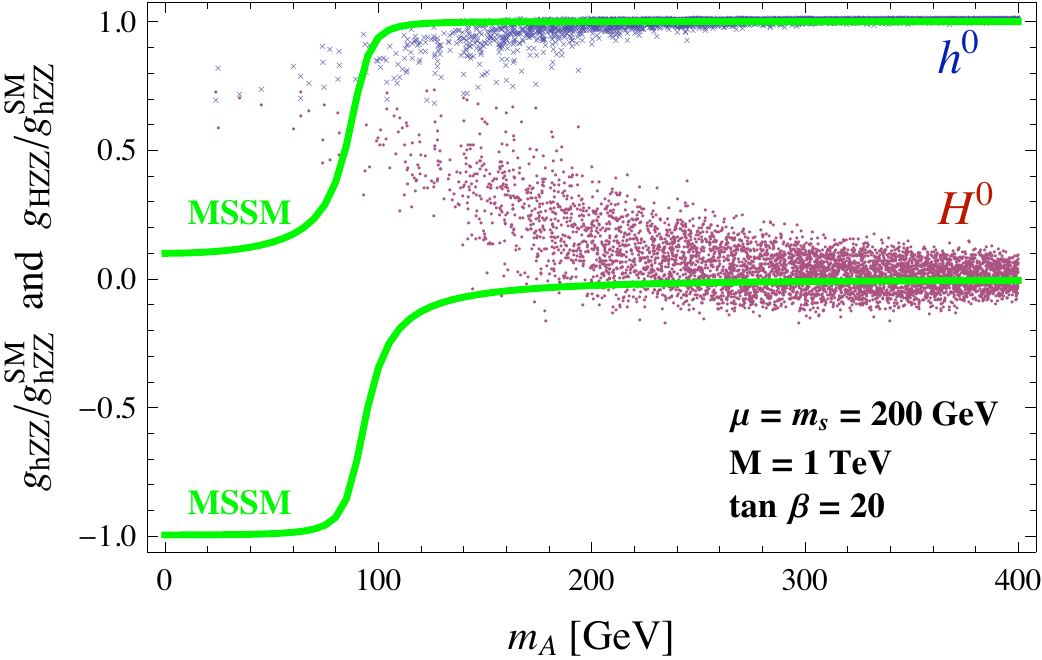}
\vspace*{8mm}
\includegraphics[height=5.2cm]{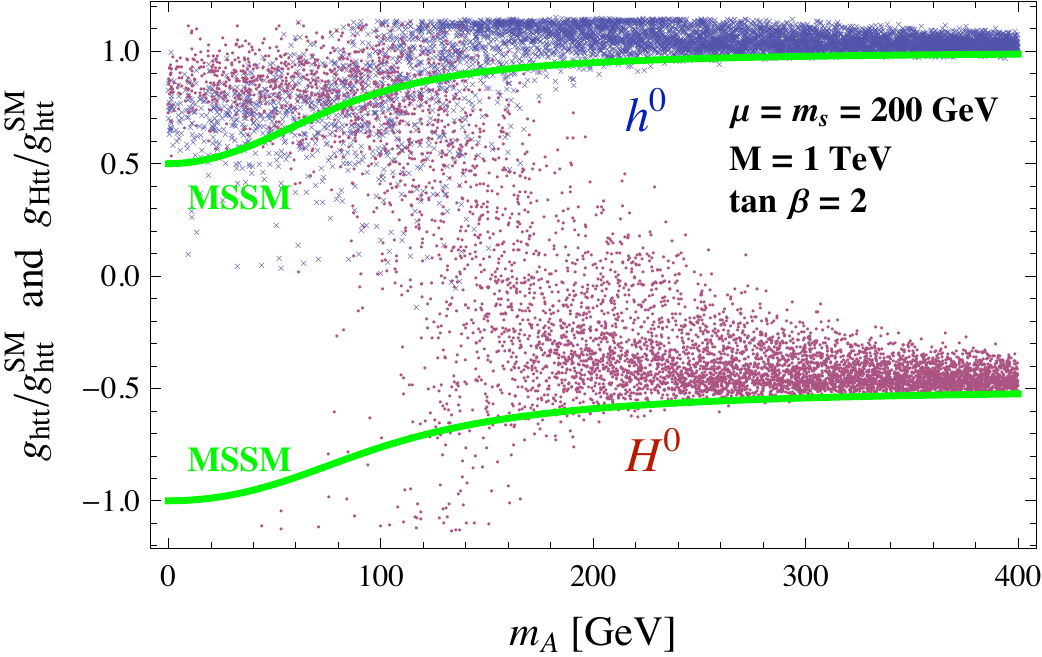}
\includegraphics[height=5.2cm]{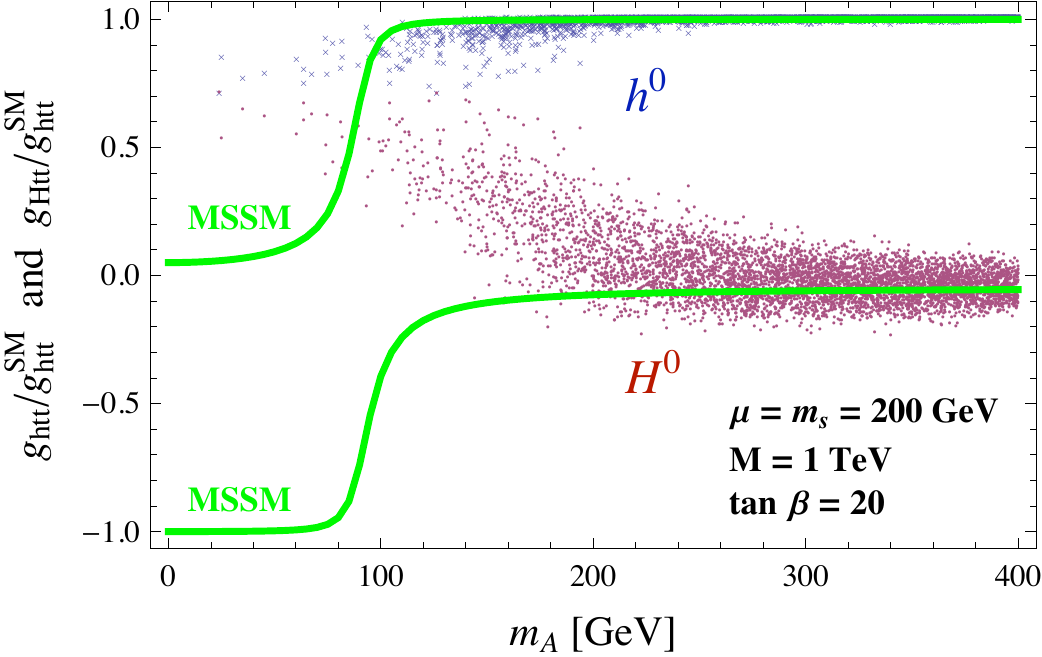}
\includegraphics[height=5.2cm]{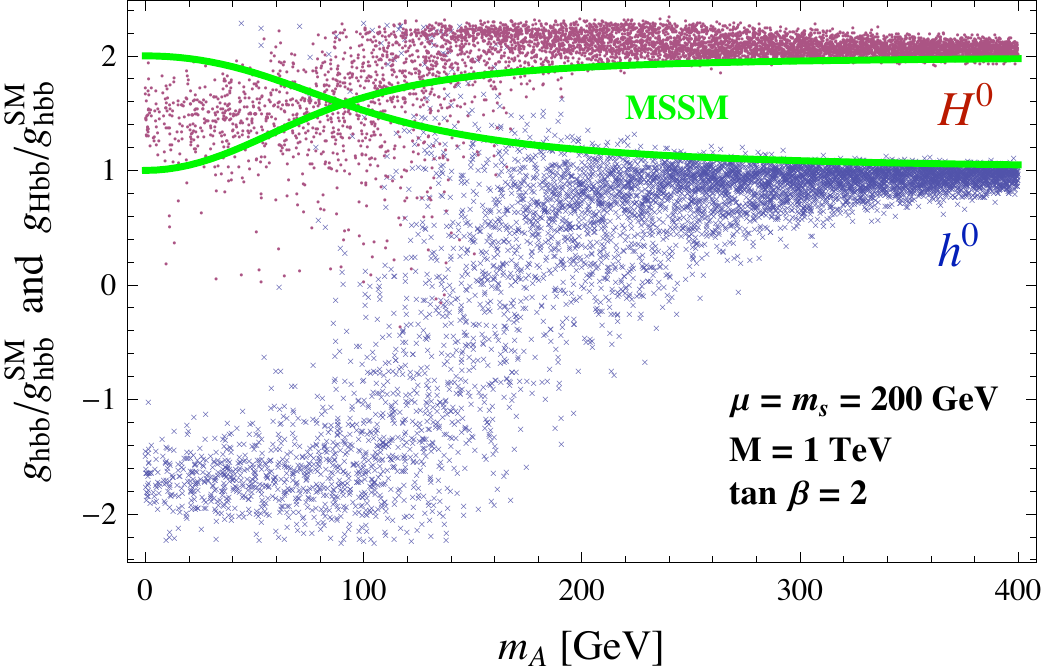}
\includegraphics[height=5.2cm]{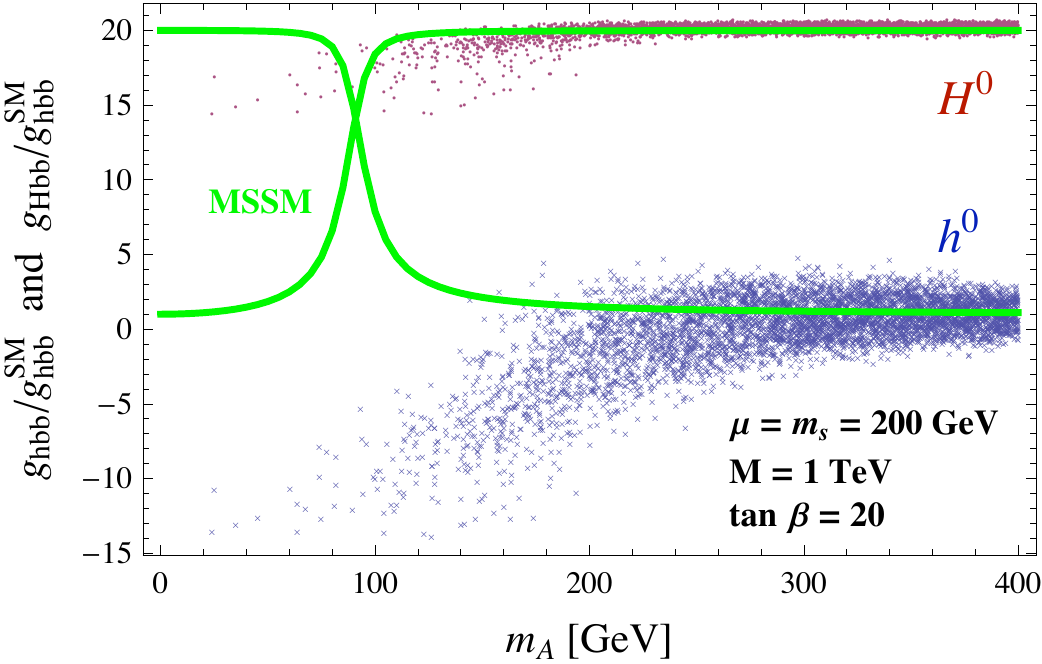}
\end{center}
\caption{\label{fig:couplings}{\em Tree-level couplings of the CP-even
Higgs states (normalized to the SM ones) to gauge bosons (upper
plots), to top pairs (middle plots) and to bottom pairs (lower plots),
as a function of $m_A$, for $\tan\beta = 2$ (left panels) and
$\tan\beta = 20$ (right panels).  The blue crosses correspond to the
couplings of $h$, and the red dots to those of $H$.  The solid green
lines correspond to the tree-level MSSM result.  The region of
parameter space in the scan is described in the main text.}}
\end{figure}

In Fig.~\ref{fig:couplings} we present the \textit{tree-level}
couplings (normalized by the SM values) of the CP-even Higgses to $Z$
pairs and up-type and down-type fermions, as a function of $m_{A}$.
The solid curves show the corresponding MSSM tree-level prediction,
making it clear that large deviations from the MSSM can be induced via
the higher-dimension operators.

The plots in the upper row of Fig.~\ref{fig:couplings} show the
couplings of the CP-even Higgses to $Z$ pairs for $\tan\beta = 2$
(left panel) and $\tan\beta = 20$ (right panel).  We will refer to the
Higgs state that has the largest coupling to the $Z$ as ``SM-like''
(i.e. $h$ is ``SM-like'' if $|g_{hZZ}| > |g_{HZZ}|$ while $H$ is
``SM-like'' if $|g_{HZZ}| > |g_{hZZ}|$).  Recall that the couplings to
$WW$ and $ZZ$ are different as a result of the corrections given in
Eq.~(\ref{deltagVV}).  The difference appears at order $1/M^2$ and is
of order a few percent when these couplings are sizable (e.g. for the
SM-like Higgs).  For a non-SM-like Higgs with a very small coupling to
$W$'s and $Z$'s, the relative difference between the two can be
significant.  We also note that in the present scenarios $g^2_{hZZ} +
g^2_{HZZ}$ need not add up to one, reflecting the fact that there are
additional (heavy) degrees of freedom that have been integrated out.
This effect also arises at order $1/M^2$ and we find that for the
chosen parameters it can be as large as $7\%$ (this can be seen more
clearly in the large $m_{A}$ region of the figures).

For large $m_{A}$, $h$ becomes SM-like to a very good approximation,
as in the MSSM. However, we also see that in this limit $H$ can have
less suppressed couplings to $Z$ pairs than in the MSSM. These are the
properties anticipated in Eq.~(\ref{hHVVLargemA}).  At low $m_{A}$ and
for $\tan\beta = 2$, we see that $H$ becomes SM-like to a better
degree of approximation than in the MSSM, while $h$ has more
suppressed couplings to $Z$ pairs than in the MSSM. There are also a
large number of models with intermediate $m_{A}$ where both CP-even
Higgses couple significantly to gauge boson pairs.  For $\tan\beta =
20$ the situation is similar in the large $m_A$ region. However, there
are very few points at low $m_{A}$ (see Subsection~\ref{robusteness}
for an explanation).

\begin{figure}[!htp]
\begin{center}
\includegraphics[height=7.5cm]{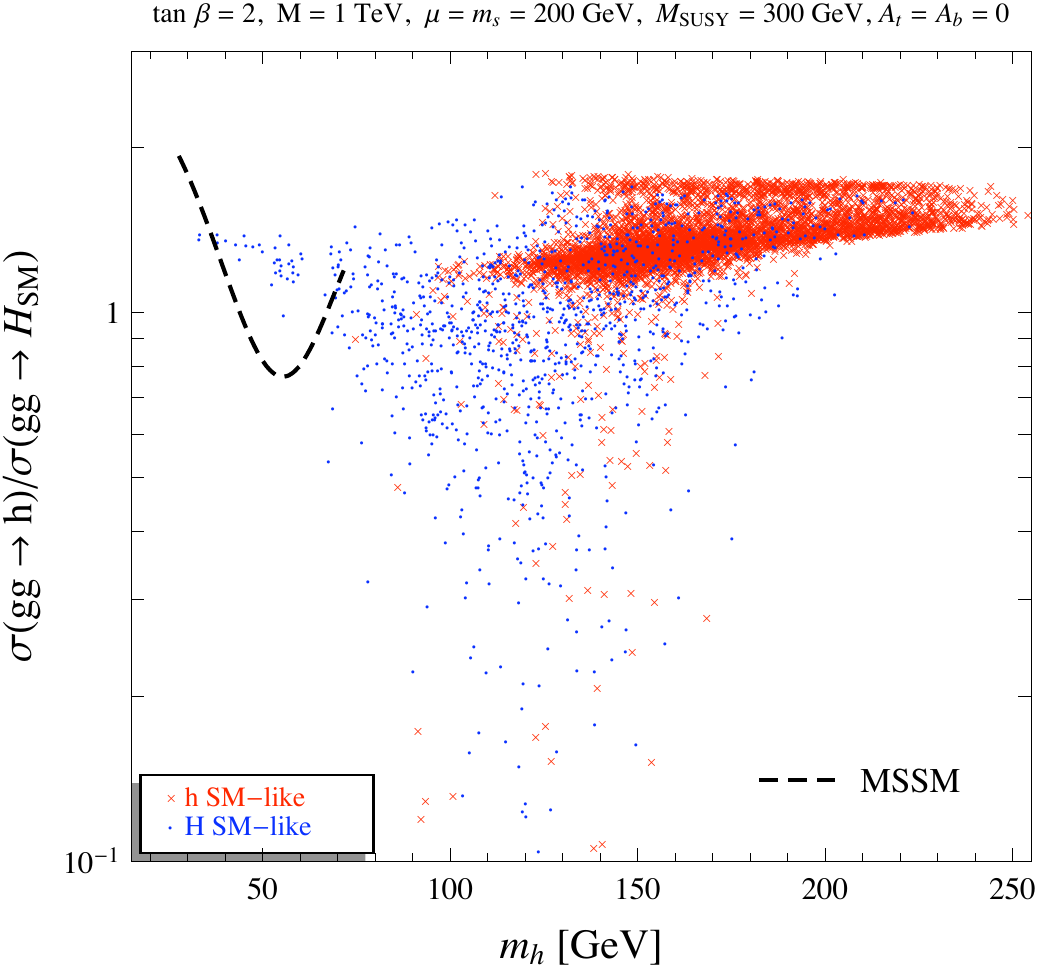}
\vspace*{8mm}
\includegraphics[height=7.5cm]{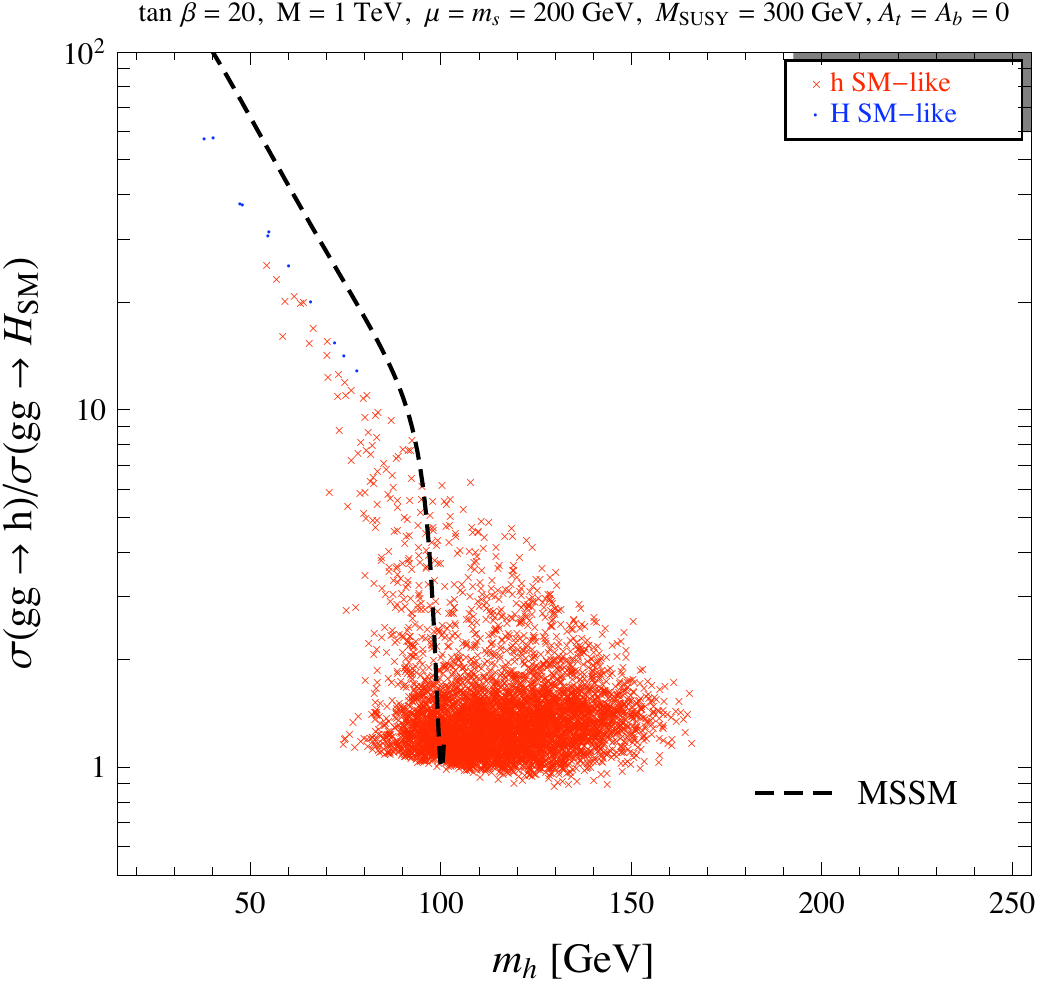}
\includegraphics[height=7.5cm]{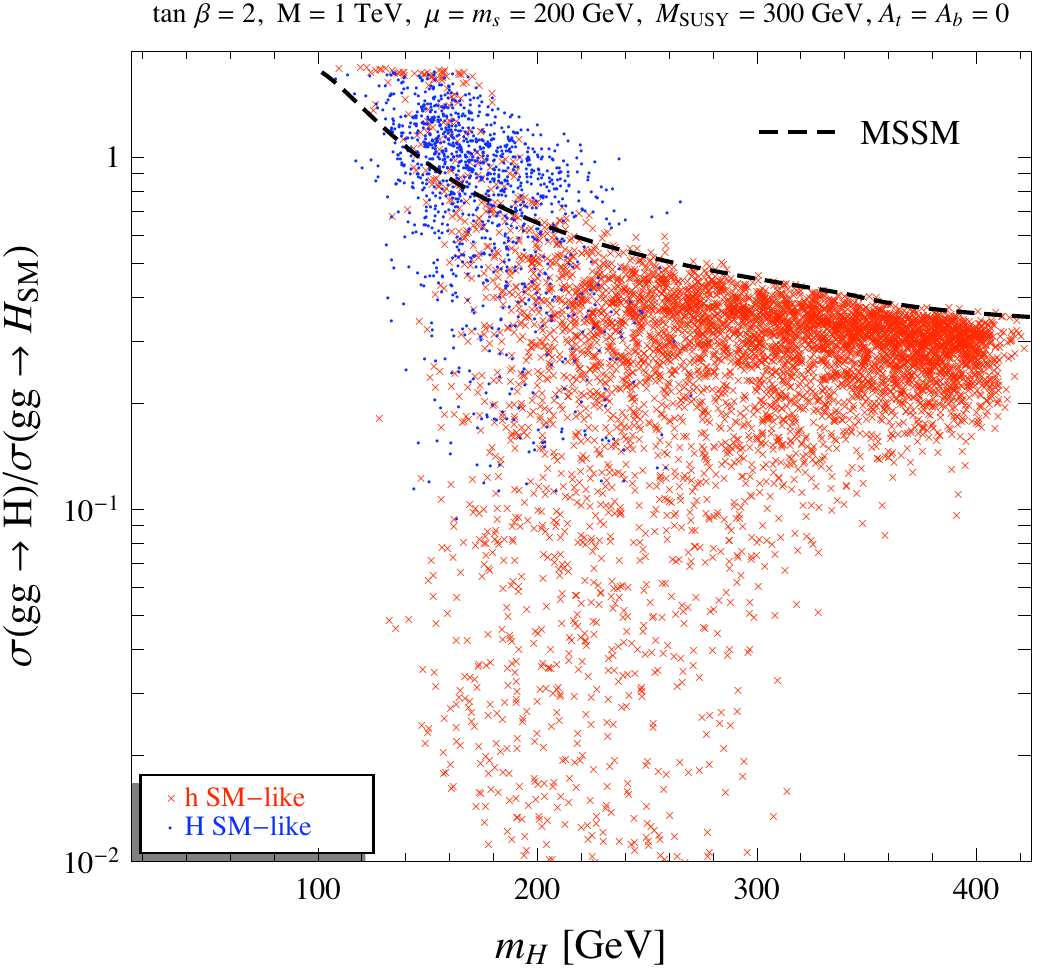}
\includegraphics[height=7.5cm]{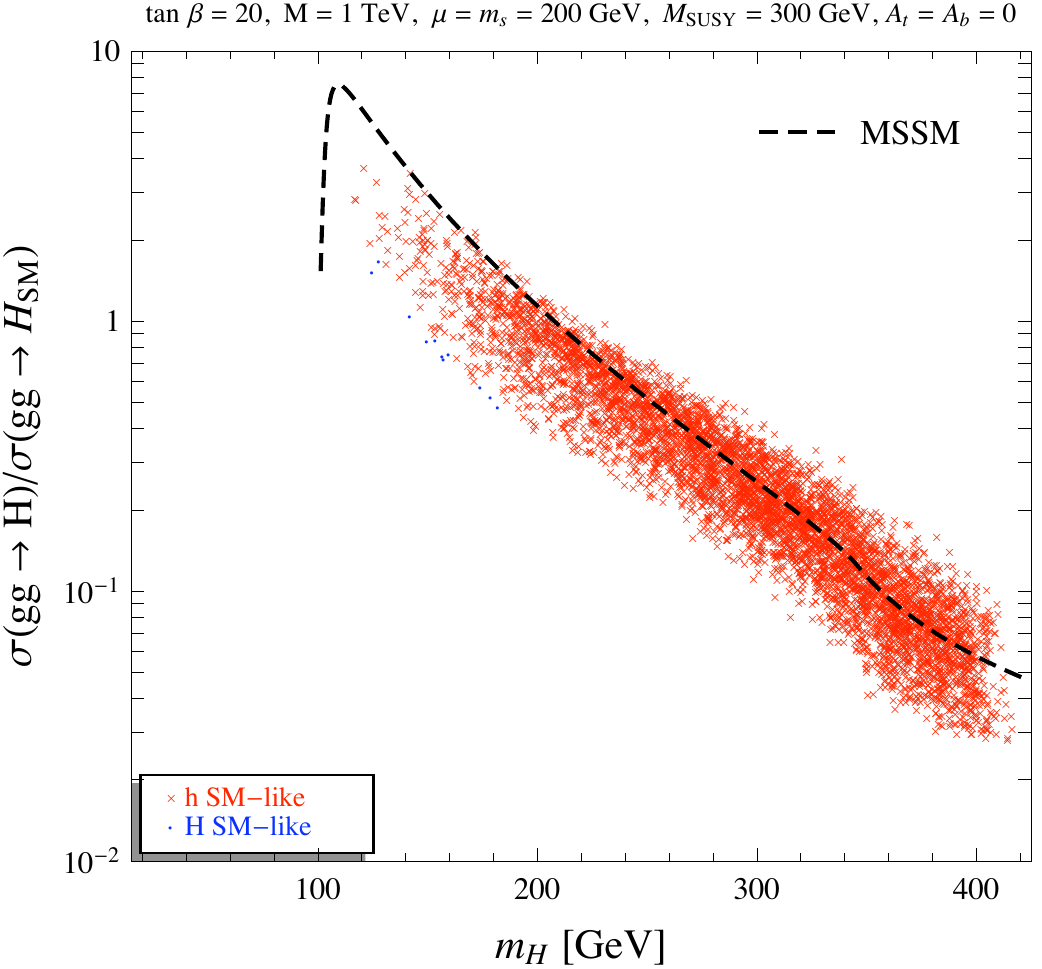}
\end{center}
\caption{\label{fig:gluonfusion}{\em Gluon fusion cross section at LO
in $\alpha_{s}$, including light SUSY particle loops and normalized
to the SM, for the light (upper panels) and heavy (lower panels)
CP-even Higgses, and for $\tan\beta = 2$ (left panels) and $\tan\beta
= 20$ (right panels).  The region of parameter space in the scan is
described in the main text.  We indicate by red crosses those points where h
is SM-like ($|g_{hZZ}| > |g_{HZZ}|$) and by blue dots those points where H
is SM-like ($|g_{HZZ}| > |g_{hZZ}|$).  The dashed line corresponds to
the MSSM result.  }}
\end{figure}

We also notice several features of the $t\bar{t}h$ and $b\bar{b}h$
($t\bar{t}H$ and $b\bar{b}H$) couplings, which are relevant for gluon
fusion induced processes.  We consider first the case of low
$\tan\beta$.  The $t\bar{t}h$ coupling can present a small enhancement
over both the SM and the MSSM of up to about $1.2$ when $m_{A} >
100~{\rm GeV}$ (for smaller $m_{A}$ it can present an enhancement of
up to $50\%$ with respect to the MSSM).  This would imply a factor of
up to $1.4$ with respect to the SM prediction for gluon fusion
production relevant for searches at the Tevatron and LHC. In addition,
given that there is no need for large radiative corrections to $m_{h}$
from the squark sector to avoid the bounds on the light Higgs, if a
light stop spectrum of order a few hundred GeV is present, additional
relevant contributions to the gluon fusion process will be expected.
These can lead to an additional enhancement of the reach in gluon
fusion induced channels~\cite{Menon:2009mz}.  The $b\bar{b}h$ coupling
is suppressed both with respect to the SM and the MSSM when $m_{A} >
200~{\rm GeV}$, and reaches a value at most equal to $\tan\beta$ in
the low $m_{A}$ region (as in the MSSM).  However, notice that there
are a number of models with suppressed couplings to $b\bar{b}$ pairs
for $m_{A} < 140~{\rm GeV}$.

For low $\tan\beta$ the contribution to gluon fusion is governed by
the top Yukawa coupling, and therefore there is a net enhancement with
respect to the SM for many models, as seen in the upper left panel of
Fig.~\ref{fig:gluonfusion}, where we present the gluon fusion cross
section for $h$ production normalized to the SM one, as a function of
$m_{h}$.~\footnote{We compute $\sigma(gg \rightarrow h)/\sigma^{{\rm
SM}}(gg \rightarrow h) \approx \Gamma(h \rightarrow gg)/\Gamma^{{\rm
SM}}(h \rightarrow gg)$, which holds at leading order in
$\alpha_{s}$~\cite{Georgi:1977gs,Spira:1995rr,Dawson:1996xz}.  The
values of $\Gamma(h \rightarrow gg)$ and $\Gamma^{{\rm SM}}(h
\rightarrow gg)$ are calculated at LO using a version of
HDECAY~\cite{Djouadi:1997yw}, modified to include the tree-level
expressions in the presence of the higher-dimension operators.  } In
this figure we include the radiative effects to the Higgs masses and
couplings assuming a SUSY spectrum as described in
Subsection~\ref{loops}.  Most of the models with $m_{h} < 115~{\rm
GeV}$ are excluded by LEP, while the Tevatron excludes some of the
models with $m_{h}$ in a window around $170~{\rm GeV}$.  A detailed
analysis of all the cross sections and branching fractions to
determine the allowed models will be presented elsewhere~\cite{CKPZ}.
In Fig.~\ref{fig:gluonfusion} we indicate by red crosses the models
where $h$ is ``SM-like'', as per the definition at the beginning of
this subsection, and by blue dots those models where $H$ is
``SM-like''.

The $t\bar{t}H$ coupling at low $\tan \beta$, shown in
Fig.~\ref{fig:couplings}, is found to be generically suppressed with
respect to both the SM and the MSSM, except for the region $m_{A} <
200~{\rm GeV}$ where some enhancement may be possible.  The
$b\bar{b}H$ coupling is generically enhanced with respect to the SM
and even with respect to the $\tan\beta$ enhanced value in the MSSM
for large $m_{A}$.  However, the bottom loop contribution to the gluon
fusion process is subdominant, as seen in the lower left panel of
Fig.~\ref{fig:gluonfusion}, where a net suppression with respect to
the SM is observed for $m_{H} > 200~{\rm GeV}$, governed by the
suppression in the $t\bar{t}H$ coupling.  In this figure we also see
an enhancement with respect to the SM for $m_{H} < 200~{\rm GeV}$,
which reflects the enhancement of the $t\bar{t}H$ coupling for $m_{A}
< 200~{\rm GeV}$ mentioned above, in addition to the enhancement due
to light stop contributions.

At large $\tan\beta$, there is no significant variation in $t\bar{t}h$
with respect to the SM or the MSSM for $m_{A} > 200~{\rm GeV}$, but
there is a small suppression for $100~{\rm GeV} < m_{A} < 200~{\rm
GeV}$, and there can be an enhancement with respect to the MSSM for
$m_{A} < 100~{\rm GeV}$.  The $b\bar{b}h$ coupling is enhanced with
respect to the SM and the MSSM in a large number of models, and
achieves the largest values for $m_{A} <100~{\rm GeV}$, although it is
smaller than the $\tan\beta$ value that occurs in the MSSM. For $m_{A}
> 200~{\rm GeV}$ there are many models where the $b\bar{b}h$ coupling
is strongly suppressed.  As shown in the upper right panel of
Fig.~\ref{fig:gluonfusion}, in many models the regions of enhanced
$b\bar{b}h$ coupling lead to a relevant enhancement of the gluon
fusion cross section with respect to the SM one.

\begin{figure}[t]
\begin{center}
\includegraphics[height=5.2cm]{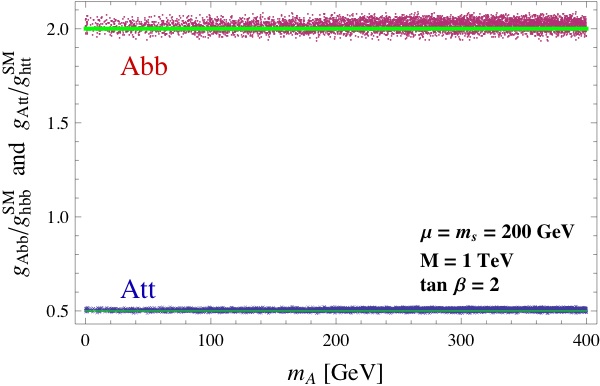}
\includegraphics[height=5.2cm]{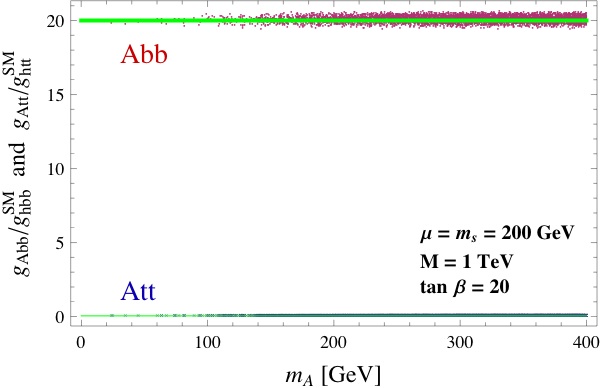}
\end{center}
\caption{\label{fig:Acouplings}{\em Tree-level couplings of the CP-odd
Higgs to fermion pairs as a function of $m_A$, for $\tan\beta = 2$
(left panel) and $\tan\beta = 20$ (right panel), and normalized to the
SM value, $g m_{f}/2m_{W}$.  The blue crosses correspond to the couplings
of $A$ to $t\bar{t}$, and the red dots to those of $A$ to $b\bar{b}$.
The solid green lines correspond to the tree-level MSSM result.  The
region of parameter space in the scan is described in the main text.}}
\end{figure}

At large $\tan\beta$, the $t\bar{t}H$ coupling can have an enhancement
with respect to the MSSM in the region $100~{\rm GeV} < m_{A} <
200~{\rm GeV}$, but is generically suppressed with respect to the SM.
The $b\bar{b}H$ coupling presents the familiar $\tan\beta$ enhancement
of the MSSM at large $m_{A}$, although there could be some suppression
in the intermediate region $100~{\rm GeV} < m_{A} < 200~{\rm GeV}$.
At low $m_{A}$ we see a few models with a significant enhancement of
the $b\bar{b}H$ coupling with respect to the SM and the MSSM.

In the large $\tan\beta$ region both top and bottom loops contribute
to the gluon fusion cross section with different weights depending on
the ratio of the fermion to Higgs masses and the specific (large) value
of $\tan\beta$.  The lower right panel of Fig.~\ref{fig:gluonfusion}
shows that for values of $m_{H}$ larger than $200~{\rm GeV}$ there is
always a suppression with respect to the SM gluon fusion cross
section for $H$, similar to the MSSM case.

\begin{figure}[t]
\begin{center}
\includegraphics[width=8cm]{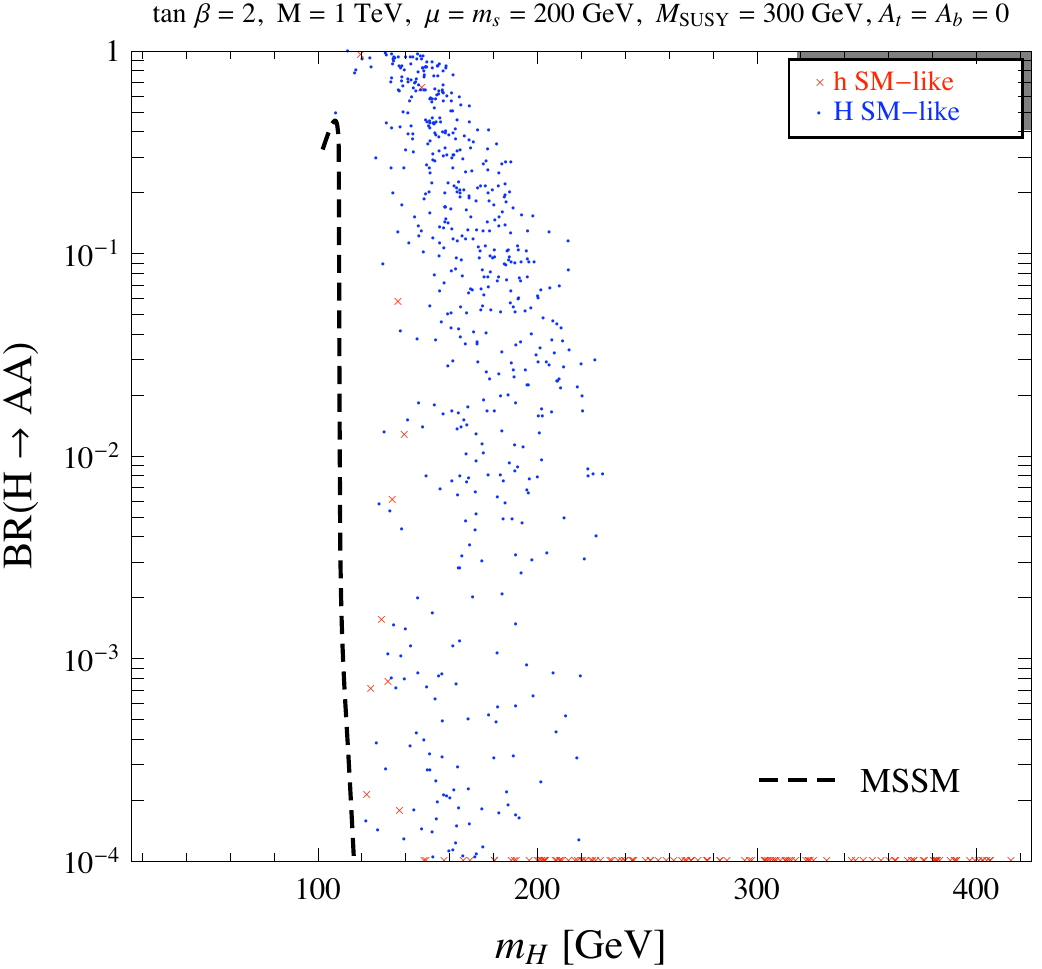}
\hspace{3mm}
\includegraphics[width=8cm]{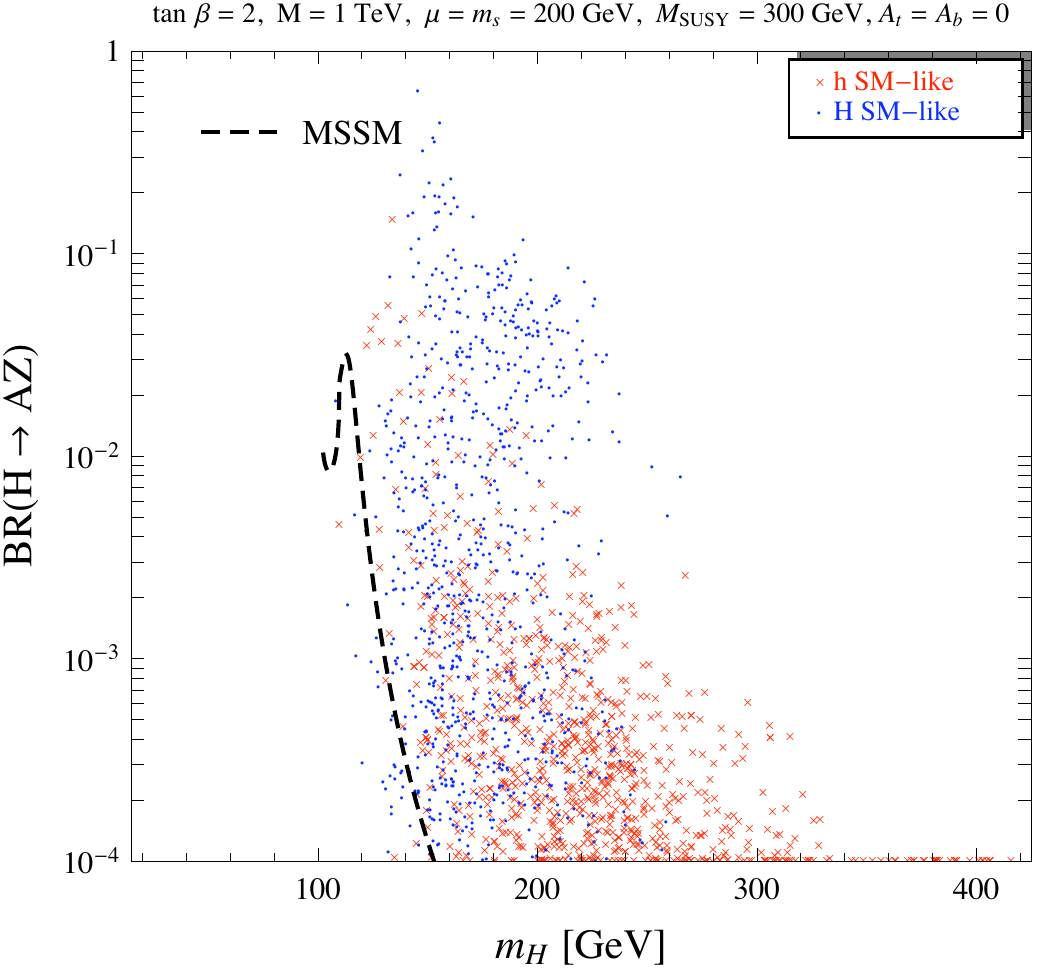}
\end{center}
\caption{\label{fig:BRHAAandHAZ}{\em Branching fractions for $H
\rightarrow AA$ (left panel) and $H \rightarrow A Z$ (right panel).
The dashed line corresponds to the MSSM result.  SUSY and QCD
radiative corrections are included.  The plots are for $\tan\beta =
2$, $M = 1~{\rm TeV}$, $\mu = m_{s} = 200~{\rm GeV}$, $M_{SUSY} =
300~{\rm GeV}$, $A_{t} = A_{b} = 0$, and a scan over the ranges
defined in Section~\ref{parameters}.  We indicate by red crosses those points
where h is SM-like ($|g_{hZZ}| > |g_{HZZ}|$) and by blue dots those points
where H is SM-like ($|g_{HZZ}| > |g_{hZZ}|$).}}
\end{figure}

The couplings of the CP-odd and charged Higgs bosons differ from those of
the MSSM only at order $1/M^2$, due to the corrections to their
kinetic terms [see comments after Eq.~(\ref{Hcouplings})].  These
deviations are far less significant than for the CP-even Higgs states.
As an example, we show in Fig.~\ref{fig:Acouplings} the couplings of
the CP-odd Higgs to up-type and down-type fermion pairs, which follows
closely the $\tan\beta$ enhancement/suppression familiar in the MSSM
[see also Eq.~(\ref{MSSMhff})].

We postpone a detailed discussion of all the relevant branching
fractions and the consequences for Higgs searches at the Tevatron and
the LHC to Ref.~\cite{CKPZ}.  Here we comment on a number of selected
``exotic'' channels that can motivate new search strategies, in
particular regarding the heavy CP-even and charged Higgs bosons.
These are mostly related to the distortion of the Higgs spectrum with
respect to the MSSM which can lead to the opening of new decay
channels, as mentioned at the end of Subsection~\ref{masses}.  For
example, we show in the left panel of Fig.~\ref{fig:BRHAAandHAZ} the
branching fraction for $H \rightarrow AA$, which can be significant
for $m_{H}$ up to about $200~{\rm GeV}$.  It can be in fact much
larger than in the MSSM, where such values of $m_{H}$ are already near
the decoupling limit so that such decays are highly suppressed by
phase space.  Similarly, there can be important branching fractions
for $H \rightarrow AZ$, as seen in the right panel of
Fig.~\ref{fig:BRHAAandHAZ}.  Note that for most of these points it is
$H$ that is SM-like.

\begin{figure}[t]
\begin{center}
\includegraphics[width=8cm]{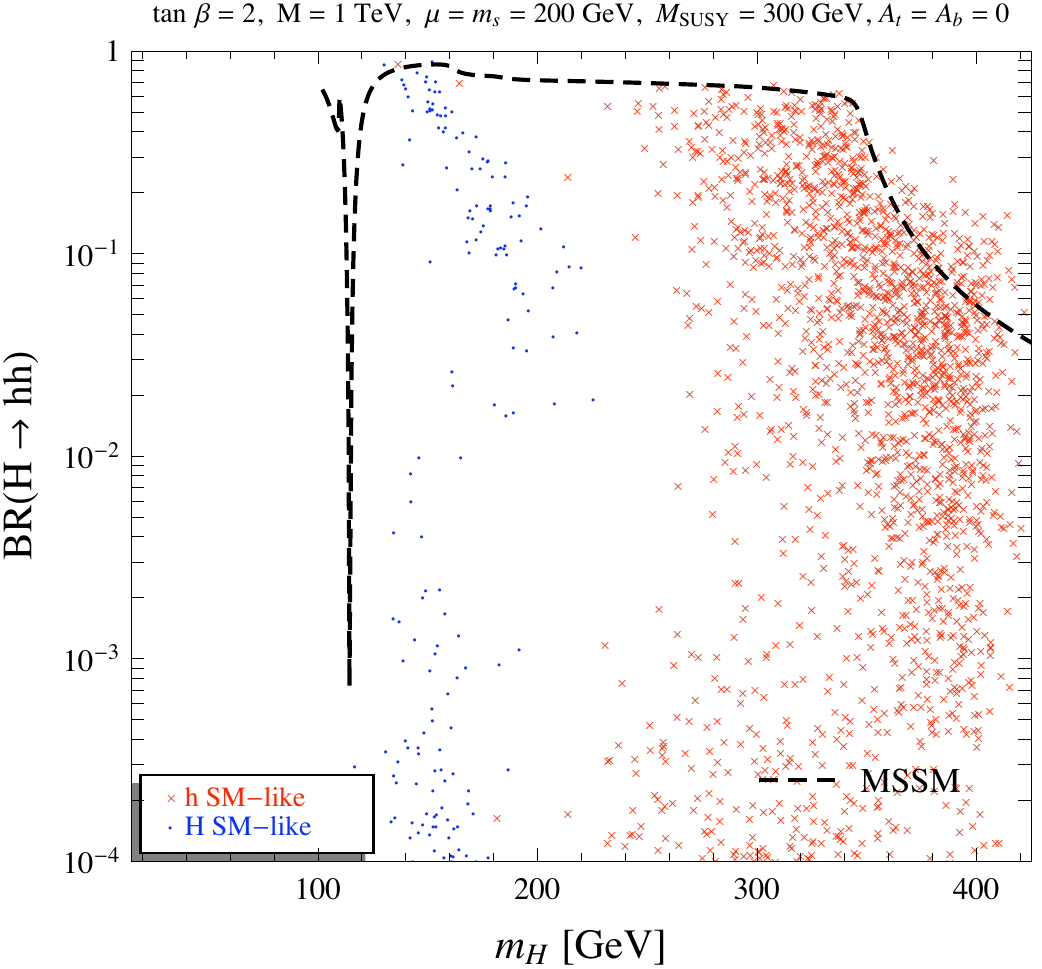}
\hspace{3mm}
\includegraphics[width=8cm]{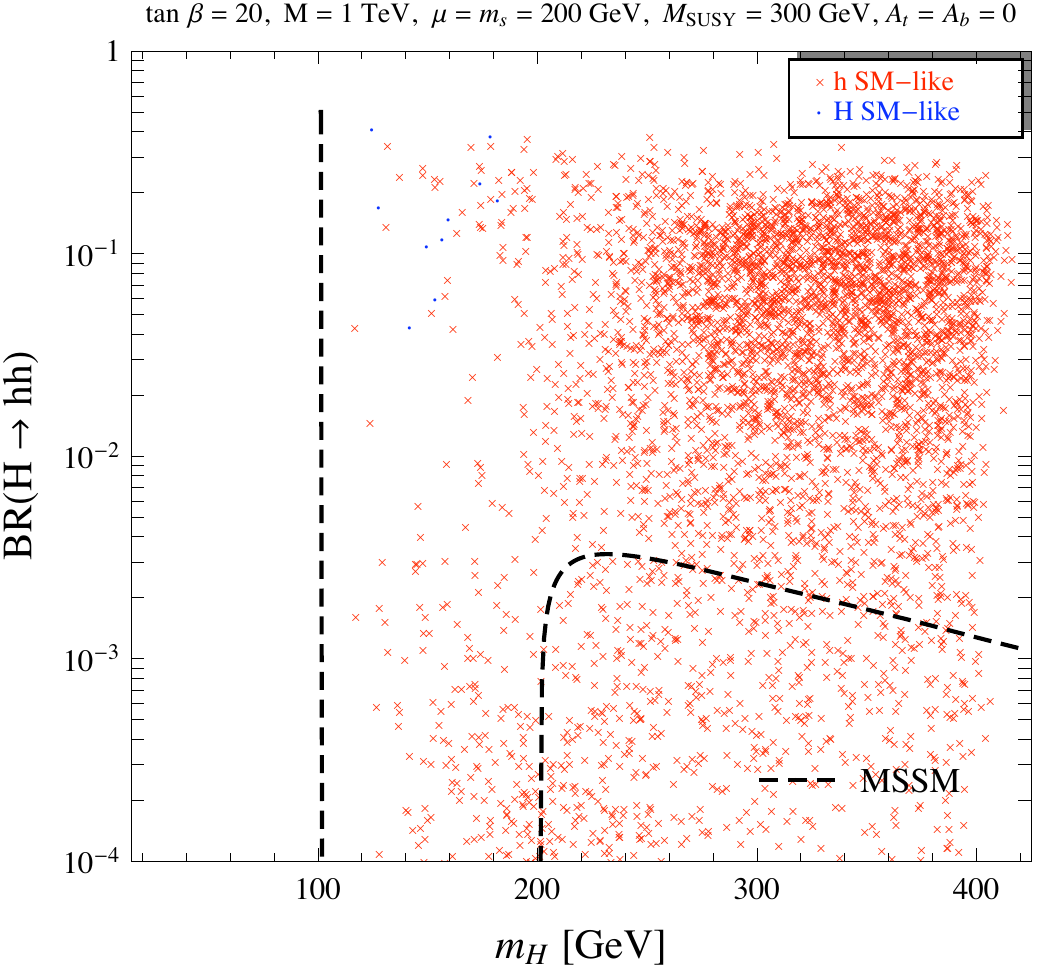}
\end{center}
\caption{\label{fig:BRHhh}{\em Branching fractions for $H \rightarrow
hh$ for $\tan\beta = 2$ (left panel) and $\tan\beta = 20$ (right
panel).  The dashed line corresponds to the MSSM result.  SUSY and QCD
radiative corrections are included.  We indicate by red crosses those points
where h is SM-like ($|g_{hZZ}| > |g_{HZZ}|$) and by blue dots those points
where H is SM-like ($|g_{HZZ}| > |g_{hZZ}|$).}}
\end{figure}

We also show in the left panel of Fig.~\ref{fig:BRHhh} that for low
$\tan\beta$ there is a generic suppression of the $H \rightarrow hh$
channel compared to the MSSM, except above the $t\bar{t}$ threshold
where some enhancement is possible (the pronounced dip at $m_{H} \sim
108~{\rm GeV}$ in the MSSM curve is due to an accidental cancellation
of the $Hhh$ coupling; in this small window H decays mostly into
$b$'s).  The above suppression can be very significant for $170~{\rm
GeV} < m_{H} < 250~{\rm GeV}$, reflecting a relatively heavy $h$ so
that the channel is closed.  At large $\tan\beta$, however, there are
large regions where ${\rm BR}(H\rightarrow hh)$ is enhanced with
respect to the MSSM, and in fact $H \rightarrow hh$ can be a
significant decay channel in this case (see right panel of
Fig.~\ref{fig:BRHhh}).

\begin{figure}[!htp]
\begin{center}
\includegraphics[height=7.5cm]{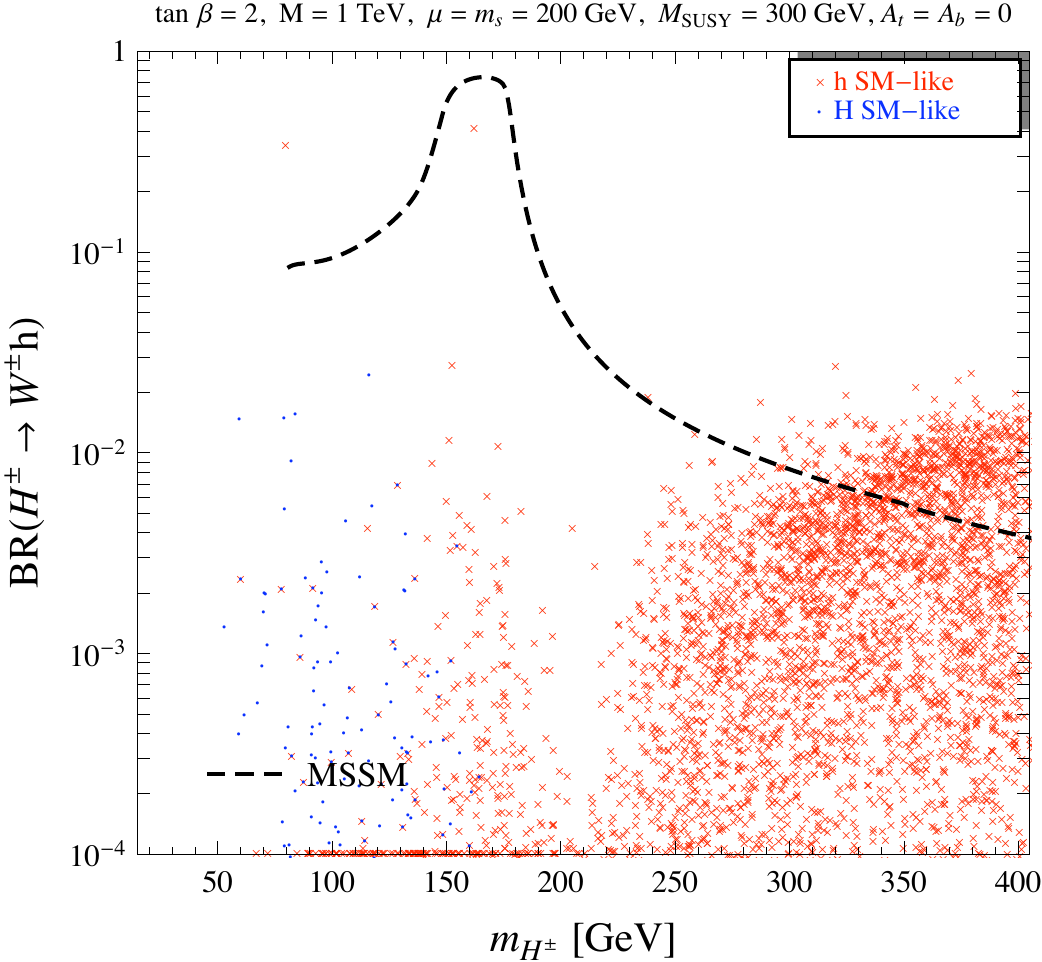}
\vspace*{8mm} 
\includegraphics[height=7.5cm]{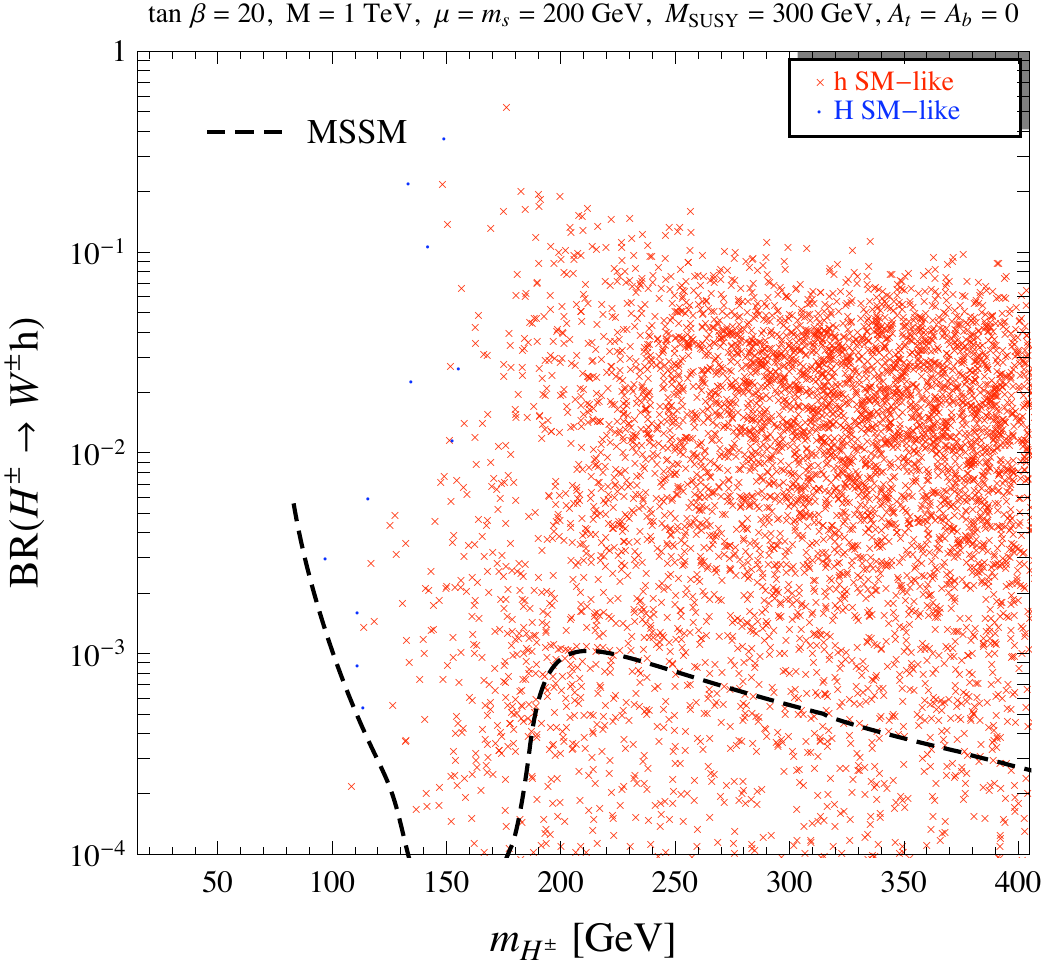}
\includegraphics[height=7.5cm]{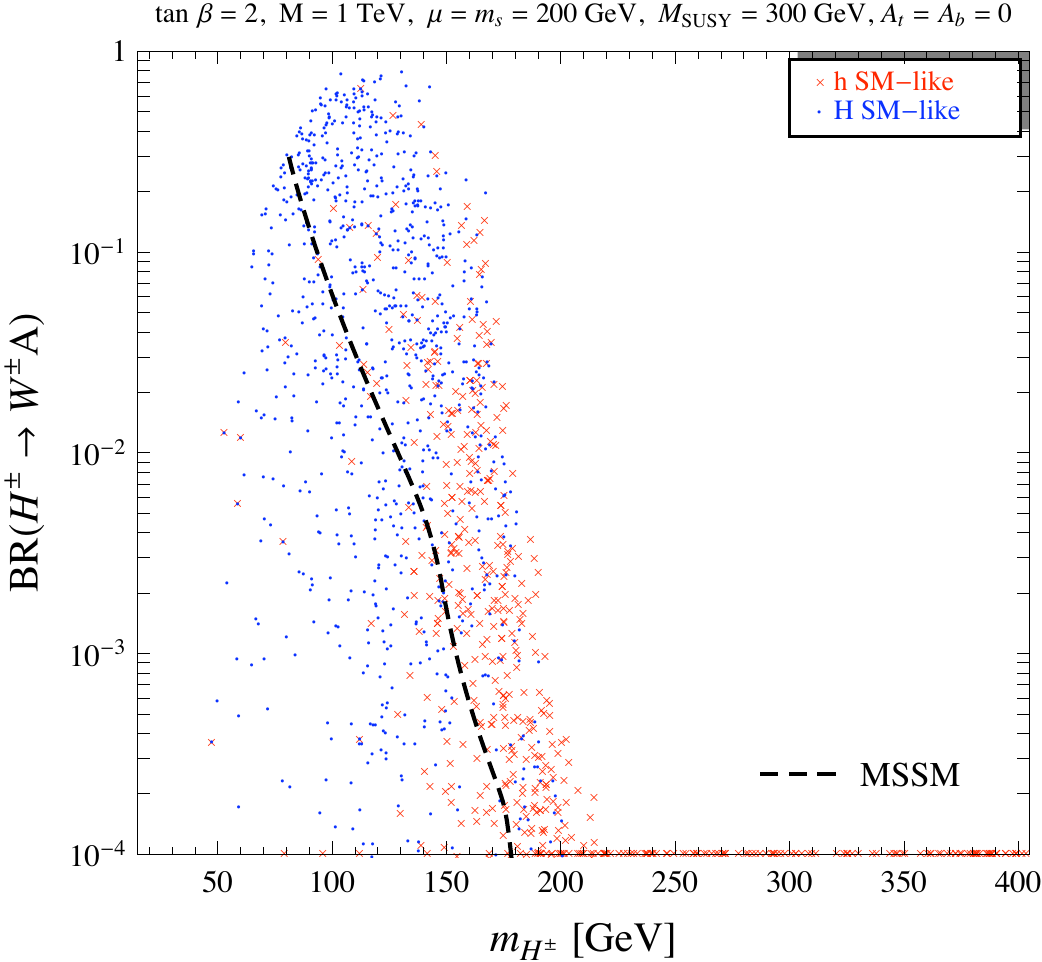}
\includegraphics[height=7.5cm]{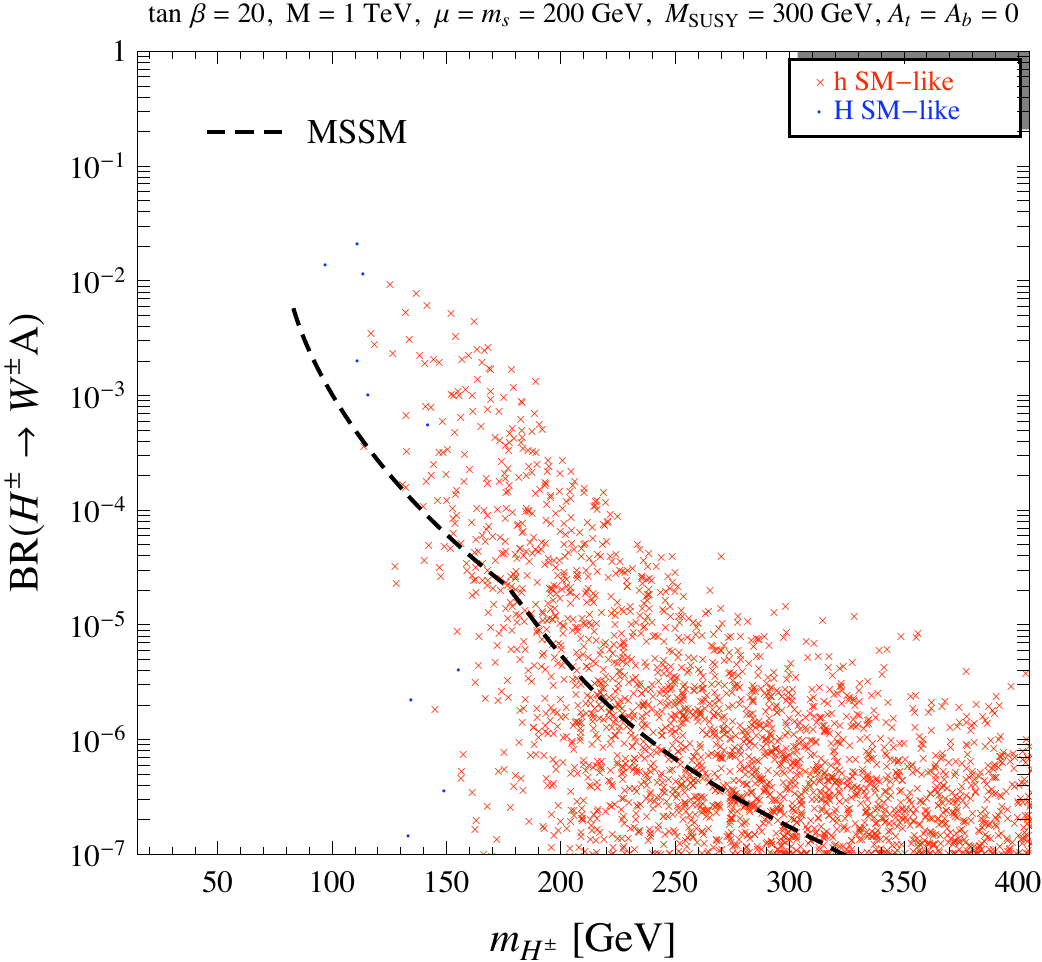}
\end{center}
\caption{\label{fig:BRCharged}{\em Branching fractions for $H^{\pm}
\rightarrow W^{\pm (*)} h$ (upper panels) and $H^{\pm} \rightarrow
W^{\pm (*)} A$ (lower panels), for $\tan\beta = 2$ (left panels) and
$\tan\beta = 20$ (right panels).  The dashed line corresponds to the
MSSM result.  SUSY and QCD radiative corrections are included.  We
indicate by red crosses those points where h is SM-like ($|g_{hZZ}| >
|g_{HZZ}|$) and by blue dots those points where H is SM-like ($|g_{HZZ}| >
|g_{hZZ}|$).}}
\end{figure}

Finally, we comment on certain charged Higgs decays.  In the upper row
plots of Fig.~\ref{fig:BRCharged} we present the branching fractions
for $H^{\pm} \rightarrow W^{\pm}h \,\, (W^{\pm *}h)$, which show a
significant suppression with respect to the MSSM at low $\tan\beta$
(except above the $t\bar{t}$ threshold, where some enhancement is
possible), while for large $\tan\beta$ there is an enhancement
compared to the MSSM in a large number of models.  In the lower row
plots of Fig.~\ref{fig:BRCharged}, we see that the branching fraction
for $H^{\pm} \rightarrow W^{\pm}A \,\, (W^{\pm *}A)$, can be sizable
at low $\tan\beta$ and for $100~{\rm GeV} < m_{H^{\pm}} < 180~{\rm
GeV}$, while it remains small for large $\tan\beta$ (although it can
still be enhanced compared to the MSSM).  Since at low $\tan\beta$,
${\rm BR}(H^{\pm} \rightarrow W^{\pm}A)$ can be close to unity, one
may therefore expect to produce a large number of CP-odd Higgs bosons
in top decays.  Note also that a large fraction of the points with
such a property present the ``inverted hierarchy'' where $H$ is
SM-like.

\section{Conclusions}
\label{sec:conclu}

We have considered a large class of supersymmetric scenarios with
physics beyond the MSSM that couples appreciably to the MSSM Higgs
sector.  Our main assumption is that the degrees of freedom beyond the
MSSM are heavier than the weak scale, and that their SUSY mass
splittings can be treated as a perturbation.  We call this
approximately supersymmetric threshold $M$.  As a result, a
model-independent analysis can be set up, based on an approximately
supersymmetric effective theory of the MSSM that includes
higher-dimension operators suppressed by powers of $1/M$.  These
higher-dimension operators can encapsulate different types of physics
beyond the MSSM, such as singlet or triplet Higgses, heavy
$Z^{\prime}$ and $W^{\prime}$, etc.  (we illustrate the detailed
connection in Appendix~\ref{app:UVcompletions}).

We argued, based on the structure of the induced Higgs quartic
couplings, that both the leading and next-to-leading order in the
$1/M$ expansion can be phenomenologically relevant, and computed the
Higgs spectrum and couplings up to this order.  This included a
careful treatment of degenerate cases and the inclusion of kinetic
term renormalization that contains information about the mixing of the
light and heavy degrees of freedom.  The most important effects of the
new physics enter through the angle $\alpha$ that characterizes the
mixing in the CP-even sector, but we have systematically included all
the effects to order $1/M^{2}$.  This allows us to include in our
analysis the recently discussed sEWSB vacua~\cite{Batra:2008rc}, which
depend crucially on certain dimension-6 operators.

We were especially careful to single out points in the effective
theory that can be expected to be reliably described by the EFT. We
also made sure that these points correspond to global minima of the
effective potential (we did not consider the possibility of long-lived
metastable minima).  In addition, we took into account the EW
precision constraints, making sure that the study points can be in
agreement with precision tests when possible effects from squarks and
sleptons not directly related to the Higgs sector are included.

The large class of SUSY models presented in this study has in part
already been explored by various Higgs searches at LEP and the
Tevatron.  We will present these constraints in the accompanying
paper~\cite{CKPZ}.  Similarly, we defer a more complete study of the
Higgs collider phenomenology to that work.  Here we simply pointed out
a few interesting features that arise from our study: a generic
enhancement of the gluon fusion production cross section of the
SM-like Higgs, the presence of new channels with significant branching
fractions, such as $H \rightarrow AA$ and $H^{\pm} \rightarrow W^{\pm}
A$, and the possible suppression of decay modes such as $H \rightarrow
hh$.

We find it interesting that a study of the light Higgs sector can
indirectly reveal the presence of new physics that either may not be
within reach, or may not be easy to produce, as was also emphasized
recently in~\cite{Mantry:2007ar,Randall:2007as}.  The measurement of
the Higgs spectrum and observation of some of their decay modes,
together with the observation of some of the SM superpartners, can
give a striking evidence for a more complicated structure beyond the
MSSM.

\section*{Acknowledgments}
We would like to thank Puneet Batra, Max Rivera and Tim Tait for
collaboration in early stages of this work, and C.~Wagner for reading
the manuscript and for discussions.  E.P. and J.Z would like to thank
the Theory Division of Fermilab for hospitality during their visit.
M.C. and E.P. also thank the Aspen Center for Physics for hospitality
while this project was being completed.  Fermilab is operated by Fermi
Research Alliance, LLC under Contract No.  DE-AC02-07CH11359 with the
U.S. Department of Energy.  K.K. is supported in part by the DOE under
contract DE-AC02-76SF00515.  E.P. is supported by DOE grant
DE-FG02-92ER40699.  The work of J.Z is supported by CONICET, 
Argentina,  and  by a Fermilab Latin American Student Fellowship.

\begin{appendix}

\section{UV completions}
\label{app:UVcompletions}

Here we consider possible UV completions to exemplify how the
operators of the effective theory can be generated.  We consider the
addition of singlets, of vector-like triplets ($Y = \pm 1$), of a
triplet with a Majorana mass and $Y=0$, and an $SU(2)$ gauge extension
under which $H_{u}$ and $H_{d}$ transform as doublets.  This
illustrates how the higher-dimension operators on which the EFT
analysis presented in the main text is based can arise, and also shows
that the coefficients of these operators can be fairly uncorrelated if
the extension is sufficiently complicated.  This justifies our
approach of scanning over these coefficients without taking into
account the correlations that may arise in a given UV model.

\subsection{MSSM + Singlet}
\label{Singlets}

The simplest model we consider shares structural features with the
NMSSM. In the NMSSM, a new singlet $S$ is added to the MSSM to solve
the $\mu$ problem, through a VEV for $S$ and the superpotential
interaction $\lambda_S S H_u H_d$.  In this model, the low energy
theory contains additional scalar states from $S$, and after EWSB the
singlet state mixes with the Higgs.  However, there is a different
region of parameter space, where the singlet $S$ has a mass somewhat
larger than the electroweak scale, and is integrated out of the
effective theory.  Although the $\mu$ problem is not addressed in this
limit, the low energy theory contains modifications in the Higgs
sector that can help solve the little SUSY hierarchy problem.

The superpotential and K\"{a}hler potential are
\begin{eqnarray} 
W &=& \mu H_u H_d + \frac{1}{2} M_S S^2 + \lambda_S S H_u H_d  - X \left( a_1 \mu H_u H_d + \frac{1}{2} a_2 M_{S} S^2 + a_3 \lambda_{S} S H_u H_d  \right)~,
\label{eq:singlet} \\[0.5em]
K&=& H_u^{\dagger} e^V H_u + H_d^{\dagger} e^V H_d + S^{\dagger}S - X^{\dagger}X \left( b_1 H_d^{\dagger} H_d + b_2 H_u^{\dagger} H_u + b_3 S^{\dagger}S\right)~,
\end{eqnarray}
where the $a_{i}$ and $b_{i}$ are dimensionless, and an $\epsilon$
tensor is understood in the contraction $H_{u} H_{d} \equiv H_{u}
\epsilon H_{d}$.  One can consider a cubic coupling $\kappa S^3$, but
it does not contribute to the effective theory up to the order $1/M^2$
we have analyzed, hence we do not write it explicitly in
Eq.~(\ref{eq:singlet}).  The parameters $b_{1}$, $b_{2}$ and $a_{1}$
map into $m^2_{H_{d}}$, $m^2_{H_{u}}$ and the $B\mu$ term,
respectively.  Integrating out $S$ at tree-level induces nonzero
$\omega_{1}$, $\alpha_{1}$, $c_{4}$, $\beta_{4}$ and $\gamma_{4}$ in
Eqs.  (\ref{genw})--(\ref{KSpurionCP}) as follows:
\be
\begin{split}
& M = M_S ~, \qquad \omega_{1} = - \lambda_S ^2 ~, \qquad  \alpha_1 = a_{2} - 2 a_{3}~,
\\
& c_4 =  |\lambda_S|^2 ~, \qquad \gamma_4 = a_{2} - a_{3}~, \qquad \beta_4 =  \left| a_{2} -  a_{3} \right|^2  - b_3~,
\end{split}
\ee
while all other EFT coefficients vanish. 

\subsection{MSSM + $SU(2)_{L}$ Higgs Triplets}
\label{Triplets}

Consider an extension with two $SU(2)_{L}$ triplets, $T$ and
$\bar{T}$, with hypercharges $Y = -1$ and $Y = +1$, respectively.  The
superpotential and \Kahler potential are:
\bea
W &=& \mu H_u H_d + M_T  T \bar{T} + \frac{1}{2} \lam_T H_u T H_u + \frac{1}{2} \lam_{\bar{T}} H_d \bar{T} H_d   \nonumber \\ 
&& \mbox{} - X \left(a_1 \mu H_{u} H_d + a_2 M_T  T \bar{T} + \frac{1}{2} a_3 \lam_T H_u T H_u + \frac{1}{2} a_4 \lam_{\bar{T}} H_d \bar{T} H_d \right)~,
\\ [0.5em]
K &=&  H_d^{\dagger} e^{2V} H_d + H_u^{\dagger} e^{2V} H_u + T^{\dagger} e^{2V} T + \bar{T}^{\dagger} e^{2V} \bar{T} 
\nonumber \\
&& \mbox{} - X X^{\dagger} \left(b_1 H_d^{\dagger}  H_d + b_2 H_u^{\dagger}  H_u+ b_3  T^{\dagger} T + b_4 \bar{T}^{\dagger} \bar{T} \right)~.
\eea
An epsilon tensor is understood in the contractions $H_u T H_u = H_u
\epsilon \, T H_u$, etc.  The parameters $b_{1}$, $b_{2}$ and $a_{1}$
map into $m^2_{H_{d}}$, $m^2_{H_{u}}$ and the $b\mu$ term,
respectively.  Integrating out the triplets, one can write the
effective Lagrangian in terms of the operators defined in Eqs.
(\ref{genw})--(\ref{KSpurionCP}), with~\footnote{Here we use the
identity $\int \!  d^4\theta \, {\cal A}^{\dagger} e^{2V} {\cal A} =
\frac{1}{2} \int \!  d^4\theta \, (H^\dagger e^{2V} H)^2$, where
${\cal A} = {\cal A}^a X^a$ with ${\cal A}^{a} = H \epsilon \, \tau^a
H$, and $X^a$ are the $SU(2)$ generators in the adjoint representation
while $H$ is in the fundamental representation of $SU(2)$.  One also
has $\int \!  d^4\theta \, {\cal A}^{\dagger} e^{2V} {\cal A} = \int
\!  d^4\theta \, \{\frac{1}{2} (H^\dagger_{u} e^{2V} H_{u})
(H^\dagger_{d} e^{2V} H_{d}) - \frac{1}{4} |H_{u} \epsilon H_{d}|^2
\}$ for ${\cal A}^{a} = H_{u} \epsilon \, \tau^a H_{d}$.}
\be
\begin{split}
& M = M_T ~, \qquad \qquad \hspace{-4.5mm} \omega_{1} = \frac{1}{4} \lambda_T \lambda_{\bar{T}} ~, \qquad  \alpha_1 = a_{2} - a_{3} - a_{4}~,
\\
& c_1 =  \frac{1}{4} \lvert \lam_{\bar{T}} \rvert ^2~, \qquad \gamma_1= a_2-a_4~, \qquad \beta_1=  |a_2-a_4|^2 -b_3~,
\\
& c_2 =  \frac{1}{4} \lvert \lam_T \rvert ^2~, \qquad \gamma_2= a_2-a_3~, \qquad \beta_2= |a_2-a_3|^2 - b_4~,
\end{split}
\ee
and all other dimensionless coefficients vanishing.

If on the other hand one considers an extension with a single
$SU(2)_{L}$ triplet with hypercharge $Y = 0$ and a Majorana mass
$M_{T}$, the superpotential and \Kahler potential are:
\bea
W &=& \mu H_u H_d + \frac{1}{2} M_T  T^2 + \lam_T H_u T H_d - X \left(a_1 \mu H_{u} H_d + \frac{1}{2} a_2 M_T  T^2 + a_3 \lam_T H_u T H_d \right)~,
\\ [0.5em]
K &=&  H_d^{\dagger} e^{2V} H_d + H_u^{\dagger} e^{2V} H_u + T^{\dagger} e^{2V} T
- X X^{\dagger} \left(b_1 H_d^{\dagger}  H_d + b_2 H_u^{\dagger}  H_u+ b_3  T^{\dagger} T \right)~,
\eea
and after integrating out the triplet, one obtains the effective
theory of Eqs.~(\ref{genw})--(\ref{KSpurionCP}), with
\be
\begin{split}
& M = M_T ~, \qquad \hspace{4mm} \omega_{1} = - \frac{1}{4} \lambda_T^2~, \qquad  \alpha_1 = a_{2} - 2 a_{3}~,
\\
& c_3 = \frac{1}{2} |\lambda_T|^2 ~, \qquad \gamma_3 = a_{2} - a_{3}~, \qquad \beta_3 =  \left| a_{2} -  a_{3} \right|^2  - b_3~,
\\
& c_4 = -\frac{1}{4} |\lambda_T|^2 ~, \qquad  \hspace{-3mm} \gamma_4 = a_{2} - a_{3}~, \qquad \beta_4 =  \left| a_{2} -  a_{3} \right|^2  - b_3~,
\end{split}
\ee
while all other EFT coefficients vanish.

\subsection{$SU(2)$ Extensions}
\label{Wprimes}

The higher-dimension operators induced by integrating out a massive
$U(1)^\prime$ were worked out in Ref.~\cite{Dine:2007xi}.  Here we consider a
product gauge group $SU(2)_{1} \times SU(2)_{2} \times U(1)_{Y}$ with
gauge couplings $g_{1}$, $g_{2}$ and $g'$, respectively.  We
concentrate on the $SU(2)_{1} \times SU(2)_{2}$ factor since the
$U(1)_{Y}$ factor enters in a trivial way.  The MSSM Higgs superfields
are assumed to transform like $(\textbf{2},\textbf{0})$ under
$SU(2)_{1} \times SU(2)_{2}$.  The gauge group is broken down to the
diagonal by a $\Sigma$ field transforming like
$(\textbf{2},\textbf{2})$ under $SU(2)_{1} \times SU(2)_{2}$, which
gets a nonzero VEV, $\langle \Sigma \rangle \propto \sigma^2$.  In
this case, the kinetic term for $\Sigma$, ${\rm Tr}(e^{2g_{2} V_{2}^T}
\Sigma^\dagger e^{2g_{1} V_{1}}\Sigma)$, leads to the structure ${\rm
Tr}(e^{2g_{2} V_{2}} e^{2g_{1} V_{1}})$, which contains a mass term for
the linear combination $V' \equiv (g_{1} V_{1} + g_{2}
V_{2})/\sqrt{g^2_{1} + g^2_{2}}$.  The orthogonal combination, $V
\equiv (g_{2} V_{1} - g_{1} V_{2})/\sqrt{g^2_{1} + g^2_{2}}$ remains
massless and can be identified with the SM $W$ gauge bosons.
Therefore, we take as the starting point the \Kahler potential
\bea
K &=& H_{u}^\dagger \, e^{2g_{1} V_{1}} H_{u} + H_{d}^\dagger \, 
e^{2g_{1} V_{1}} H_{d} + \frac{M^2_{V'}}{2(g^2_{1} + g^2_{2})} 
{\rm Tr}\left[ e^{2g_{2} V_{2}} e^{2g_{1} V_{1}} \right]~,
\label{KSU2}
\eea
where $g_{1} V_{1} = \tilde{g} V' + g V$, $g_{2} V_{2} =
(g^2_{2}/g^2_{1}) \tilde{g} V' - g V$ and $g = g_{1}
g_{2}/\sqrt{g^2_{1} + g^2_{2}}$ is the SM $SU(2)_{L}$ gauge coupling
while $\tilde{g} = g_{1}^2/\sqrt{g^2_{1} + g^2_{2}}$ is the coupling
of the massive $V'$.  The term of second order in $V'$ in the last
term of Eq.~(\ref{KSU2}) identifies $M_{V'}$ as the gauge boson mass.
In order to integrate out $V'$ to order $1/M^2_{V'}$, it is sufficient
to keep terms up to quadratic order in $V'$ in the ``mass term", while
keeping up to linear order in the terms not proportional to
$M^2_{V'}$.  We can also assume that $V$ is in the Wess-Zumino gauge
so that we can expand to quadratic order in $V$ everywhere.  Note that
the term proportional to $M^2_{V'}$ contains terms in addition to the
pure mass term for $V'$:
\bea
\frac{M^2_{V'}}{2(g^2_{1} + g^2_{2})} \int \! d^4\theta \, 
{\rm Tr}\left[ e^{2[(g^2_{2}/g^2_{1}) \tilde{g} V' - g V]} e^{2(\tilde{g}V' + gV)} \right] &=& 
\nonumber \\
& & \hspace{-5cm}
\frac{1}{2} M^2_{V'}
\int \! d^4\theta \, \left\{ V^{\prime a} V^{\prime a} - \frac{g^2}{3} \left[ \left( V^a V^{\prime a}\right)^2 -\left( V^a V^{a} \right) \left( V^{\prime b} V^{\prime b} \right)   \right] \right\}~. 
\label{WprimeMass}
\eea
The terms quartic in the vector multiplets are essential for
maintaining the low-energy gauge invariance.\footnote{These terms are
not present in the $U(1)^\prime$ case.  For instance gauging only the
$U(1)_{2} \subset SU(2)_{2}$ generated by $\tau^3$, i.e. setting
$V^{1}_{2} = V^{2}_{2} = 0$, the last two terms in
Eq~(\ref{WprimeMass}) cancel out.} The equation of motion for $V'$, to
order $1/M^2_{V'}$, gives
\bea
V^{\prime a} = - \frac{2\tilde{g}}{M^2_{V'}} \sum_{i = u,d} H^\dagger_{i} 
\left\{ \left( 1-\frac{g^2}{6} V^b V^b \right) \tau^{a} + 
\frac{1}{6} g V^{a} \left( 3 + 2g V^{a} \tau^{a} + 2g V^b \tau^{b} \right) \right\} H_{i}~,
\eea
where there is no summation over $a$ (but there is over $b$), and
$\tau^{a}$ are the $SU(2)$ generators in the fundamental
representation.  Replacing back in the action leads to the effective
\Kahler potential
\bea
K_{\rm eff} &=&H^\dagger_{u} \, e^{2gV} H_{u} + H^\dagger_{d} \, 
e^{2gV} H_{d} - \frac{\tilde{g}^{2}}{2M^2_{V'}} 
\left\{ \left( H^\dagger_{u} \, e^{2gV} H_{u} + H^\dagger_{d} \, e^{2gV} H_{d} \right)^2 
- 4 \left| H_{u} \epsilon H_{d} \right|^2 \right\}~,
\eea
where we restored the terms in $V$ into the exponential form, and the
hypercharge vector multiplet can be put back in a trivial way.
Therefore, the massive $W'$ induces the operators of
Eqs.~(\ref{genw})--(\ref{KSpurionCP}) with coefficients
\be
\begin{split}
& c_1 =  - \tilde{g}^{2}~, \qquad c_2 =  - \tilde{g}^{2}~, 
\qquad c_3 =  -\tilde{g}^{2}~,\qquad c_4 =  2 \tilde{g}^{2}~,
\end{split}
\ee
$M = M_{V'}$, and all other EFT coefficients vanishing.

There are also excitations of the $\Sigma$ field (a SM singlet and
triplet) with masses proportional to $\langle \Sigma \rangle$.  These
may be split by SUSY breaking and generate further contributions to
the dimension-6 operators in the effective \Kahler potential.  These
were considered in Ref.~\cite{SU2} in the opposite limit that
interests us here, namely when SUSY breaking is comparable to the SUSY
preserving contribution.

For completeness, we also record here the SUSY-preserving operators
induced by integrating out a massive $U(1)^\prime$ gauge field, i.e.
integrating out $V^\prime$ from
\bea
K &=& H_{u}^\dagger \, e^{2Q_{u} g^\prime V^\prime} H_{u} + H_{d}^\dagger \, 
e^{2Q_{d} g^\prime V^\prime} H_{d} + \frac{1}{2} M^2_{V'} V^{\prime 2}~,
\label{KU1}
\eea
where $g^\prime$ is the $U(1)^\prime$ gauge coupling and $Q_{u,d}$ are
the $U(1)^\prime$ charges of $H_{u,d}$, respectively (for simplicity,
we have omitted the SM gauge factors, which are implicitly
understood).  Unlike in the non-abelian case, the $c_{4}$ operator is
not induced, while
\be
\begin{split}
& c_1 =  -4Q^2_{d} g_{1}^{\prime 2}~, \qquad c_2 =  -4Q^2_{u} g^{\prime 2}~, 
\qquad c_3 =  -4Q_{u} Q_{d} g^{\prime 2}~.
\end{split}
\ee
From the examples above, it should be clear that MSSM extensions
including singlet and triplet Higgses, plus abelian or non-abelian
gauge factors can generate all of the operators we considered in the
main text, with essentially arbitrary coefficients, except for the
$c_{6}$ and $c_{7}$ operators.  We have nevertheless included the latter
operators in our phenomenological analysis, since they are allowed by
supersymmetry.

\section{Comments on Custodial Symmetry}
\label{app:custodial}

Neglecting the Yukawa couplings, the Higgs sector in the MSSM can be
written in terms of a chiral superfield
\bea
\Phi =
\left(
\begin{array}{ccc}
H^{0}_d & H^+_{u}   \\
H^-_{d} & H^{0}_u          
\end{array}
\right)
\nonumber
\eea
as 
\bea
K &\supset& H^\dagger_{u} \, e^{2gW+g^{\prime}B} H_{u} + H^\dagger_{d} \, e^{2gW-g^{\prime}B} H_{d} 
\,\,=\,\,  {\rm Tr} \left\{ \Phi^{\dagger} e^{2gW} \Phi
\left( \!\!
\begin{array}{cc}
e^{-g^{\prime}B} & 0 \\
0 & e^{g^{\prime}B} \\  
\end{array}
\!\!\right) 
\right\}~,
\nonumber \\
W &\supset& \mu H_{u} \epsilon H_{d} \,\,=\,\, \frac{1}{2} {\rm Tr} \, \epsilon^{T} \Phi^{T} \epsilon \Phi~,
\nonumber
\eea
where $W$ and $B$ are the $SU(2)_{L}$ and $U(1)_{Y}$ vector
superfields, while $\epsilon$ is the antisymmetric 2-index tensor.
This shows that in the limit $g^{\prime} \rightarrow 0$ (and
neglecting Yukawa couplings), the theory is invariant under $SU(2)_{L}
\times SU(2)_{R}$ where $\Phi \rightarrow U_{L} \Phi U^{\dagger}_{R}$.
The $SU(2)_{L}$ factor is gauged and can be complexified.  The global
$SU(2)_{R}$ implies that in the limit $v_{u} = v_{d}$ the theory has a
custodial $SU(2)_{L+R}$ global symmetry.  This symmetry has a somewhat
limited use since it is broken away from $\tan\beta =1$.

However, at tree-level (and for $g^{\prime} = 0$) the Higgs (scalar)
sector of the MSSM exhibits an $SU(2)^{\rm local}_{L} \times
SU(2)^{\rm global}_{R}$ symmetry such that the Higgs scalar components
\bea
\phi_{u} =
\frac{1}{\sqrt{2}}
\left(
\begin{array}{ccc}
H^{0*}_u & H^+_{u}   \\
-H^-_{u} & H^{0}_u          
\end{array}
\right)~,
\hspace{1cm}
\phi_{d} =
\frac{1}{\sqrt{2}}
\left(
\begin{array}{ccc}
H^{0}_d & -H^+_{d}   \\
H^-_{d} & H^{0*}_d          
\end{array}
\right)~,
\nonumber
\eea
transform as $\phi_{u,d} \rightarrow U_{L} \phi_{u,d}
U^{\dagger}_{R}$.  In fact, if the Higgsinos are taken to be singlets
under the $SU(2)_{R}$ transformation, only the Higgs-Higgsino-gaugino
couplings (apart from the superpotential Yukawa terms) violate this
global symmetry.  Thus, for arbitrary expectation values, $v_{u}$ and
$v_{d}$, contributions to the $T$ parameter enter only at one loop
level.

The fact that in the MSSM the tree-level gauge boson masses satisfy
the SM relation $\rho = 1$ can be seen directly from the identities
$H^\dagger_{u} \, (2gW)^{2} H_{u} = {\rm Tr} \, \Phi^{\dagger}_{u}
(2gW)^{2} \Phi_{u}$ and $H^\dagger_{d} \, (2gW)^{2} H_{d} = {\rm Tr}
\, \Phi^{\dagger}_{d} (2gW)^{2} \Phi_{d}$, which lead to the gauge
boson mass terms in the \Kahler potential after replacing the chiral
superfields by their scalar components, $\Phi_{u,d} \rightarrow
\phi_{u,d}$.  This observation also implies that higher-dimension
operators such as $(H^\dagger_{u} \, e^{2V} H_{u}) (H_{u} \epsilon
H_{d} + {\rm h.c.})$ or $(H^\dagger_{d} \, e^{2V} H_{d}) (H_{u}
\epsilon H_{d} + {\rm h.c.})$, i.e. those in Eq.~(\ref{DeltaKCP}),
lead to $\rho = 1$ at tree-level.  On the other hand, the
higher-dimension operators of Eq.~(\ref{DeltaKCV}) contribute to the
gauge boson mass terms through the linear term in $e^{2V} = 1 + 2V +
\cdots$, and these terms do \textit{not} respect a custodial symmetry.
This was seen explicitly in Eq.~(\ref{alphaT}).  As shown above, an
exception is the operator $(H^\dagger_{u} \, e^{2V} H_{u} +
H^\dagger_{d} \, e^{2V} H_{d})^{2}$, in the limit $v_{u} = v_{d}$.
This operators is naturally induced by massive $W$ primes, as shown in
\ref{Wprimes}.

\section{Charginos and Neutralinos}
\label{sec:inos}

Although our focus is on the scalar Higgs sector, neutralinos and
charginos can provide important decay channels, and the effects of the
higher-dimension operators on their masses play an important role.  We
have implemented in HDECAY the corresponding mass formulas to order
$1/M$, as computed in \cite{Batra:2008rc}.  At order $1/M^2$ there are
additional contributions to the mass matrix as well as to the kinetic
terms, that are needed to compute the physical chargino/neutralino
masses.  Given that these states tend to be near the experimental
limits, it would be interesting to compute these next order
corrections, but we postpone it to future work.  For completeness, we
collect the mass matrices to order $1/M$.

The chargino mass matrix is
\bea
\left(\tilde{W}^+, \tilde{H}_u^+ \right) \left( \begin{array}{cc} M_2 & \sqrt{2} m_W c_{\beta} \\ \sqrt{2} m_W s_{\beta} & \mu \left( 1 - \rho s_{2 \beta} \right) \end{array} \right) \left(\begin{array}{c} \tilde{W}^- \\ \tilde{H}_d^- \end{array} \right)~,
\label{eq:charginomass}
\eea
where $\rho \equiv \omega_{1} v^2/(4\mu M)$ takes into account the
effects from the heavy physics.  The neutralino mass matrix is
\bea
\half \left(\tilde{B},\tilde{W}^3,   \tilde{H}_d^0, \tilde{H}_u^0\right) \left( \begin{array}{cccc} M_1 &  &- m_Z s_W c_{\beta} &m_Z s_W s_{\beta} \\ & M_2   & m_Z c_W c_{\beta} &-m_Z c_W s_{\beta}\\  -m_Z s_W c_{\beta}  &m_Z c_W c_{\beta}& 2 \mu \rho s_\beta^2 & -\mu \left( 1-2 \rho s_{2 \beta}\right) \\ m_Z s_W s_{\beta} &-m_Z c_W s_{\beta}&  -\mu \left( 1-2 \rho s_{2 \beta}\right) & 2 \mu \rho c_\beta^2 \end{array} \right)   \left(\begin{array}{c}\tilde{B}\\\tilde{W}^3\\   \tilde{H}_d^0\\ \tilde{H}_u^0 \end{array} \right), \nn \\
\label{eq:neutralinomass}
\eea
where $c_W$ stands for the weak-mixing angle $\cos \theta_W$.  $M_{1}$
and $M_{2}$ are the $SU(2)_{L}$ and $U(1)_{Y}$ gaugino soft breaking
parameters.

\section{Higgs couplings}
\label{app:couplings}

The couplings between the $Z$ gauge boson and two Higgs fields,
defined in Eq.~(\ref{eq:trilinear}), are given by
\bea
\eta_{ZhA} &=& c_{\alpha - \beta} \left\{ 1 - \frac{1}{2} \left[ \left(A_{1} + E_{1} \right) c_{\gamma} - B_{1} s_{\gamma}  \right] \right\} + \frac{1}{2} \left[ \left(D_{1} + E_{1} \right) s_{\gamma} - B_{1} c_{\gamma} \right] s_{\alpha - \beta}
\nonumber \\ [0.5em]
&& \mbox{} + c_{\gamma} \delta \eta_{ZhA} -s_{\gamma} \delta \eta_{ZHA}~, 
\nonumber \\ [0.5em]
\eta_{ZAh} &=& c_{\alpha - \beta} \left\{ 1 - \frac{1}{2} \left[ \left(A_{1} + E_{1} \right) c_{\gamma} - B_{1} s_{\gamma}  \right] \right\} + \frac{1}{2} \left[ \left(D_{1} + E_{1} \right) s_{\gamma} - B_{1} c_{\gamma} \right] s_{\alpha - \beta}
\nonumber \\ [0.5em]
&& \mbox{} + c_{\gamma} \delta \eta_{ZAh} -s_{\gamma} \delta \eta_{ZAH}~, 
\\ [0.5em]
\eta_{ZHA} &=& s_{\alpha - \beta} \left\{ 1 - \frac{1}{2} \left[ \left(D_{1} + E_{1} \right) c_{\gamma} + B_{1} s_{\gamma}  \right] \right\} - \frac{1}{2} \left[ \left(A_{1} + E_{1} \right) s_{\gamma} + B_{1} c_{\gamma} \right] c_{\alpha - \beta}
\nonumber \\ [0.5em]
&& \mbox{} + c_{\gamma} \delta \eta_{ZHA} + s_{\gamma} \delta \eta_{ZhA}~, 
\nonumber \\ [0.5em]
\eta_{ZAH} &=& s_{\alpha - \beta} \left\{ 1 - \frac{1}{2} \left[ \left(D_{1} + E_{1} \right) c_{\gamma} + B_{1} s_{\gamma}  \right] \right\} - \frac{1}{2} \left[ \left(A_{1} + E_{1} \right) s_{\gamma} + B_{1} c_{\gamma} \right] c_{\alpha - \beta}
\nonumber \\ [0.5em]
&& \mbox{} + c_{\gamma} \delta \eta_{ZAH} + s_{\gamma} \delta \eta_{ZAh}~, 
\nonumber \\ [0.4em]
\delta \eta_{ZH^+H^-} &=& 1 - F_{1} + \delta\eta_{ZH^+H^-}~,
\nonumber
\eea
where $\gamma$ is defined by Eq.~(\ref{gamma}), $A_{1}$, $B_{1}$,
$D_{1}$, $F_{1}$ and $E_{1}$ are defined in Eq.~(\ref{ABDEF}), and
\bea
\delta \eta_{ZhA} &=& \frac{v^{2}}{M^{2}} s_{\beta} \left\{ 
3 \tb^{-1} \sb^2 \left( c_{2} \ca + c_{1} \tb^{-1} \sa \right) 
+ \tb^{-1} \sb^2 \left[ c_{6} (2 \sa - \tb^{-1} \ca)  - c_{7} (\sa - 2 \tb^{-1} \ca) \right] \rule{0mm}{6mm} \right\}~, 
\nonumber \\ [0.4em]
\delta \eta_{ZAh} &=& \frac{v^{2}}{M^{2}} s_{\beta} \left\{ 
\tb^{-1} \sb^2 \left( c_{2} \ca + c_{1} \tb^{-1} \sa \right) 
+ \frac{1}{2} \, c_{6} \tb^{-1} (3 - c_{2\beta}) \sa  + \frac{1}{2} \, c_{7} (3 + c_{2\beta}) \ca \rule{0mm}{6mm} \right\}~, 
\nonumber \\ [0.5em]
\delta \eta_{ZHA} &=& \frac{v^{2}}{M^{2}} s_{\beta} \left\{ 
3 \tb^{-1} \sb^2 \left( c_{2} \sa - c_{1} \tb^{-1} \ca \right) 
- \tb^{-1} \sb^2 \left[ c_{6} (2 \ca + \tb^{-1} \sa)  - c_{7} (\ca + 2 \tb^{-1} \sa) \right] \rule{0mm}{6mm} \right\}~, 
\nonumber \\ [0.4em]
\delta \eta_{ZAH} &=& \frac{v^{2}}{M^{2}} s_{\beta} \left\{ 
\tb^{-1} \sb^2 \left( c_{2} \sa - c_{1} \tb^{-1} \ca \right) 
- \frac{1}{2} \, c_{6} \tb^{-1} (3 - c_{2\beta}) \ca  + \frac{1}{2} \, c_{7} (3 + c_{2\beta}) \sa \rule{0mm}{6mm} \right\}
\\[0.4em]
\delta\eta_{ZH^+H^-} &=& \frac{v^{2}}{M^{2}} \sb^4 \left\{ 
\frac{1}{4} \frac{c^2_{W}}{c_{2W}} c_{3} \left( 1 + 3 \sb^{-4} - 6 \tb^{-2} + \tb^{-4} \right)
+ \tb^{-1} \left[ c_{6} - \frac{s^2_{W}}{c_{2W}} (c_{1} + c_{2}) \tb^{-1} + c_{7} \tb^{-2} \right]
\right\}
~.
\nonumber
\eea
Similarly, the couplings between the $W$ and two Higgs fields, defined
in Eq.~(\ref{eq:trilinear}), are given by
\bea
\eta_{W^\pm h H^\mp} &=& c_{\alpha - \beta} \left\{ 1 - \frac{1}{2} \left[ \left(A_{1} + F_{1} \right) c_{\gamma} - B_{1} s_{\gamma}  \right] \right\} + \frac{1}{2} \left[ \left(D_{1} + F_{1} \right) s_{\gamma} - B_{1} c_{\gamma} \right] s_{\alpha - \beta}
\nonumber \\ [0.5em]
&& \mbox{} + c_{\gamma} \delta \eta_{W^\pm h H^\mp} -s_{\gamma} \delta \eta_{W^\pm H H^\mp}~, 
\nonumber \\ [0.5em]
\eta_{W^\pm H^\mp h} &=& c_{\alpha - \beta} \left\{ 1 - \frac{1}{2} \left[ \left(A_{1} + F_{1} \right) c_{\gamma} - B_{1} s_{\gamma}  \right] \right\} + \frac{1}{2} \left[ \left(D_{1} + F_{1} \right) s_{\gamma} - B_{1} c_{\gamma} \right] s_{\alpha - \beta}
\nonumber \\ [0.5em]
&& \mbox{} + c_{\gamma} \delta \eta_{W^\pm H^\mp h} -s_{\gamma} \delta \eta_{W^\pm H^\mp H}~, 
\\ [0.5em]
\eta_{W^\pm H H^\mp} &=& s_{\alpha - \beta} \left\{ 1 - \frac{1}{2} \left[ \left(D_{1} + F_{1} \right) c_{\gamma} + B_{1} s_{\gamma}  \right] \right\} - \frac{1}{2} \left[ \left(A_{1} + F_{1} \right) s_{\gamma} + B_{1} c_{\gamma} \right] c_{\alpha - \beta}
\nonumber \\ [0.5em]
&& \mbox{} + c_{\gamma} \delta \eta_{W^\pm H H^\mp} + s_{\gamma} \delta \eta_{W^\pm h H^\mp}~, 
\nonumber \\ [0.5em]
\eta_{W^\pm H^\mp H} &=& s_{\alpha - \beta} \left\{ 1 - \frac{1}{2} \left[ \left(D_{1} + F_{1} \right) c_{\gamma} + B_{1} s_{\gamma}  \right] \right\} - \frac{1}{2} \left[ \left(A_{1} + F_{1} \right) s_{\gamma} + B_{1} c_{\gamma} \right] c_{\alpha - \beta}
\nonumber \\ [0.5em]
&& \mbox{} + c_{\gamma} \delta \eta_{W^\pm H^\mp H} + s_{\gamma} \delta \eta_{W^\pm H^\mp h}~, 
\nonumber \\ [0.5em]
\eta_{W^\pm H^\mp A} &=& 1 - \frac{1}{2} \left(E_{1} + F_{1} \right) 
+ \delta \eta_{W^\pm H^\mp A}~,
\nonumber \\ [0.5em]
\eta_{W^\pm A H^\mp} &=& 1 - \frac{1}{2} \left(E_{1} + F_{1} \right) 
+ \delta \eta_{W^\pm A H^\mp}~,
\nonumber
\eea
where 
\bea
\delta \eta_{W^\pm h H^\mp} &=& \frac{v^{2}}{M^{2}} 
\sb^3 \left\{ \frac{1}{8} c_{3} \left[ \left(3 + \sb^{-2} - 9\tb^{-1} \right) \sa - \left( \tb^{-1} (9 - \sb^{-2}) - 3 \tb^{-3} \right) \ca \right] \right.
\nonumber \\[0.4em]
&& \hspace{1.4cm} \left. \mbox{}
+ \frac{3}{2} \tb^{-1} \left( c_{1} \tb^{-1} \sa + c_{2} \ca \right)
\rule{0mm}{6mm} \right\}~, 
\nonumber \\ [0.5em]
\delta \eta_{W^\pm H^\mp h} &=& \frac{v^{2}}{2M^{2}} 
\sb^3 \left\{ c_{3} \left( \sa + \tb^{-3} \ca \right) + \tb^{-1} \left( c_{1} \tb^{-1} \sa + c_{2} \ca \right) \right\}~,
\nonumber \\ [0.5em]
\delta \eta_{W^\pm H H^\mp} &=& \frac{v^{2}}{2M^{2}} 
\sb^3 \left\{ \frac{1}{4} c_{3} \left[ \left(3 + \sb^{-2} - 9\tb^{-2} \right) \ca + \left( \tb^{-1} (9 - \sb^{-2}) - 3 \tb^{-3} \right) \sa \right] \right.
\\[0.4em]
&& \hspace{1.4cm} \left. \mbox{}
+ 3 \tb^{-1} \left( c_{1} \tb^{-1} \ca - c_{2} \sa \right)
\rule{0mm}{6mm} \right\}~, 
\nonumber \\ [0.5em]
\delta \eta_{W^\pm H^\mp H} &=& \frac{v^{2}}{2M^{2}} 
\sb^3 \left\{ c_{3} \left( \ca - \tb^{-3} \sa \right) + \tb^{-1} \left( c_{1} \tb^{-1} \ca - c_{2} \sa \right) \right\}~,
\nonumber \\ [0.5em]
\delta \eta_{W^\pm H^\mp A} &=& \frac{v^{2}}{2M^{2}} 
\sb^4 \left\{ 3(c_{1} + c_{2}) \tb^{-2} - \frac{1}{4} c_{3} \left( 1 - 5 \sb^{-4} - 6 \tb^{-2} + \tb^{-4} \right) \right\}~,
\nonumber \\ [0.5em]
\delta \eta_{W^\pm A H^\mp} &=& \frac{v^{2}}{8M^{2}} 
\sb^4 \left\{ 4 \tb^{-2} (c_{1} + c_{2}) + c_{3} \left( 1 + 3 \sb^{-4} - 6 \tb^{-2} + \tb^{-4} \right) \right\}~.
\nonumber
\eea

\end{appendix}



\end{document}